\documentclass[fleqn,10pt]{wlscirep}
\usepackage[utf8]{inputenc}
\usepackage{etoolbox}
\usepackage{graphicx}
\usepackage{float}
\usepackage{geometry}
\usepackage{csvsimple}
\usepackage{booktabs}
\usepackage{subcaption}
\usepackage{longtable}
\usepackage{caption}
\usepackage{afterpage}
\usepackage{authblk}
\usepackage{mwe}
\usepackage{adjustbox}
\usepackage{multirow}
\usepackage[bottom]{footmisc}
\usepackage{subfiles}
\usepackage{xurl}
\usepackage{setspace}

\usepackage{dcolumn}
\usepackage{amsmath}
\usepackage[T1]{fontenc}
\usepackage{color}
\usepackage{bbm}
\usepackage{booktabs} 
\usepackage{babel}

\makeatletter
\expandafter\let\csname ver@cite.sty\endcsname\relax
\let\citestyle\@gobble
\makeatother

\usepackage[maxbibnames=150,style=numeric-comp,sorting=none]{biblatex}
\bibliography{reference.bib}

\title{\vspace{-2.5em} 
The Innovation Recognition Paradox: How Science Undervalues the Boundary-Crossing Work Women Produce
}

\author[1]{Carolina Biliotti}
\author[2,3]{Massimo Riccaboni}
\author[4]{Jeffrey W. Lockhart}
\author[5,6]{James A. Evans}

\affil[1]{\small Department of Economics, European University Institute, San Domenico di Fiesole (FI), Italy}
\affil[2]{\small AXES, IMT School for Advanced Studies Lucca, Italy}
\affil[3]{\small IUSS, Pavia, Italy}
\affil[4]{\small Department of Sociology, University of California, Berkeley, USA}
\affil[5]{\small Department of Sociology \& Knowledge Lab, University of Chicago, Illinois, USA}
\affil[6]{\small Santa Fe Institute, New Mexico, USA}

\newcommand{\abs}[1]{\lvert #1 \rvert}
\DeclareUnicodeCharacter{0301}{*************************************}
\newcommand{\rvs}[1]{{\textcolor{black}{#1}}}
\newcommand{\rvss}[1]{{\textcolor{black}{#1}}}

\usepackage{graphicx} 
\usepackage[maxbibnames=150,style=numeric-comp,sorting=none]{biblatex}
\bibliography{reference.bib}

\usepackage{url}
\usepackage{comment}
\usepackage{setspace}
\usepackage{multirow}
\usepackage{booktabs}
\usepackage{threeparttable}
\usepackage{csquotes}
\usepackage[utf8]{inputenc}
\usepackage[T1]{fontenc}
\usepackage{color}
\usepackage{bbm}

\date{}

\begin{abstract}  

Women and men pursue different but complementary forms of scientific innovation. Analyzing 261,452 solo-authored papers by U.S. scholars, with patterns confirmed by millions of multi-authored articles, we show that women more often bridge distant disciplines through novel reference combinations, while men more often recombine concepts within fields. Women's interdisciplinary innovations prove more disruptive and more prescient—yet science penalizes them for it. For equally innovative work, women's papers land in lower-prestige journals and tend to receive less downstream citation credit, though their disruptive impact is greater. These gaps narrow only at extreme levels of novelty, suggesting women must produce exceptionally surprising work to achieve parity. Men's within-field concept innovations, by contrast, attract recognition from disciplinary gatekeepers who control careers. The asymmetry reveals not a deficit in women's contributions but a reward structure that systematically undervalues the boundary-crossing work most likely to transform fields.

\end{abstract}

\begin{document}
\flushbottom
\maketitle 

\section{Introduction}


Scientific innovation requires novelty, ideas and insights that diverge from established knowledge \supercite{Zhao2025ARO}. Yet novelty itself is not monolithic. Innovation in science operates along two fundamental dimensions, each representing a distinct pathway to advancing knowledge. \textit{Content or concept novelty} involves recombining concepts, methods, and findings \emph{within} research domains \supercite{Shi2023, doi:10.1142/S0219525921500168} -- the kind of within-field advances that earn status and awards from disciplinary communities. \textit{Context or reference novelty} connects disparate disciplines by drawing on literature from distant fields \supercite{doi:10.1126/science.1240474, WANG20171416} -- the interdisciplinary bridging that originates from the periphery yet proves most disruptive to established paradigms \supercite{Shi2023}. 


These represent fundamentally different innovation strategies with distinct implications for scientific progress. Content novelty reflects mastery within disciplinary boundaries, signaling deep expertise through new connections among ideas that disciplinary gatekeepers are most likely to recognize and reward \supercite{Shi2023, szell2018nobel}. Context novelty, by contrast, signals boundary-crossing that can read as lack of focus to traditional, disciplinary evaluators, even as it produces the disruptive insights that transform and spawn new fields \supercite{Foster2021SurpriseMN}. Novel context combinations reflect violations of boundaries between fields and disciplines \supercite{LEE2015684, doi:10.1073/pnas.2200841119}, indicating interdisciplinary work \supercite{10.1162/qss_a_00191, doi:10.1177/0001839216665364}. Crucially, these two forms of innovation attract different audiences: content novelty garners citations from within a field, while context novelty draws attention from across disciplinary boundaries \supercite{Shi2023}. They also yield different impacts: content novelty advances within domains; context novelty reorganizes relationships between domains. 

While both approaches to innovation may coexist in some work, separating these two kinds of innovation gives us a more complete picture of how innovation operates in scientific research \supercite{Shi2023, doi:10.1177/01655515231161133}. If men and women differ in the strategies they pursue, and rewards favor one type over the other, persistent gender gaps may reflect institutional and cultural values rather than differences in productivity or quality, with implications for equality and fostering innovation.

Existing literature on gender and scientific innovation documents persistent disparities in the kinds of ideas pursued, the reception of those ideas, and the career returns to innovation. On average, men and women adopt different approaches to innovation \supercite{RePEc:nat:nathum:v:1:y:2017:i:11:d:10.1038_s41562-017-0235-x, doi:10.1073/pnas.2113067119}, influenced by gender expectations \supercite{https://doi.org/10.1002/job.2500}
 and fear of social backlash, which, under some circumstances, can lead women to avoid selection of overly unusual ideas \supercite{doi:10.1287/orsc.2022.16176}. Women's novel contributions in science often experience lower uptake \supercite{doi:10.1073/pnas.1915378117, doi:10.1086/725551}, likely resulting in slower career advancement \supercite{Ginther2004WomenIE, Skibba2019WomenIP} and under-representation in academic leadership \supercite{Huang2019HistoricalCO, Lerchenmuellerl6573, 10.1093/ej/ueac032, Bendels2017GenderEI}. Ill-structured reward systems may further reduce creativity, sometimes disproportionately for women \supercite{doi:10.5465/AMPROC.2024.12bp}.

Much of the literature on gender gaps in innovation assesses aggregate indicators such as total citations \supercite{Ross2022}, disruption \supercite{ZHANG2024101520}, or journal placement \supercite{10.1371/journal.pbio.2004956, 10.1371/journal.pone.0189136}, which can obscure heterogeneity in contributions. Evidence on novel content combinations is mixed: women show higher novelty in patents \supercite{4df1dc57518f47f18789e3b499c1cfcc} and U.S. PhD theses \supercite{doi:10.1073/pnas.1915378117}, but lower rates in biomedical dissertations \supercite{LIU2024103743}. Other research examines novel combinations of references across fields \supercite{doi:10.1073/pnas.2200841119, 10.1162/qss_a_00335}. 


Some of these innovation measures are bespoke, and most research does not evaluate comparable forms. The most comprehensive prior attempt to link diversity, innovation, and scientific rewards is Hofstra et al. \supercite{doi:10.1073/pnas.1915378117}, who analyzed $\sim$1.2 million U.S. doctoral dissertations and showed that underrepresented groups introduce novel concept pairs at higher rates, yet their innovations receive less uptake and yield lower career returns—a ``diversity–innovation paradox.'' This landmark finding established the broad phenomenon, but its design leaves critical questions unresolved. Hofstra et al.\ measure only one form of novelty---new concept co-occurrences extracted from dissertation abstracts---conflating what we distinguish as content and context innovation. Their measure of adoption is raw future co-occurrence counts, with a separate measure of how `distal' those connections are, rather than a unified model-based estimate of how surprising a combination was relative to evolving scientific expectations. And their outcome is career attainment (continued publishing, becoming a PhD supervisor), not paper-level recognition, leaving open whether the paradox operates through journal placement, citation credit, disruptive impact on specific contributions, or some other pathway. As we show, distinguishing content from context novelty does not merely refine the paradox---it reverses one of its key premises: women do not uniformly innovate more than men. They innovate \emph{differently}, contributing more boundary-spanning reference novelty while men contribute more within-field concept novelty. The reward asymmetry is thus not a blanket penalty for innovation by women but a structural devaluation of the specific kind of innovation women disproportionately produce.

\subsection{Our Framework}

We analyze per-paper gender differences in investment and returns to scientific innovation by proposing a unified approach that seeks to address these limitations. First, we distinguish between two dimensions of novelty: content novelty that involves combining research concepts, versus context novelty that involves bridging disciplines by connecting distant venues. A published discovery involving a new combination of contents may surprise because it has never before succeeded, even if previously considered and attempted \supercite{rzhetsky2015choosing, foster2015tradition}. A discovery drawing upon a divergent combination of sources may surprise because it has neither been attempted nor imagined. Considering both offers a more complete picture of how innovation operates \supercite{Shi2023, doi:10.1177/01655515231161133}. Prior work that measures only concept-level novelty \supercite{doi:10.1073/pnas.1915378117} necessarily attributes all innovation differences to a single dimension, obscuring the possibility that demographic groups specialize in complementary forms of novelty with different reward profiles.

\begin{figure}[H]
\includegraphics[width=\linewidth]{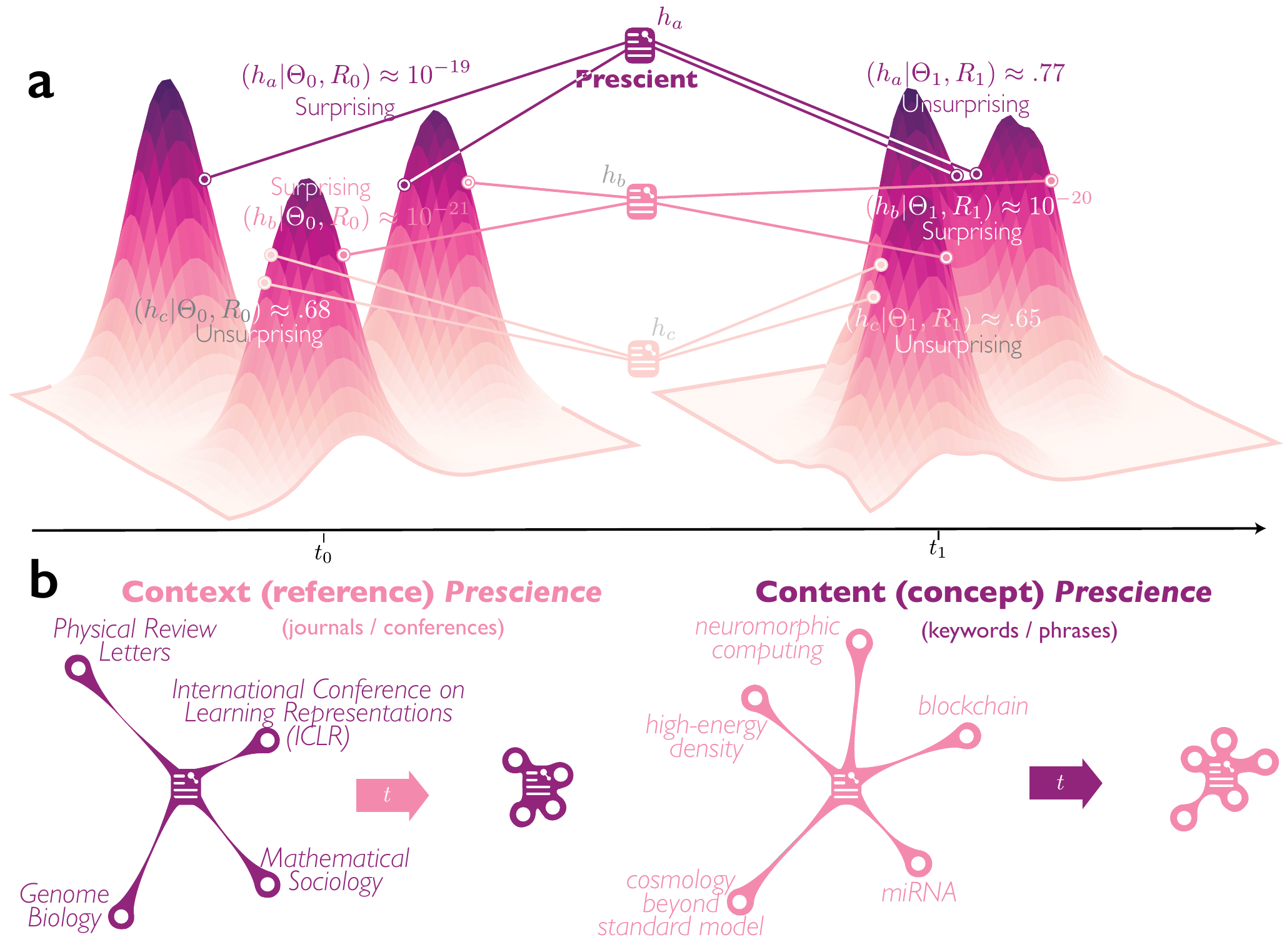}
\caption{\textbf{Illustration describing and contrasting context and content prescience. a}, Contrast between three papers (represented as hypergraphs), $h_a$, $h_b$, and $h_c$. $h_c$ is unsurprising in the first time period, $t_0$, but also in $t_1$ because its components are close together in the manifold ($\Theta$) and both are highly visible ($R$) to scientists in both periods. $h_b$, by contrast, is surprising in $t_0$ (its components are distant in $\Theta$), but it becomes no less surprising in the next period. $h_a$, on the other hand, is surprising in $t_0$, but it changes the scientific world by $t_1$ to make it normal or common; its components are distant in $\Theta$ at $t_0$ but become close by $t_1$. Only $h_a$ is \textit{prescient}. \textbf{b}, Illustration of the difference between context prescience, which involves the shift from a surprising journal and conference combination within a paper's reference list to a routine one, and content prescience, which involves the shift from divergent concepts or keywords to familiar ones over time.} 
\label{fig:GFig1.pdf}
\end{figure}

Second, we consider the innovative influence of novel contributions, distinguishing between those that are merely \textbf{surprising} at the time of publication -- introducing combinations that defy current trends in science \supercite{Shi2023} -- and those that prove \textbf{prescient}, driving or anticipating future directions of research. A paper is surprising to the extent it introduces connections between empirically distant concepts or journals in ways that differ from trajectories at time of publication. Surprising ideas may prove prescient by anticipating future research trends, either through their innovative influence on others or the foresight of their originators \supercite{Vicinanza2022ADM}. This distinction separates the adoption of ideas from citation-based credit, allowing for the possibility that the first people working on an idea may not be credited or even remembered when it is later popularized. It further allows us to distinguish between surprising research that is ephemeral from that which changes or anticipates the course of science. Surprise and prescience are illustrated both for novel contexts and contents in Figure \ref{fig:GFig1.pdf}.

By combining these dimensions, we can evaluate whether published papers by men and women differ in content and context novelty, whether their contributions to each are equally prescient, and whether they receive comparable rewards for similar levels of innovation.

\subsection{Why Innovation Approaches and Rewards May Be Gendered}

Gender differences in investment and returns to novelty arise from multiple factors. Under some circumstances, women manifest greater risk-aversion than men \supercite{nelson_are_2015}. Novel content combinations within a discipline are distinct from novel connections across disciplines \supercite{FONTANA2020104063}, and these two kinds of novelty may entail different risks, which could explain gendered differences in innovation investment. Social sanctions and cultural expectations framing creativity as masculine \supercite{doi:10.1287/orsc.2022.16176, https://doi.org/10.1002/job.2500} may discourage competition for unusual ideas within fields, whereas interdisciplinary work may avoid such disciplinary scrutiny and sanction \supercite{RHOTEN200756}. Access to diverse scientific networks may further shape opportunities for women to pursue and be recognized for innovative research \supercite{10.1371/journal.pone.0238229}. 
 
 In terms of scientific recognition, women consistently receive less credit for comparable work—across citations \supercite{Huang2019HistoricalCO, Larivire2011SexDI, Ross2022}, authorship \supercite{Lerchenmuellerl6573, Filardoi847}, journal placement \supercite{10.1371/journal.pbio.2004956, 10.1371/journal.pone.0189136}, and the adoption of ideas \supercite{doi:10.1086/725551}. Team composition and network position can mitigate some gaps: higher shares of women in teams can boost disruption \supercite{ZHANG2024101520, doi:10.5465/AMBPP.2022.10167abstract}, and women occupying structural holes benefit more when producing novel and disruptive work \supercite{Wang2023Sc}. Yet innovative work remains particularly vulnerable to being overlooked or misattributed \supercite{WANG20171416, Johnson2024GreaterVI, https://doi.org/10.1002/job.2500}, suggesting gender gaps may be amplified for highly novel contributions.

\subsection{Methods and Contributions}

We analyze 261,452 solo-authored papers published between 2000 and 2023 by 141,786 U.S.-based scholars, using data from OpenAlex \supercite{priem2022openalex}. Focusing on solo-authored work allows us to attribute innovation to individual scholars without ambiguity from team composition, collaboration norms, or gendered credit allocation \supercite{rossiter_matthew_1993}. Because no large-scale self-reported gender database exists, we impute gender from names using \emph{Genderize.io} and follow best-practice recommendations by restricting to high-confidence cases and demographic groups where name–gender signals are most reliable (white and Hispanic authors in the U.S. context) \supercite{Lockhart2023NamebasedDI}. In supplementary analyses, we relax these restrictions and include multi-authored papers. These findings remain broadly consistent, though interpretation requires caution given assumptions about author order and the well-documented tendency for men to receive disproportionate credit for team creativity \supercite{doi:10.1177/0956797615598739,https://doi.org/10.1002/job.2219}.

We examine women's participation in surprising science, measured as how unlikely or unexpected a combination of concepts (i.e., content novelty) or cited sources (i.e., context novelty) is at the time of publication, given current trends \supercite{Shi2023}. Unlike novelty measures based only on whether something appears first or how it deviates from a random or typical work \supercite{doi:10.1126/science.1240474, WANG20171416}, our approach accounts for the overall distance of ideas and the evolution of their relationships over time across the scientific literature. We use trained classifiers on hypergraphs of concepts and citations to dynamically learn what to expect in a given year's publications from recent historical trends \supercite{Shi2023}. 

This model-based approach to measuring surprise contrasts with approaches that identify novelty through first co-occurrences of concept pairs and measure adoption through raw reuse counts \supercite{doi:10.1073/pnas.1915378117}. By estimating the probability of a combination given the evolving geometry of scientific knowledge, our surprise scores distinguish genuinely unexpected connections from merely rare ones, and our prescience scores capture whether ideas were ahead of their time rather than merely repeated. Moreover, we analyze published papers rather than dissertation abstracts, and measure paper-level recognition—journal placement, disruption, and downstream citation credit that illuminate prior work on career outcomes \supercite{doi:10.1073/pnas.1915378117}.

We then examine the role of gender in \emph{prescient} work, where scientists introduce ideas (concept or source combinations) before they become widely adopted. Such papers are \emph{ahead of their time}, making connections unusual at publication but common in subsequent science \supercite{lockhart2025china}. 

\begin{figure}[H]
\includegraphics[width=\linewidth]{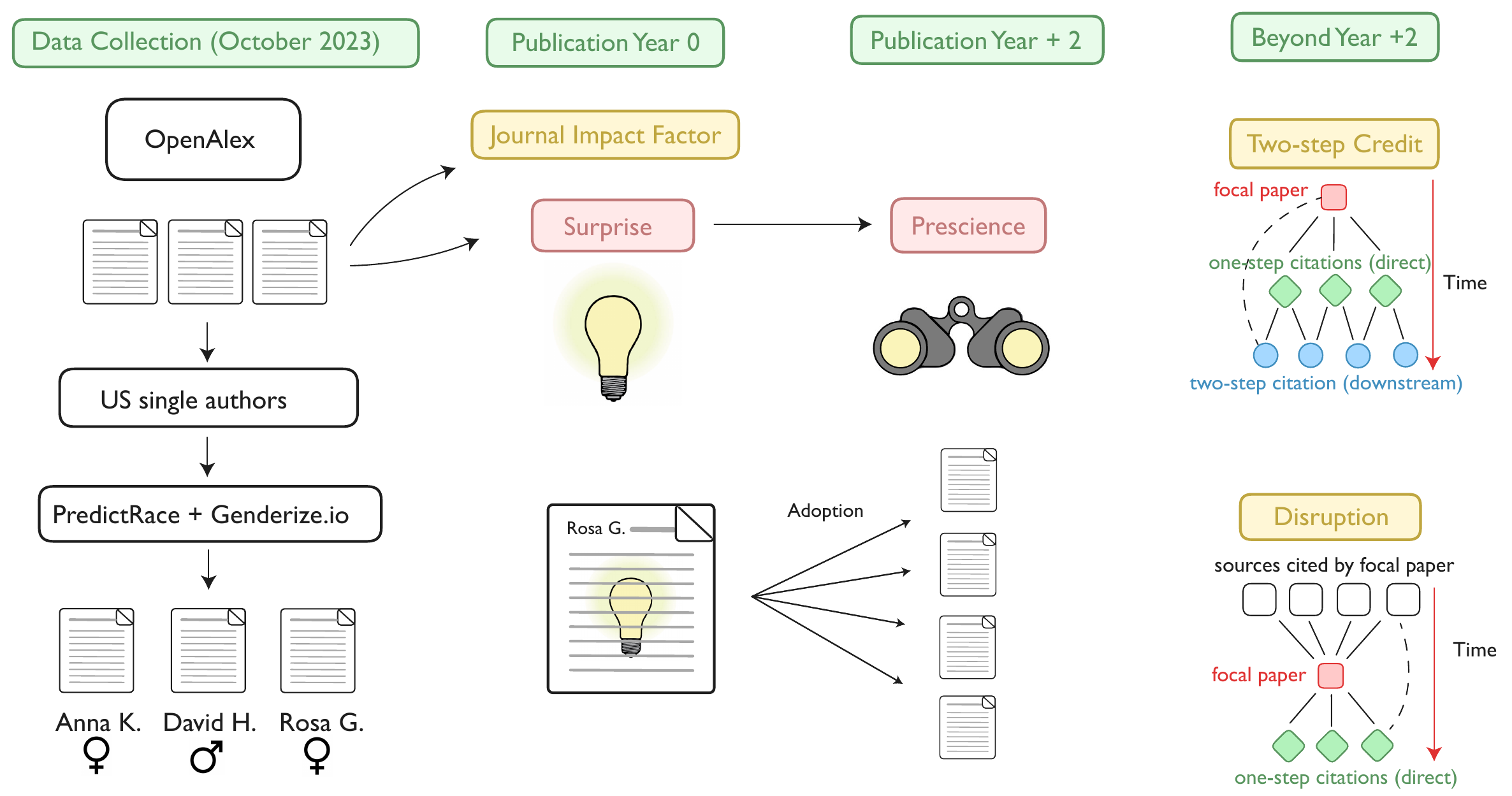}
  \caption{Summary of empirical framework. We collect a snapshot of data from OpenAlex \supercite{priem2022openalex} as of October 2023 on single-authored papers published by authors with affiliations in the US. Authors' names in papers are used to select names that appear to be white/hispanic. We then employ Genderize.io to infer gender with high probability. We measure surprising science, "unexpected" at time of publication, and prescient contributions that drive the future or are "ahead of their time". A paper can propose innovative combinations of (i) concepts (level-three topics) or (ii) references. We evaluate gender differences in returns by considering (i) two-step credit, (ii) five-year disruption score, and (iii) journal impact factor.}
  \label{fig:fig_0}
\end{figure}

We assess the recognition received by women's contributions to surprising and prescient publications using three metrics: (1) two-step citation credit, which measures the extent to which a paper continues to be credited not just by papers that cite it, but papers that cite those papers as well; (2) five-year disruption score \supercite{doi:10.1287/mnsc.2015.2366}, which evaluates how well a focal paper overshadows prior literature; and (3) journal placement, assessed by the two-year journal impact factor. Together, these metrics capture both the ability of innovative research to displace prior work and the degree to which it retains credit over time. Our research design is summarized in Figure \ref{fig:fig_0}.

\textbf{Our study makes three significant contributions.} First, we document per-paper gendered investments in different types of novelty: papers authored by women are more likely to bring together innovative combinations of sources in their reference lists, while men's works are more likely to combine concepts within fields. Second, we examine the degree to which such work is prescient -- driving or anticipating future research trends -- and find gender variation in which ideas "catch on". Third, we assess whether equally innovative papers receive equal recognition, finding consistent evidence of gender disparities in returns to innovation. 

Taken together, our analysis provides a unified framework for measuring scientific innovation and demonstrates how gender shapes both the production of innovative papers and the rewards they generate. By moving beyond within-field concept combinations alone, we show that women's contributions to innovation through boundary-spanning reference combinations and broader disciplinary integration would otherwise remain overlooked. We further demonstrate that gender differences persist even when comparing papers with comparable levels of innovation.

\section{Results}
\subsection{Gendered Approaches in Innovation Strategies}

We begin by examining whether solo-authored papers by men and women differ in scientific novelty, measured with surprise, and innovation, measured with prescience. Using paper-level regressions (equation \ref{eq_1}, "Materials and Methods"), we examine how an author's gender relates to these novelty measures while controlling for publication year and research field, as well as the author's career stage and department size; novelty scores are normalized within fields to ensure comparability. Controlling for these covariates allows us to estimate gender gaps in novelty and scholarly rewards without conflating them with structural differences in field composition, academic age, institutional resources, or time trends in publishing \supercite{doi:10.1177/21582440231184847, 10.1162/qss_a_00117}.

Focusing on solo-authored work ensures that the gender coefficient reflects paper-level differences between women and men. Standard errors are clustered at the author level for solo-authored papers, and heteroskedastic errors are used when comparing solo- and multi-authored work, where clustering is less appropriate.

\begin{figure}[H]
\includegraphics[width=\textwidth]
{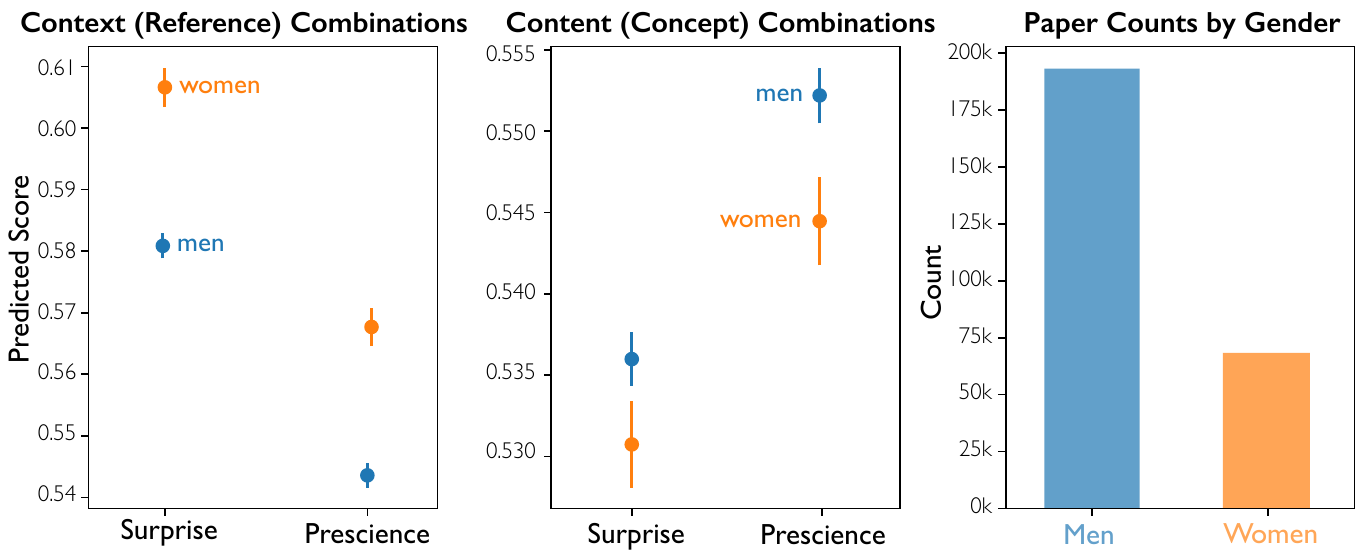}
  \caption{Gender difference in innovation strategies for solo-authored papers, alongside 95\% confidence intervals, and sample distribution of gender in counts (thousands), with clustered standard error at the author level (117\,583 clusters). Women engage in more context novelty (first panel) -- bridging disciplines through reference combinations -- while men engage in more content novelty (second panel) -- recombining concepts within fields. Both surprise (how unexpected at publication) and prescience (how much ideas catch on) show this gender difference in investment. Dots indicate predicted values, averaged over the observed distributions of covariates (career age, department size, publication year, and level-one research field). Solo-authored papers are disproportionately authored by men (third panel). 
  }
  \label{fig:fig_2}
\end{figure}

Figure \ref{fig:fig_2} reveals striking gender differences in innovation patterns. Women's work tends to be more innovative in context novelty---producing reference combinations that are significantly more surprising (b=.0258, \emph{P}<.001) and prescient (b=.0241, \emph{P}<.001) than men's. This means women's solo-authored papers draw on combinations of sources across disciplines that are more unexpected at the time of publication and more likely to anticipate future interdisciplinary trends.

Conversely, papers by men are more likely to develop content novelty, producing concept combinations within fields that are more surprising (b=.00525, \emph{P}<.001) and prescient (b=.00773, \emph{P}<.001) than women's work. This indicates men's papers more successfully combine previously disconnected concepts in ways that presage future developments within their fields. See Table 4 of the SI file for full regression estimates.

\textbf{This is not a deficit on either side---it represents divergent, complementary approaches to advancing science}. This finding revises a central claim of the diversity–innovation paradox \supercite{doi:10.1073/pnas.1915378117}: that underrepresented groups uniformly innovate at higher rates. When innovation is measured along a single dimension—concept novelty—men's papers are, on average, more innovative. Women's greater innovation emerges when we consider context novelty, the dimension that prior work did not measure. The paradox is thus not that women innovate more and benefit less, but that the form of innovation women disproportionately pursue is the form scientific fields most undervalue. Woman-authored papers are more likely to transgress disciplines, while man-authored papers remix concepts within them. 
And as we demonstrate next, these strategies face profoundly unequal recognition.


Overall, papers by women tend to entail combinations of sources that are more unexpected at the time of publication than papers from men. Further, women's interdisciplinary innovations experience higher uptake in the future, foreshadowing emerging trends more than men's innovations of the same kind. On the other hand, papers by men are more likely to combine concepts that were previously distant. In this kind of innovation, men's papers are more likely to presage the future of the field. This finding extends to multi-authored papers, where the main variable of interest is the proportion of female authors (Table 28, Section I of the SI file). Testing how gender differently predicts surprise and prescience of new science among solo and multi-authored papers, the same pattern holds, although it is attenuated in multi-author work relative to solo author work (Sections J, Table 32).

Gender differences in scientific novelty remain robust in unconditional models without covariates (Section D, SI) and when controlling individually for career age, department size, and field (Section E.1, SI), suggesting that these factors explain only a small portion of the observed gender differences in our solo-authored setting.


Novelty in women's solo-authored papers benefits relatively more from an increasing proportion of women in their field (a "critical mass" effect; \supercite{price_little_1986}, Section F.1, SI) and from larger department size (Section F.2, SI). Across career cohorts (Section F.3), women's work leads in prescient and surprising references but lags in concept novelty, although gender differences in concept science shrink as careers progress. Fields such as economics, mathematics, and philosophy — where solo-authored work is highly valued and women face greater career penalties \supercite{10.1257/aer.p20171126, Ginther2004WomenIE, Skibba2019WomenIP, Wang2017GenderGI} — do not show lower levels of innovation among women-authored papers (Section F.5).

While contributions by women may feature highly unusual idea combinations, potentially contributing to slower uptake \supercite{doi:10.1073/pnas.1915378117}, gender differences in prescience diminish as the initial surprise of a solo-authored paper increases (Section G, Table 21). Reference novelty is more closely related to citations, whereas content novelty is associated with status and awards \supercite{Shi2023}. Using past citations as a proxy for author status, we find a positive and significant gender–status interaction for concept prescience (Section F.4), suggesting that women benefit relatively more from their reputation in terms of uptake when producing work in which they are typically underrepresented.

\subsection{Content and Context Novelty Attract Different Forms of Recognition}

Before examining gender differences in rewards, we first establish that content and context novelty attract distinct forms of recognition, reflecting different innovation strategies within the prestige economy of science. Awards and community recognition tend to favor novel combinations of concepts from within disciplines, whereas novel reference combinations transgress disciplinary boundaries but lead to more outsized citations across the fields of science \supercite{Shi2023}. To examine this, we analyze how surprise scores relate to the share of citations coming from \emph{outside} versus \emph{inside} a paper's field, using Web of Science subfield classifications.

\begin{figure}[H]
    \includegraphics[width=\textwidth]{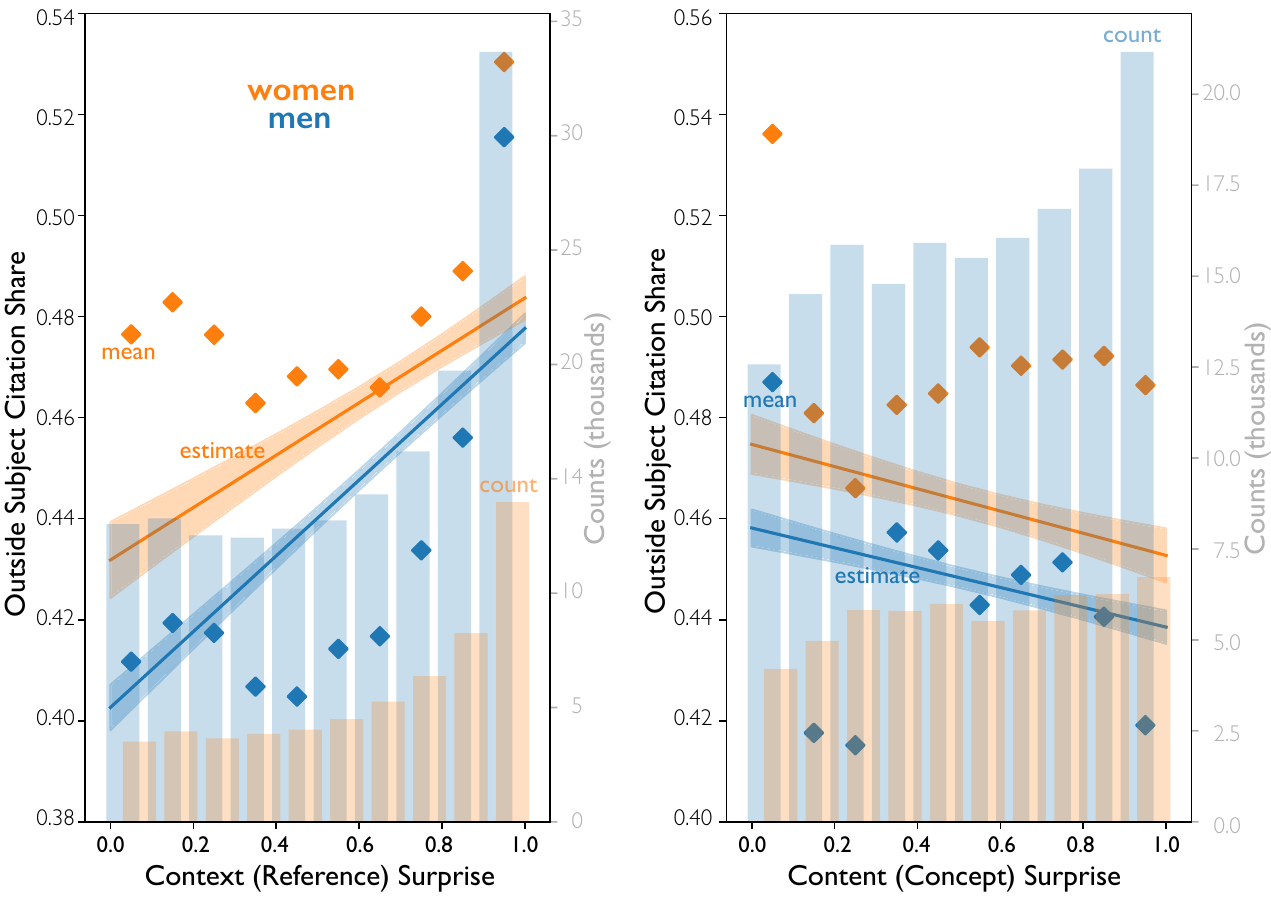}
  \caption{Content and context novelty attract different audiences. Surprising reference combinations (context novelty, left) correlate with more outside-subject citations and fewer within-field citations. Surprising concept combinations (content novelty, right) show the opposite pattern. Women's solo-authored work attracts more cross-field attention than men's for both innovation types, while men's work dominates within-field citations. The $y$ axis traces the outside-subject citation share, computed as the number of citations from outside subject over total citations. Estimated relations are inverted for inside citation share, which is equal to $(1-\textit{Outside Subject Citation Share)}$. Predicted values are marginalized over the observed distribution of covariates (department size, career age, level-one field, open access status, and publication year). Lines represent fitted linear regression models with clustered standard errors at the author level, plotted over side-by-side histograms of surprise by gender (counts in thousands). Diamonds indicate the sample mean of scholarly rewards within each surprise interval, by gender.}
  \label{fig:fig_inout_1_g_c}
\end{figure}

Figure \ref{fig:fig_inout_1_g_c} shows linear predictions of outside-field citation shares by gender as a function of concept and context surprise. Context surprise attracts more outside-field citations, while content surprise draws attention predominantly from within-field peers. Across all levels of surprise, women's papers receive a higher share of outside-field citations, whereas men's papers receive more within-field citations (inside-field share = 1 - outside-field share). Because awards and prestige are largely conferred by disciplinary insiders, content novelty is more directly rewarded in academic evaluation systems, while context novelty may allow women to produce trans-discplinary and potentially disruptive work without equivalent recognition from disciplinary gatekeepers. These patterns suggest that equally innovative contributions may be rewarded differently by gender (Section H, SI).

\subsection{Asymmetric Rewards for Equally Innovative Science}

Having established gender differences in innovation types and shown that these types attract different audiences, we now examine whether equally innovative contributions receive equal rewards. We estimate paper-level linear regression models with interactions between surprise or prescience scores and author gender, examining three measures of recognition: journal placement (log two-year journal impact factor), downstream citation credit (two-step credit), and disruptive impact (five-year disruption score) -- see equation (\ref{eq_2}), and details in `Materials and Methods'. 

\subsubsection{Rewards to Surprising Science}

We begin by discussing gender differences in how surprise is related to rewards, with regression estimates reported in Table S5. 

Journal placement tends to disadvantage women's contributions. Reference surprise has no significant association with journal prestige ($\beta$=.004, \emph{P}=.48), while concept surprise is negatively associated with journal impact ($\beta$=-.03, \emph{P}<.001). Women's papers secure lower journal placement than men's for both reference ($\beta$=-.066, \emph{P}<.001) and concept surprise ($\beta$=-.065, \emph{P}<.001).

\begin{figure}[H]
 \includegraphics[width=\linewidth]{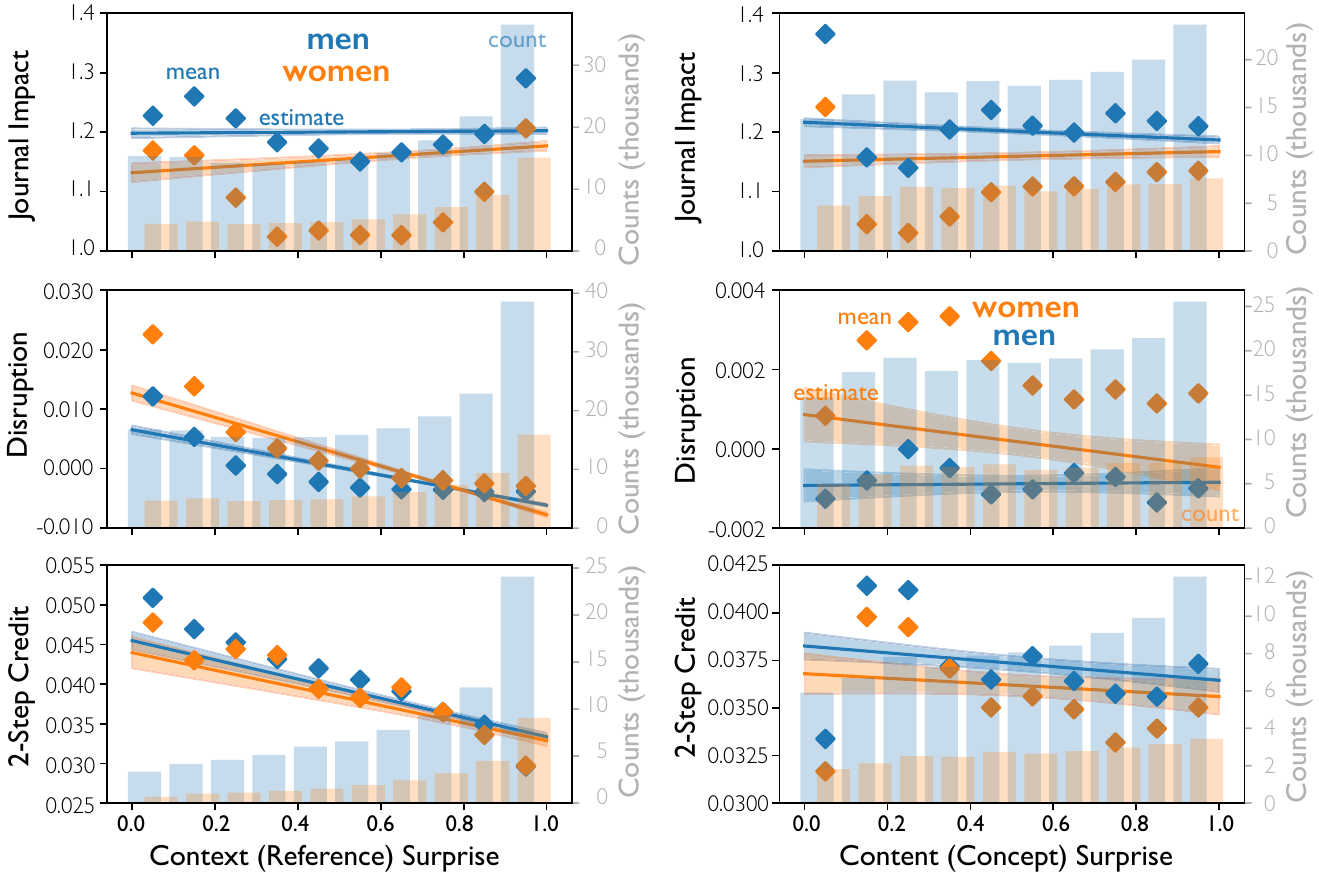}
  \caption{Rewards to surprising science by gender in solo-authored papers, alongside 95\% confidence intervals, with clustered standard errors at the author level. For equally surprising science, women's work lands in lower-prestige journals (row 1) and receives less citation credit for concept combinations (row 3, right panel). Women's work is more disruptive (row 2), but this advantage narrows as surprise increases. Gender gaps in journal placement narrow at high surprise levels, indicating women must produce exceptionally surprising work to achieve parity. Predicted outcomes are marginalized over the observed distribution of covariates (department size, career age, level-one field, open access status, and publication year -- plus citations and JIF for disruption). Lines represent fitted linear regression models, plotted over side-by-side histograms of surprise by gender (counts in thousands). Diamonds indicate sample means.}
  \label{fig:fig_4a}
\end{figure}

Nevertheless, women's papers gain relatively more in journal prestige for increasing surprise of contributions: as reference or concept surprise rises, women's work climbs in journal prestige faster than men's (reference interaction $\beta$=.041, \emph{P}<.001; concept interaction $\beta$=.046, \emph{P}<.001). When papers are authored by women, the negative association between concept surprise and journal placement reverses (Figure \ref{fig:fig_4a}, first row). This suggests that papers by women who commit to higher signaling efforts -- as by producing more surprising science -- can achieve parity in treatment \supercite{doi:10.5465/AMBPP.2021.10557abstract}.
Multi-authored patterns show that teams with higher proportions of women also end up in less impactful journals, with women's contributions having lower marginal returns with increasing reference surprise (Section I, Table 31) -- which differs from the solo-authored case. We address this difference later, in the \emph{Robustness and Sensitivity Analyses} section. 

Disruptive impact tells a different story. Both reference innovation ($\beta$=.006, \emph{P}<.001) and concept innovation ($\beta$=.0018, \emph{P}<.001) are more disruptive in papers authored by women. Disruption decreases with increasing reference surprise ($\beta$=-.0127, \emph{P}<.001), however, with women-authored papers penalized more steeply (interaction $\beta$=-.0078, \emph{P}<.001). For concept surprise, we find no significant effect on disruption at baseline ($\beta$ = .00008, \emph{P} = .8), but gender gaps in disruption narrow as reference surprise increases (interaction $\beta$=-.0014, \emph{P}=.024) -- papers by women start with higher disruption but converge with men's at extreme novelty (Figure \ref{fig:fig_4a}, second row). We find similar patterns in multi-authored papers (Section I, Table 30).


Two-step credit declines with increasing surprise in both references ($\beta$ = -.012, \emph{P} < .001) and concepts ($\beta$ = -.0018, \emph{P} = .001), showing that the effect is much stronger for reference (context) novelty. This indicates that highly unusual work is less likely to be cited two steps downstream, with context novelty particularly penalized. Women's work earns significantly less credit than men's for equal content novelty ($\beta$ = -.0146, \emph{P} = .026), but not for reference novelty ($\beta$ = -.001, \emph{P} = .20). Gender differences in downstream credit show no statistically significant interaction with content novelty ($\beta$ = .0006, \emph{P} = .55) or reference novelty ($\beta$ = .001, \emph{P} = .47). Figure \ref{fig:fig_4a} (third row) presents predicted two-step credit by surprise across genders. In multi-authored papers, a higher share of women is associated with lower downstream credit for equal concept novelty and reference novelty, similar to the solo-authored case, while simultaneously yielding significantly higher marginal returns for surprising reference combinations. These effects are larger for context than for content novelty, consistent with the solo-authored results (Section I, Table 29).


\subsubsection{Rewards to Prescient Science}

Next, we estimate the gender gap in how prescience in solo-authored papers gets rewarded in terms of two-step credit and five-year disruption (Table 6 of the SI file). We do not include journal impact factor, as its timing precedes prescience -- prescience is computed considering how much less unexpected combinations become two years after publication, while journal impact factor reflects how papers cite past work in the journal at the year of publication.

\begin{figure}[H]
 \includegraphics[width=\linewidth]{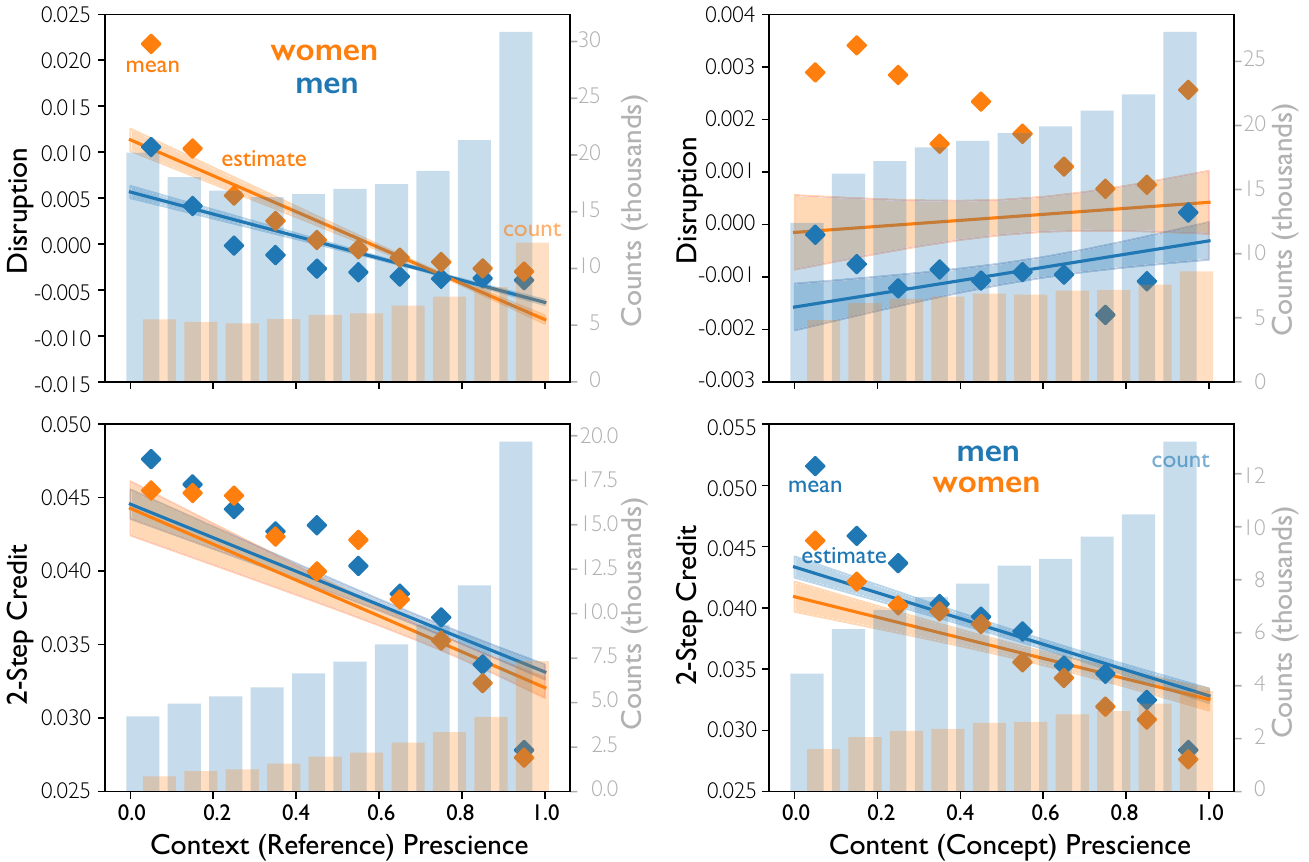}
  \caption{Rewards to prescient science by gender in solo-authored papers, alongside 95\% confidence intervals, with clustered standard errors at the author level. Women earn relatively more credit for increasing prescience in concept combinations (row 2, right panel), suggesting that women who commit to higher signaling efforts by investing in more surprising science can achieve treatment parity in areas where they are typically fewer. Women's disruption advantage erodes as reference prescience increases (row 1, left panel). Predicted outcomes are marginalized over observed distribution of covariates (department size, career age, level-one field, open access status, and publication year -- and 2-years citations and JIF for disruption). Lines represent fitted linear regression models, plotted over side-by-side histograms of prescience by gender (counts in thousands).  Diamonds indicate sample means.}
  \label{fig:fig_4b}
\end{figure}

Disruptive impact depends critically on innovation type (Figure \ref{fig:fig_4b}, row one). Higher prescience in concepts increases disruption ($\beta$=.0012, \emph{P}<.001), but higher prescience in references reduces it ($\beta$=-.012, \emph{P}<.001). This latter effect is expected mechanically: papers with prescient combinations of sources by definition are followed by others repeating those combinations of sources, lowering disruption metrics but not the innovative value of the work.

Women's solo-authored papers are more disruptive than comparable men's work by both measures (references $\beta$=.0057, \emph{P}<.001; concepts $\beta$=.0014, \emph{P}=.001), but face a steeper decline in disruption as reference prescience rises (interaction $\beta$=-.0075, \emph{P}<.001), narrowing the gender gap at high novelty levels. 
Instead, no significant gender difference emerges for concept prescience (interaction $\beta$=-.0006, \emph{P}=.3). Multi-author teams with more women also experience greater disruption losses with increasing reference prescience (Section I, Table 30).


Two-step credit declines with prescience in both references ($\beta$ = -.011, \emph{P} < .001) and concepts ($\beta$ = -.01, \emph{P} < .001). This pattern is expected to some extent: when science follows new trends, attention tends to concentrate on the latest literature. However, it also indicates that early work on a popular innovation is less likely to receive credit than more incremental, "normal science". Across all levels of concept prescience, women's papers are less likely to be cited two steps out ($\beta$ = -.0024, \emph{P} = .002), whereas no gender difference is observed for reference prescience ($\beta$ = -.0003, \emph{P} = .79). Once again, conditional gender differences in two-step credit show no statistically significant interaction with reference prescience (interaction $\beta$ = -0.0007, \emph{P} = .56) and are weakly significant when interacted with concept prescience (interaction $\beta$ = .002, \emph{P} = .052). Figure \ref{fig:fig_4b}, second row, reports the predicted values of two-step credit by prescience. 
Papers by mixed-gender teams display patterns consistent with solo-authored papers (see Table S29).

\subsubsection{Robustness and Heterogeneity Analysis}

We conduct a series of robustness checks. Estimates of gender differences in returns to surprise and prescience in solo-authored papers are stable across alternative outcome measures, including three-year disruption and five-year journal impact (SI, Sections K–L), when conditioning on funding listed in papers or authors' past citations (SI, Section M), and when considering both types of innovation simultaneously (Section N). Generalized additive models confirm that the linear specification captures most of the relationship between novelty, prescience, and rewards, but perhaps understates gender differences at the extremes (SI, Section P).

Controlling for baseline surprise, gender differences in rewards to prescience vanish for the most unexpected science but emerge for papers with low to moderate initial surprise, across both reference and concept novelty (SI, Section G). As surprise and prescience increase, women's work receives relatively lower marginal returns in direct citations, measured by two-year forward counts, except for surprise in concept combinations, where no citation penalty is evident (SI, Section O).

Prior work suggests that estimating gender effects requires accounting for differences in academic age, research field, and other compositional factors, because aggregated \emph{apples-to-oranges} comparisons can differ substantially from comparisons of similar researchers in similar contexts \supercite{doi:10.1177/21582440231184847,10.1162/qss_a_00117, CondeRuiz2025CitationGG}. In our data, unconditional estimates of gender differences in journal placement and disruption closely match the fully specified model (SI, Section D, Tables S8 and S10), and controlling for individual covariates -— career age, time trends, department size, or research field —- does not alter these results (SI, Section E.2, Tables S12–S13).

By contrast, gender gaps in downstream citations are sensitive to specification: conditioning on field effects alone reverses the unconditional gender coefficient, reflecting women's greater representation in lower-citation-intensity fields, while fully specified models recover a negative gender coefficient, indicating aggregate disadvantages that outweigh within-field effects (Sections D–E.2, SI; Tables 9 and 14).

Finally, gender gaps in journal placement and disruption narrow more rapidly with increasing surprise and prescience in solo-authored than in multi-authored work (Section J), consistent with collaboration-related penalties \supercite{10.1257/aer.p20171126}. Considering gender-homogeneous teams and non-linear effects of women's shares in multi-authored papers, all-women papers earn higher marginal returns in journal prestige to increasing surprise than all-men papers, consistent with patterns observed in solo-authored work; other outcomes are discussed in the SI (SI, Section Q). 


\section{Discussion}

Our findings reveal a troubling paradox at the heart of scientific reward systems. Women and men tend to pursue different---but equally valuable---innovation strategies, yet face profoundly asymmetric rewards. Women's works engage in more discipline-spanning work that proves most disruptive to established knowledge. Men's papers tend more to recombine content within-domains, which earns recognition from disciplinary gatekeepers. For equally innovative contributions, women's works land in lower-prestige journals, tend to receive less downstream citation credit, while manifesting greater disruptive impact---especially at lower levels of novelty where gender gaps are widest.

Critically, these gaps tend to narrow---sometimes even reverse---at the highest levels of surprise and prescience, suggesting that \textbf{extreme} signals of innovation can override gender penalties. 
The appropriate interpretation is not that women succeed at the highest levels, but rather that their contributions persistently face disadvantage at all other levels representing the vast majority of scientific production.

\subsection{Differences Reflect Structural Position}

The patterns we document likely do not reflect innate differences in capability or preference, but rather structural biases in how innovation is recognized and how scientists are positioned within knowledge networks. Content novelty aligns with traditional markers of disciplinary mastery and signals belonging within a field's intellectual tradition. Context novelty, while potentially more transformative, signals outsider status and can be read as a lack of focus by evaluators embedded in single disciplines \supercite{Foster2021SurpriseMN}. 

Women's greater investment in context novelty likely stems from multiple, mutually reinforcing mechanisms. Women may have restricted access to central network positions within disciplines \supercite{10.1371/journal.pone.0238229}, making cross-boundary connections more feasible than deep within-field integration. Women may face exclusion from within-field status hierarchies, reducing returns to within-domain innovation \supercite{RHOTEN200756}. They may strategically avoid highly unusual ideas within their home field due to risks of social backlash \supercite{doi:10.1287/orsc.2022.16176}, finding it safer to import legitimated ideas from other fields. Gender expectations about creativity itself may code within-field innovation as more masculine \supercite{https://doi.org/10.1002/job.2500}. Instead, interdisciplinary research may attract women, as cross-field scientific contributions are rarely constrained by established social structures or scientific communities, as with domain publications \supercite{RHOTEN200756}. 

Regardless of origin, the consequence is clear: the work women disproportionately produce---disruptive, interdisciplinary, prescient context-bridging innovation---is the work science most undervalues. Context novelty attracts cross-field citations but not the within-field recognition that determines careers. Indeed, women's solo-authored papers receive a higher share of citations from outside their field at every level of surprise (Figure \ref{fig:fig_inout_1_g_c}), a pattern that aligns with---but is not mechanically entailed by---their greater investment in field-bridging reference novelty. This cross-field attention, while reflecting genuine impact, falls outside the within-field recognition systems that determine tenure, promotion, and awards. Women's prescient innovations that draw ideas from multiple disciplines, anticipating interdisciplinary trends, go under-credited as later adopters receive the recognition they deserve.

\subsection{Self-Selection Cannot Explain Asymmetric Rewards}

Could self-selection explain these patterns? Women pursuing context-spanning work might be concentrated in lower-prestige positions, which could affect recognition. We find this explanation unlikely. Variations in department size, past citations, and career stage do not eliminate gender gaps in novelty or prescience (SI Sections D–F). If self-selection were driving the results, we would expect similar patterns across all women's work. Instead, we observe penalties specifically for context novelty and convergence at extreme levels of surprise. Gender differences also vary by outcome. Women's papers gain relatively more journal prestige from concept surprise than men's (Figure \ref{fig:fig_4a}, row 1), contrary to what a general self-selection account would predict. Moreover, self-selection into lower-citation fields is unlikely to drive our results, as we control for field differences and normalize novelty and prescience scores to ensure comparability across disciplines. Solo-authored patterns align with multi-authored work, indicating that results are not confined to specific team structures (Section I, SI).

Gendered preferences alone are unlikely to explain these patterns. Women's content novelty increases disproportionately with department size and with women's share in the field. For indicators of status and experience (career age and prior citations), effects are not statistically significant 
(Section F, SI). As women accumulate seniority, status, and institutional resources, they tend to engage more in within-field novelty, consistent with an account acknowledging agency and institutional positioning. 

Overall, the patterns we observe are more consistent with the devaluation of boundary-crossing innovations than with individual sorting. Evaluators appear to respond differently to observably identical innovations based on author gender and innovation type,  discounting the interdisciplinary, boundary-crossing work in which women disproportionately invest energy.

\subsection{Implications for Understanding Gender Inequality in Science}

Our unified framework clarifies conflicting findings in prior literature. Studies that focus solely on concept-based novelty miss women's substantial contributions to context-bridging innovation—precisely the work that proves most transformative for science. Studies examining only aggregate citations or journal placements would overlook the undervaluation of interdisciplinary work. By distinguishing content from context novelty and measuring both surprise and prescience, we uncover gendered patterns in innovation strategies and the asymmetry in how these contributions are rewarded.

The narrowing of gender gaps at extreme novelty levels helps explain why some studies find gender parity among elite scientists: those studies examine only the most exceptional work, where women have overcome disadvantage through extraordinary innovation \supercite{doi:10.1073/pnas.1915378117}. Our findings also clarify why Hofstra et al.\ observed a blanket innovation premium for underrepresented groups: their single novelty dimension---new concept pairs in dissertation abstracts---blends what we show are two distinct phenomena with opposite gender associations. By separating content from context novelty, we reveal that the diversity–innovation paradox is more precisely a \emph{diversity–recognition} paradox: women's boundary-crossing innovations are not merely discounted in aggregate but are structurally undervalued by the disciplinary gatekeepers who allocate prestige, such as journal placement and editorial access. Furthermore, by analyzing paper-level outcomes rather than career trajectories, we identify the specific mechanisms---lower journal placement, reduced two-step credit at baseline, and steeper penalties in disruptive impact for increasing novelty---through which the paradox operates on individual contributions. The paradox is thus sharper and more actionable than previously understood: it is not simply that innovative outsiders have worse careers, but that specific papers producing specific kinds of innovation receive less recognition, paper by paper, systematically.




Conditioning on initial surprise, men's and women's highly unexpected, prescient papers receive similar recognition, indicating that gender matters less for the most innovative work. We find no evidence that gender differences are concentrated in fields. Instead, larger departments and critical mass enhance novelty and uptake of women's papers, and past citation-based prestige encourages investment in concept-based innovation. 

Evaluators may interpret surprising work differently across genders due to uncertainty and hindsight bias \supercite{doi:10.1287/orsc.2021.1567}, and prior studies suggest women face harsher scrutiny \supercite{doi:10.1073/pnas.1211286109, https://doi.org/10.1111/0022-4537.00237}. Future research should develop measures to assess how conflicting prior evidence affects evaluations of solo-authored work by men and women.

\subsection{Policy Implications}

Our findings suggest several concrete interventions. First, evaluation criteria should explicitly value interdisciplinary impact---citations from outside one's field, disruption scores, cross-boundary influence---not only within-field recognition. Tenure and promotion committees should be trained to recognize that context-bridging work may not accumulate traditional prestige markers even when highly influential \supercite{gerow2018measuring}. 

Second, because gender gaps narrow at extreme surprise, perhaps double-anonymous review or requirements for "novelty impact statements" could help evaluators focus on work quality rather than author identity. Anonymization alone is insufficient if evaluators systematically discount interdisciplinary work regardless of author, however, or if other signals, such as writing style, correlate with both gender and scientific rewards \supercite{lerchenmueller_gender_2019}.

Third, journals and funding agencies could track and report their own patterns of gender bias in recognizing different innovation types. Making institutions accountable for whether they systematically undervalue context-bridging work could catalyze reform.

Fourth, mentoring and advising should acknowledge the distinct challenges discipline-bridging work faces. Rather than coaching women toward more "recognized" innovation styles (i.e. those typical of men), we should reform reward structures to appropriately value the interdisciplinary work women already produce.

\subsection{Strong Warning Against Discriminatory Measurement}

We caution strongly against interpreting these findings as grounds for creating gender-specific "novelty accounting systems". Some might propose adjusting how we measure women's contributions to account for differences in approaches and strategies in novelty 
---for example, weighting interdisciplinary citations more heavily for women's works or using different evaluation criteria by gender. This would be misguided.

The problem is not that we are measuring the wrong things -- it is that we are \emph{valuing} the wrong things. Context-bridging innovation drives paradigm shifts, reorganizes relationships between fields, and produces long-term impact that within-field citations fail to capture. These contributions should be valued equally regardless of who produces them. Similarly, most scientific work is not radical innovation, but normal science contributions \supercite{kuhn1997structure} to a robust knowledge base. This work is also systematically undervalued when authored by women. Creating separate evaluation systems by gender would institutionalize rather than remedy discrimination, implicitly accepting that women's work deserves different (lesser) treatment.

The appropriate response is to reform reward structures for all scientists, explicitly recognizing that multiple pathways to innovation exist and that interdisciplinary, boundary-crossing work merits equal recognition to within-field advances. Any policy response must center on changing how science values different types of innovation, not on adjusting measurements to account for gender differences in what gets produced.

\subsection{Limitations and Future Directions}

In order to have high confidence in our results, given the dangers of imputation over heterogeneous populations \supercite{Lockhart2023NamebasedDI}, we focus on solo-authored papers from U.S. institutions with high-certainty gender inference. This emphasis on quality introduces sampling bias that limits generalization. To improve external validity, we relax our constraints and find that patterns remain consistent when extending to multi-authored papers and across multiple robustness checks, suggesting that the findings are not artifacts of sample restrictions. We cannot rule out all confounding influences, but the pattern of interactions provides strong evidence against simple omitted variable explanations. Moreover, defining novelty as surprise in the combination of contents and contexts poses limits what is identified as novelty, failing to capture novelty that builds incrementally, or appears elsewhere in a paper (e.g., a dataset or dataset combination \supercite{yu2024does}). 
Finally, our study focuses on gender differences in novelty, prescience, and rewards at the paper level, meaning we do not examine how these factors translate into long-term career trajectories or cumulative scientific influence. 

Future research should investigate mechanisms of innovation production and evaluation more deeply. Do evaluators consciously discount interdisciplinary work, or does the structure of peer review disadvantage context-bridging innovations? Do women's peripheral network positions limit their ability to cite and be cited by central figures? How do gender expectations about what constitutes "serious" versus "peripheral" science shape evaluations of novelty? How do race, ethnicity, institutional prestige, and other dimensions of marginalization interact with gender \supercite{doi:10.1073/pnas.2113067119}?

\subsection{Conclusion}

Women-authored papers are more likely to bridge disciplines through context-spanning innovation; men's production is more likely to advance within domains through content-recombining innovation. Both strategies advance science, but they are not given equal recognition. The work women disproportionately produce---disruptive, interdisciplinary, prescient---is  undervalued by existing reward structures across contexts, even as it drives paradigm shifts and long-term impact. Women must produce exceptionally surprising or prescient work to overcome this disadvantage, and even then, they can face discounting in journal placement and allocation of credit at the paper level.

These patterns reveal more than gender gaps---they expose fundamental tensions in how science recognizes innovation. Our institutions reward disciplinary mastery over boundary-crossing \supercite{szell2018nobel}, within-field status over cross-field influence \supercite{foster2015tradition}, and conventional markers of prestige over disruptive impact. As science grows increasingly specialized and interconnected, this misalignment between what we value and what drives progress will become more consequential \supercite{Shi2023}.

Reforming these structures requires more than addressing explicit bias. It demands rethinking what counts as a valuable scientific contribution, how we measure impact, and whose work we center in narratives of innovation. Until reward systems appropriately value multiple pathways to advancing knowledge, gender inequality in science will persist---not despite women's contributions, but because of them.

\printbibliography

\section{Materials and Methods}\label{data3}

\subsection{Data collection and gender inference}
We collect data from OpenAlex \supercite{priem2022openalex} on single authored papers published by authors with affiliation in the US, using a snapshot of the data as of October 2023 (1,774,303 unique authors).  

We obtain accurate disambiguation of authors by using the OpenAlex identification codes of the OpenAlex v2 Author Name Disambiguation (AND) System.\footnote{For more information, see \url{https://github.com/ourresearch/openalex-name-disambiguation/tree/main/V2}.} 
We employ the Genderize API to assign genders to first and middle names of authors. We infer the ethnicity of first names and surnames of authors using R package \textit{predictrace} \supercite{Tzioumis2018DemographicAO}. Then, we select the subset of authors from the US with an ethnicity where we know the Genderize error rates are low -- white and latin/hispanic \supercite{Lockhart2023NamebasedDI}. We are left with a subset of 1,105,268 unique authors, with 3,444 unique first names and 30,087 unique middle names. 

Following the need for a conservative test, we consider an inferred gender reliable if the associated name is of high confidence, that is, if either the Genderize count is higher than 100, or the Genderize probability of assignment is above 0.90. We identify 22,747 first names of low confidence and a total of 460,902 middle names with low confidence. These names are set as missing. The distribution of Genderize counts and probabilities for first and middle names are reported in the SI file, Section A.

Next, we remove names given by single characters or initials, for which Genderize produces unreliable inferences. Middle names given by initials are also set as missing. Disagreeing cases of different inferred genders for first and middle are dropped from the sample. Authors whose first and middle names both resulted in missing values after the conservative processing procedure described above are also removed from the sample. We are left with a cleaned, reliable sample of 1,074,062 authors.

After merging the data on solo-authored papers, we obtain a sample of 713,684 solo-authored papers by 301,585 authors. In order to compare estimates of the gender differences on innovation and recognition on comparable samples, we focus exclusively on papers for which we observe all four measures of innovation -- surprise and prescience in concepts and references -- we are left with a final solo-authored sample of 226,208 unique papers by 141,786 unique authors (31\% women and 69\% men). Section B.1 of the SI file reports paper-level sample statistics. 

Sample statistics at the author level indicate that men dominate mid- to late-career solo-authored contributions, whereas women are relatively well-represented in multi-authored work, particularly at early-career stages (SI file B.2, Figures 7–8). Solo-authors tend to come from smaller departments and, on average, have longer career experience than team-authors (SI file B.2, Figure 10). Because solo-authored papers are concentrated among mid- to late-career researchers and women are underrepresented in this group, analyses of gender differences in solo-authored work should account for career age. Conditioning on career stage ensures that observed differences in outcomes reflect gender rather than experience.

Solo-authors publish across all disciplines, with Biology, Medicine, and Computer Science as the main publication areas for both solo- and multi-authoring scholars (SI file B.2, Figure 9). As expected, solo-authorship is relatively more common in the humanities and social sciences, and the pattern of differences between solo and multi-authorship is similar for men and women (SI file B.2, Figure 10). Overall, these patterns underscore the importance of covariate adjustment to isolate gender-specific differences. 


\subsection{Surprise and Prescience Measures}
In OpenAlex, scientific topics of papers are not chosen by authors, as OpenAlex assigns topics to publications using an automated system, providing a fine-grained representation of concepts and identifying research at different levels: from broader fields like art, computer science, or economics (level-zero concepts), to more specific fields, such as machine learning, visual arts or ecology (level-one), and even more fine-grained topics resembling concepts, such as social identity theory, ecosystem services, biodiversity hotspot, or social benefits (level-three concepts). The higher the level, the more fine-grained the concept is. 

We look at novelty of a paper in terms of combinations of (i) journals in its reference list (contexts), or (ii) concepts as fine-grained concepts of level three (contents). 

\subsection*{Formal Model of Surprise and Prescience}

We construct a dynamic hypergraph $\mathcal{G}_t$ at each year $t$, 
in which hypernodes are either scientific concepts (level-three 
OpenAlex topics) or publication venues (journals), and hyperedges 
are papers, each connecting the set of concepts or venues 
co-occurring within it. For content novelty, nodes are concepts; 
for context novelty, nodes are cited journals.

\paragraph{Latent embedding and salience.}
For each node $i$, the model learns a latent vector 
$\boldsymbol{\theta}_i^{(t)} \in \mathbb{R}^D$, where each entry 
$\theta_{id}^{(t)} \geq 0$ and $\sum_d \theta_{id}^{(t)} = 1$, 
representing the probability that node $i$ belongs to latent 
dimension $d$ at time $t$. These dimensions naturally recover 
scientific fields or disciplinary communities 
\supercite{Shi2023}. The model additionally assigns each node a 
scalar salience parameter $r_i^{(t)} > 0$, capturing its cognitive 
accessibility to scientists through cumulative frequency in the 
literature. The embedding is estimated from the sequence of 
hypergraphs $\{\mathcal{G}_{t'}\}_{t' \leq t}$, allowing the 
latent space to evolve as the structure of science changes.

\paragraph{Expected propensity of a combination.}
Given a paper $h$ (a hyperedge connecting a set of concept or venue 
nodes), the model's expected propensity for that combination to 
appear in newly published work is:
\begin{equation}\label{eq:propensity}
    \lambda_h^{(t)} 
    = \underbrace{\left(\sum_d \prod_{i \in h} \theta_{id}^{(t)} 
      \right)}_{\text{coherence}} 
    \times 
    \underbrace{\prod_{i \in h} r_i^{(t)}}_{\text{salience}},
\end{equation}
where the coherence term measures the probability that all nodes in 
$h$ load onto the same latent scientific dimension, and the salience 
term captures the joint cognitive accessibility of those nodes to 
working scientists. The number of publications realizing combination 
$h$ in year $t$ is modeled as a Poisson random variable with mean 
$\lambda_h^{(t)}$, and model parameters are estimated by maximum 
likelihood over the observed hypergraph. Across biomedical science, 
physics, and patents, this model distinguishes a combination that 
appeared in a real publication from a randomly drawn combination in 
more than 95\% of cases (AUC $> 0.95$) \supercite{Shi2023}.

\paragraph{Surprise.}
The surprise of paper $h$ at time $t$ is the negative 
log-coherence of its combination:
\begin{equation}\label{eq:surprise}
    S^{(t)}(h) = -\log \sum_d \prod_{i \in h} \theta_{id}^{(t)}.
\end{equation}
A high value of $S^{(t)}(h)$ indicates that the combination was 
improbable given the prevailing geometry of scientific knowledge---a 
low-probability event in the information-theoretic sense 
\supercite{Shi2023}. The salience term $\prod_i r_i^{(t)}$ is 
excluded from the surprise score to isolate the structural 
improbability of the combination from the mere obscurity of its 
constituent nodes; rare nodes that are unsurprisingly paired yield 
low surprise, while common nodes combined across distant conceptual 
territory yield high surprise. We compute two distinct surprise 
scores for each paper: \emph{content surprise}, $S^{(t)}_{\rm 
con}(h)$, using combinations of level-three concept nodes, and 
\emph{context surprise}, $S^{(t)}_{\rm ctx}(h)$, using combinations 
of cited journal nodes. These two scores are empirically 
uncorrelated, confirming that they capture orthogonal dimensions of 
novelty.

\paragraph{Prescience.}
Prescience measures the degree to which a paper's combination, 
surprising at publication, became routine in subsequent science. We 
compute two surprise scores for each paper: one using the embedding 
from the publication year $t_0$ and one using the embedding from two 
years later, $t_0 + 2$. Prescience is then defined as the decrease 
in surprisal over this interval:
\begin{align}
    P_{\rm con}(h) &= S^{(t_0)}_{\rm con}(h) 
                     - S^{(t_0+2)}_{\rm con}(h), \label{eq:pcon}\\
    P_{\rm ctx}(h) &= S^{(t_0)}_{\rm ctx}(h) 
                     - S^{(t_0+2)}_{\rm ctx}(h). \label{eq:pctx}
\end{align}
A large positive value of $P(h)$ indicates that the paper was ahead 
of its time: its combination was unusual at publication but became 
predictable as subsequent science reorganized around it 
\supercite{lockhart2025china}. Crucially, prescience reflects the 
uptake of the \emph{ideas} proposed in $h$, not citation-based 
credit accruing to $h$ itself. Early contributions to a popular 
direction may receive little downstream citation even as their 
combinations become routine---precisely the dynamic motivating our 
separate examination of prescience and scholarly credit.

As this measure captures what is expected and what is unexpected based on current trends, surprising combinations are those unlikely to occur in the given year of publication, according to the model \supercite{Shi2023}. In other words, a surprising work is a publication that makes 'weird', or improbable, combinations of contents or contexts.

Because the model can capture the improbability of published combinations each year, we can measure the change in papers' surprise scores after publication. A decrease in surprise scores between the year of publication and two years after defines \emph{prescience} scores -- our second measure of interest. A prescient publication is one that was unusual at the time of publication but becomes less unlikely and more commonly used, expected, or even assumed over time \supercite{lockhart2025china}. Importantly, prescience does not automatically imply the uptake of the paper, but rather reflects the uptake of ideas \textit{proposed} in the paper. Ideas may become popular in the future, but prescient papers that drove or anticipated this evolution may receive different levels of adoption and credit. 

As indicated by \supercite{FONTANA2020104063}, previous measures of innovation derived from average or random combinations of references may dangerously blur the lines between novelty and interdisciplinarity of a given paper. The measure of surprise in combinations of references we use in this paper captures how ideas distinctively connect diverging contexts in \emph{unexpected} ways. Surprising science may well capture creativity and non-conventionality \supercite{Simonton01042012}, as the more surprising an idea is, the more it perceived as creative in its embedded context and therefore less conventional. 
This suggests that surprise blends novelty with \emph{unconventionality}, or \emph{non-obviousness} (as described by \supercite{doi:10.1287/orsc.2021.1567}), instead of blending novelty and conventionality to reflect innovative science \supercite{https://doi.org/10.1111/jpim.12294, doi:10.1126/science.1240474}. This allows us to capture how combinations of references are \emph{non-obvious} and \emph{surprising}, rather than merely serving as a proxy for interdisciplinarity (i.e., the distance between connected fields) or merely capturing new or rare connections.

The two measures of surprise reflect different dimensions: reference and concept surprise show no correlation, and the same holds for prescience in concept and source combinations — see Figure 5 in the SI file. \supercite{Shi2023} report \supercite{10.1093/nar/25.17.3389} as an example in which context and content surprise diverge, as the paper introduced a computer system to search DNA and protein databases, proposing an innovative tool used across computational and biomedical sciences, rather than producing a discovery in the biomedical or the computational science  -- achieving high surprise in references (97th percentile) but low in concepts (15th percentile). 
Surprise scores are capable of distinguishing creative science from dissemination and reviews that share the same topics: \supercite{RePEc:oup:qjecon:v:121:y:2006:i:2:p:351-397.} significantly influenced the study of global income distribution and poverty, with a surprise score in the top of the distribution (99th percentile of reference surprise), while \supercite{10.1257/jel.50.1.115} provides a review of existing literature on global poverty and poverty economics, and  is associated with very low surprise (10th percentile of reference surprise)  -- as expected.


After computing these scores, papers as concept (references) combinations will have one \emph{raw} surprise score (at time of publication) and one \emph{raw} prescience score (two years after the year of publication). But papers touch on multiple topics and multiple fields (i.e., fine-grained concepts of level one), and different research areas have different normal ranges of surprise scores. As every paper has multiple fields, it will have a rank in surprise with respect to each field that it belongs to. For example, if the focal paper is related to both sociology and computer science, that paper's novelty score would put it in a different spot in each field's ranking. To allow for comparison of novelty score across fields, we compute within-field percentile ranks of novelty scores for papers in our analysis.\footnote{The percentiles in each field are computed considering the full snapshot of OpenAlex data from October 2023 (more than 80 million unique papers), including non-US and multi-authored papers.} The percentile-ranked version of scores forces them between 0 and 1 -- so the lowest observed surprise (prescience) value within a field will be 0.  

As each paper has multiple novelty scores—one for each scientific field (level-one concepts) it addresses—we take the maximum surprise score and the maximum prescient score for the paper. This means that each paper will be uniquely identified by one surprise score and one prescient score, each corresponding to the paper's maximal rank within related fields. 
The variables \emph{Surprise (References)} and \emph{Prescience (References)} denote the maximum surprise and prescience scores, respectively, with respect to the combination of journals in the reference list of the focal paper. \emph{Surprise (concepts)} and \emph{Prescience (concepts)} indicate the maximum surprise and prescience scores in the contents of the paper, using fine-grained concepts of level-three as contents. 


We compute prescience as the change in a paper's raw surprise score between its publication year and two years later. As a result, papers with very different initial surprise can have the same prescience score if their change in surprise places them similarly within the field-specific distribution. Converting prescience into percentile ranks allows us to compare papers across fields and provides insight into how papers by men and women are rewarded in science, as they occupy equal ranks within their distributions. Nevertheless, equal prescience ranks could reflect different initial surprise. To account for this, we examine gender differences in prescience conditional on increasing ranks of initial surprise (SI, Section G). In future work, we plan to explore this further, including extending the time horizon beyond two years to compute prescience.

\subsection{Rewards Measures}
For each publication, we collect information regarding the research areas of interest as concepts fine-grained of level one, and disciplines related to the paper (concepts fine-grained of level zero), public accessibility of the paper, number of publications from the institution in the same field and year as the focal paper (\emph{Department Size})\footnote{As papers belong to multiple fields, we compute the number of publications from the institution in the same fields as the focal paper and we take the maximal value across fields. 
}, year and journal of publication, journal impact factor, and number of citations. We compute \emph{Career Age} of an author as the number of years between the focal publication and the author's first publication. It is set to missing if the career age exceeds 60 or if they have only one paper in their lifetime, as both are usually due to author disambiguation errors. 

Following \supercite{doi:10.1287/mnsc.2015.2366}, we compute paper-level disruption scores, calculated five years after publication. Disruption varies between -1 and 1, where 1 indicates that the paper completely overshadows the previous literature, while a value equal to -1  indicates that the paper is only cited alongside the literature on which it builds. 

We also compute the \emph{two-step credit}, which is obtained as the ratio between the number of papers two-steps out that directly cite the focal paper and the number of papers that cite a paper that cites the focal paper.  
We compute the two-step credit from citation information collected up until October 2023. 
Two-step credit in the citation network is fairly difficult to secure (mean = .036, median = .025,  75th percentile = .043).
As being able to retain credit down the citation cascade and building saliency is less obvious, we compute two-step credit for papers with more than five two-step citations. Two-step credit reflects how much future publication building on the focal paper will \emph{disrupt} the focal paper: a value close to zero means that future work is overshadowing the contribution of the focal paper, a value close to one means that the paper is still being cited two-steps out in the citation network alongside the literature building upon it. 

Finally, we retrieve the two-year impact factor, which reflects how papers \emph{today} (i.e., at the time of publication of the focal paper) view the journal, considering citations to the journal's publications from two years prior. This means it will not have citations to our focal paper, but represent how other scholars treat the journal in the same year that the focal paper is being published, considering publications in the journal from two years prior.

\subsection{Multi-authored sample}
To quantify gender gaps in investment and rewards to surprising and prescient science by teams with varying female membership, we compute, for each paper, the share of women on the team. Because we can confidently infer the gender only for a subset of the authors (US with high confidence), we compute the share of women among team members with high-confidence inferred gender. Rather than identifying the overall female share in teams using low-confidence inferred genders, we focus on those team members for which gender is identified with higher confidence, i.e., for which an assumption on binary gender classes can be more \emph{easily} formed. In other words, we interpret this measure as an indicator of how strongly teams \emph{signal} a clear and unequivocal female presence to readers approaching the publication. For example, a publishing team whose members are John, Sam and Leslie (where both Leslie and Sam are common names for men and women in the US) has a much more vague female presence with respect to a team composed by Anna, Barbara and Sam. We compute the average department size and average career age of clearly gender-coded male and female authors within each publishing team.

In the SI file, we extend and compare the findings among solo-authored papers with a sample of multi-authored papers --   4\,085\,543 
unique multi-authored papers from 2000 onwards -- but the interpretation of these findings is limited by strong assumptions of a bi-univocal relation between the share of women as authors and women's contribution to novelty, without accounting for authorship order. 
We find 246 observations with more than a 1000 authors -- this is a negligible quantity to influence the regression estimates. Table 2 in the SI file reports the summary sample statistics for the multi-authored sample of papers.

\subsection{Regression Models}

To estimate gender differences in surprising and prescient science, we use cross-sections of papers observed over time and linearly regress innovation scores on the author's gender for solo-authored papers:

\begin{equation}\label{eq_1}
    NoveltyScore_{i} = \alpha + \beta Female_{i} + \sum_{j} \delta_{j} x_{i,j} + \gamma_{year(i)} + \epsilon_{i}
\end{equation}

where $NoveltyScore_{i}$ denotes one of four combination-based innovation scores for paper $i$: reference (concept) surprise and reference (concept) prescience. $Female_{i}$ is a binary indicator for the gender of the author of paper $i$. 


The vector $x_{i,j}$ represents paper- and author-level controls: year of publication, author's department size (i.e., number of publications from the institution in the same field as the focal paper), research area (concepts at fine-grained level 1, capturing scientific fields within disciplines, such as economic policy or biochemistry), and career age. Year effects capture temporal shifts in creativity and gender participation; department size captures variations in institutional environment, resources, and prestige, accounting for differences across institutions without requiring numerous institution-level controls that would reduce degrees of freedom. Research field controls for disciplinary differences in innovation practices. Career age accounts for compositional differences in experience and career stage, which are correlated with both gender and scholarly recognition \supercite{Huang2019HistoricalCO}. Unconditional estimates reported in Section D of the SI show that including these covariates does not substantially alter observed gender differences in novelty or prescience. We estimate four separate regressions, one for each innovation measure, treating each solo-authored paper as a distinct observation. Standard errors are clustered at the author level to account for within-author correlation in residuals. To ensure comparability across disciplines, novelty scores are normalized by within-field percentile ranks.

For science that is equally unexpected at the time of publication or equally 'ahead of its time', women and men-authored papers may receive different returns in terms of credit and recognition within scientific academia. To investigate gender differences in the scholarly returns to surprising and prescient science, we estimate the following model on cross-sections of papers observed over time:

\begin{equation}\label{eq_2}
\textit{ScholarlyReward}_{i} = \alpha + \beta Female_{i} + \gamma_{year(i)} + \lambda NoveltyScore_{i} + \tau (Female_{i} \times NoveltyScore_{i}) + \sum_{j} \delta_{j} x_{i,j} + \epsilon_{i}
\end{equation}

where $\textit{ScholarlyReward}_{i}$ represents one of our measures of credit and recognition for paper $i$: two-step credit, journal placement (natural logarithm of two-year Journal Impact Factor), or five-year disruption score. $NoveltyScore_{i}$ indicates one of the four combination-based innovation measures of paper $i$: context surprise, context prescience, content surprise, and content prescience. The interaction term between gender and novelty captures differences in the returns to surprising or prescient science.

Across all models, we include a common set of controls—year of publication, department size, career age, open-access status, and scientific field—to ensure that observed gender differences in scholarly rewards are not driven by differences in institutional context, disciplinary composition, or career stage. For the disruption model, we additionally include two-year citations and the journal's two-year impact factor, to account for variation in journal placement and early citations that could influence both uptake and disruptiveness. We consider the log-transformation of two-year journal impact to approximate normality. Because prescience reflects changes in surprise occurring after publication — and therefore after journal placement is determined — we do not analyze gender gaps in the relationship between prescience and journal placement, and focus solely on surprise at the time of publication. Standard errors are clustered at the author level.





\section{Supporting Information}\label{data4}

\renewcommand{\thetable}{S\arabic{table}}
\renewcommand{\thefigure}{S\arabic{figure}}
\maketitle 
\appendix
\section{Genderize inference results on first and middle names}\label{genderize}

\begin{figure}[H]
\includegraphics[width=\linewidth]{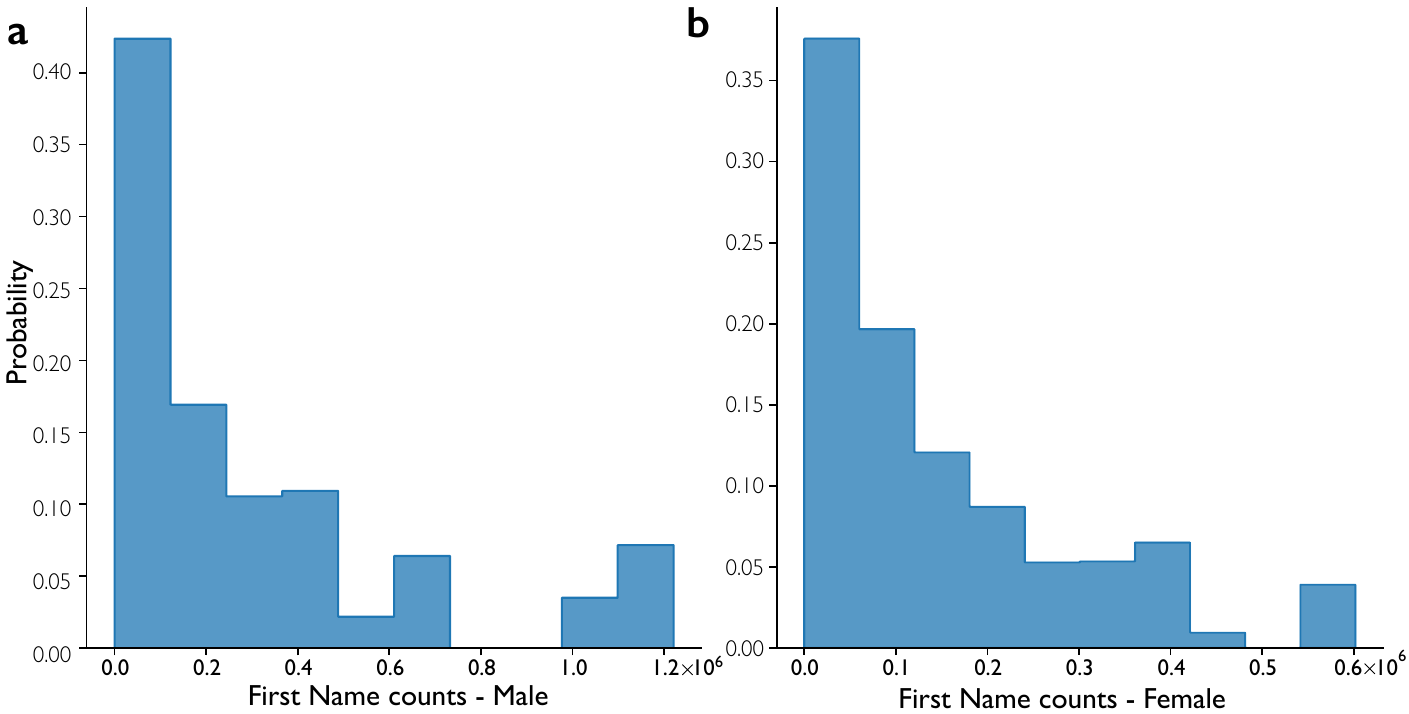}
\caption{Sample distribution of counts of first names for (a) Male and (b) Female assigned genders.}
  \label{fig:descr_fig_2}
\end{figure}

\begin{figure}[H]
\includegraphics[width=\linewidth]{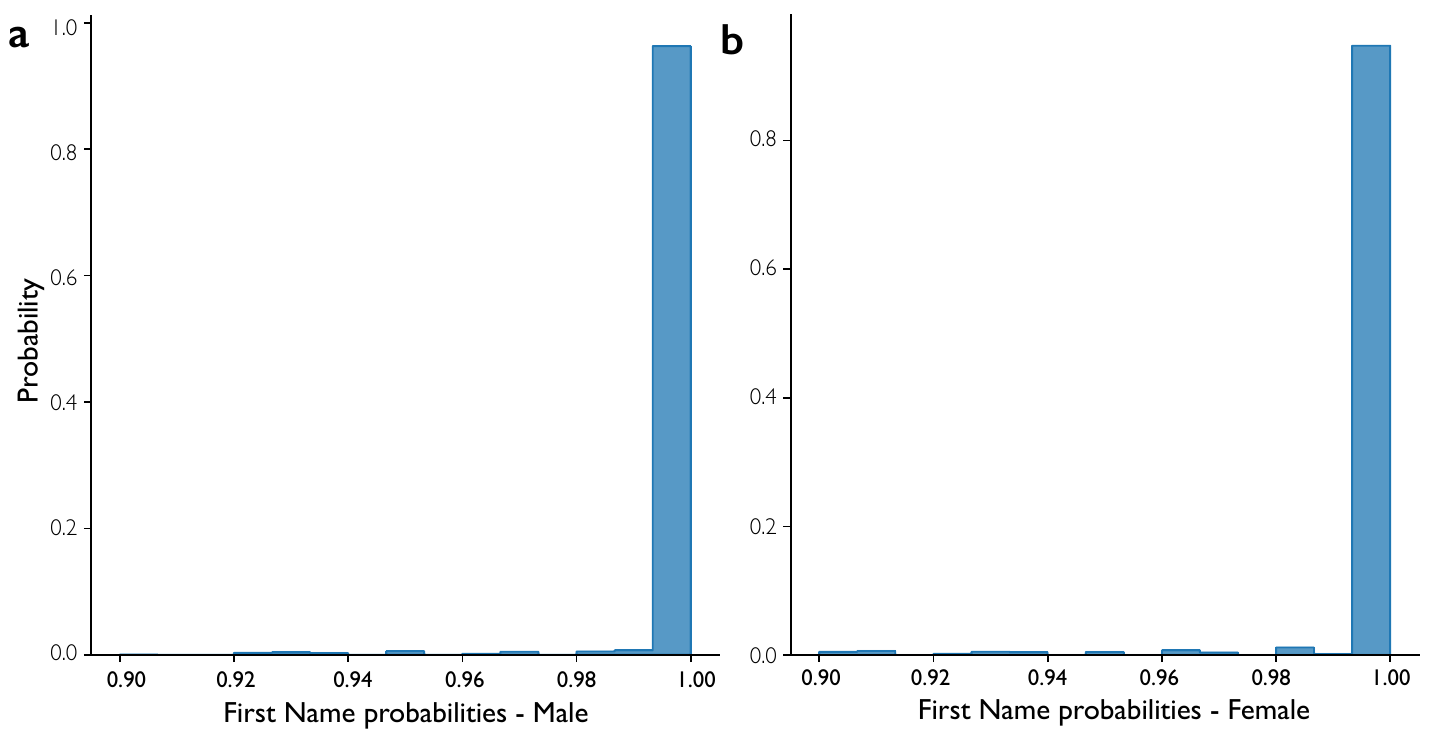}
  \caption{Sample distribution of probabilities for first names, for (a) Male and (b) Female assigned genders.}
  \label{fig:descr_fig_2}
\end{figure}

\begin{figure}[H]
\includegraphics[width=\linewidth]{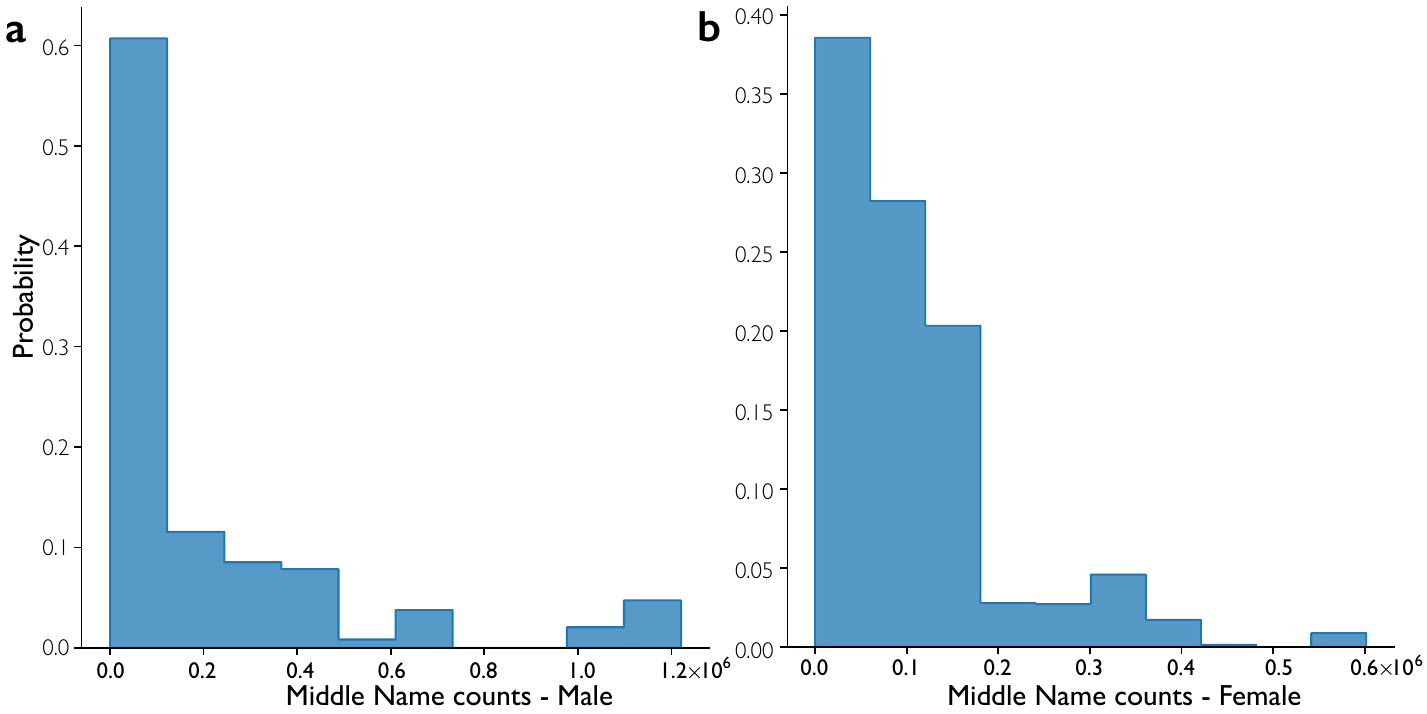}
  \caption{Sample distribution of counts of middle names for (a) Male and (b) Female assigned genders.}
  \label{fig:descr_fig_2}
\end{figure}

\begin{figure}[H]
\includegraphics[width=\linewidth]{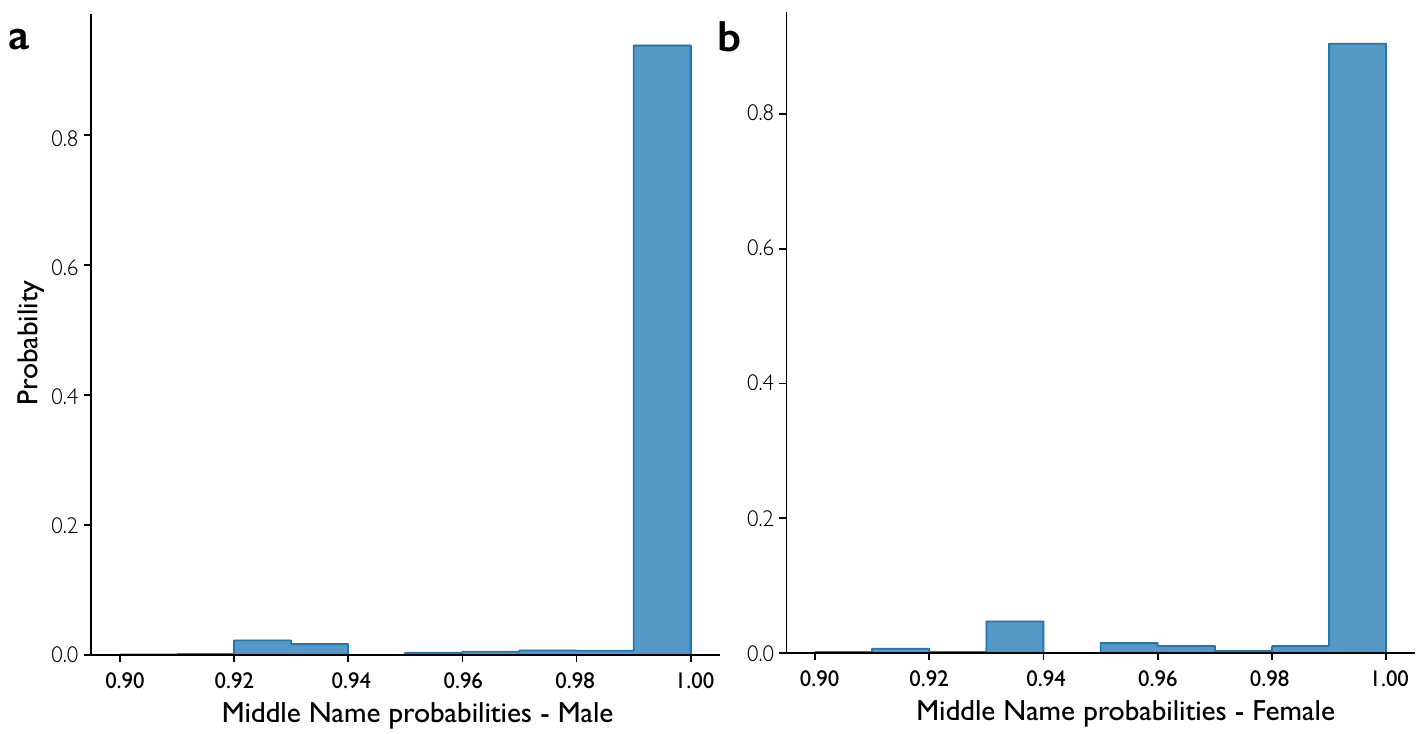}
  \caption{Sample distribution of probabilities for middle names, for (a) male and (b) female assigned genders.}
  \label{fig:descr_fig_2}
\end{figure}
\section{Descriptive statistics}\label{app:descr_stats}
\subsection{Paper level statistics}


\begin{figure}[H]
\includegraphics[width=0.5\linewidth]{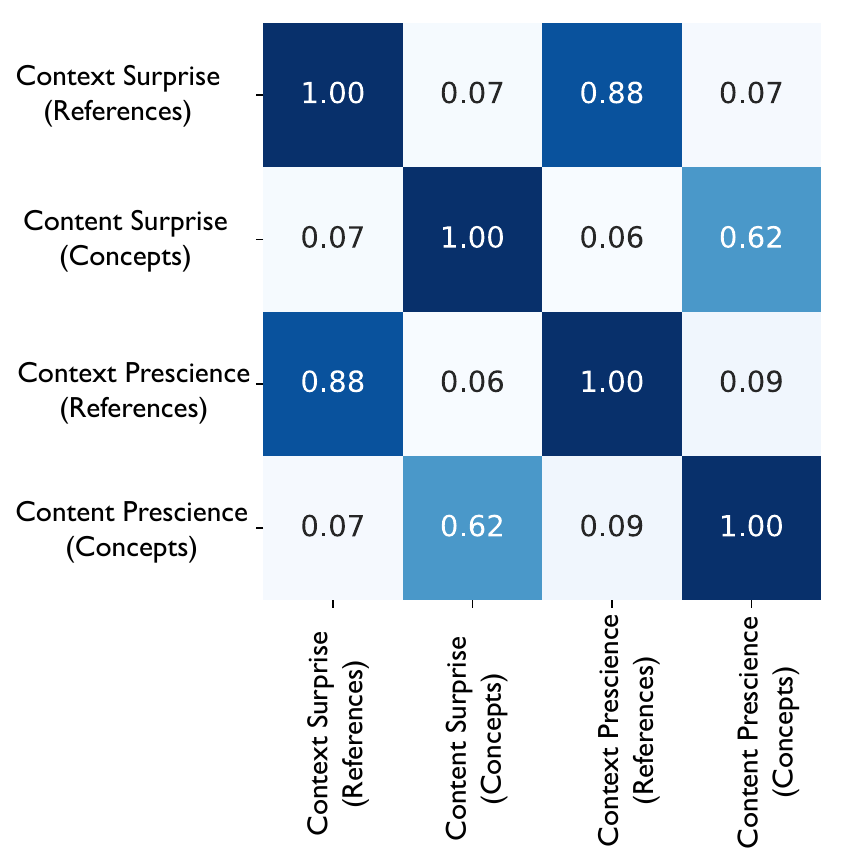}
  \caption{Correlation Plot of sample surprise and prescience in references and concepts.}
  \label{fig:fig_corrplot}
\end{figure}

\begin{table}[H]\centering
\def\sym#1{\ifmmode^{#1}\else\(^{#1}\)\fi}
\caption{Summary statistics of variables in our sample of solo-authored papers (226\,208 publications).}\label{tab:descr_stats}
\begin{adjustbox}{width=\textwidth}
\begin{tabular}{l*{1}{ccccccc}}
\toprule

                    &\multicolumn{1}{c}{Mean}&\multicolumn{1}{c}{Standard Deviation}&\multicolumn{1}{c}{25th perc.}&\multicolumn{1}{c}{50th perc.}&\multicolumn{1}{c}{75th perc.}&\multicolumn{1}{c}{Min value}&\multicolumn{1}{c}{Max value}\\

\midrule 
Surprise (References)    &    .5878723&    .3068257&     .321875&     .636409&    .8734216&           0&           1\\
Prescience (References)  &    .5501471&    .3054902&    .2816802&    .5752494&    .8305932&    7.27e-06&           1\\
Surprise (concepts)   &    .5345609&     .288986&     .284888&    .5426767&    .7905379&    .0000121&           1\\
Prescience (concepts) &    .5501005&      .28599&    .3087056&    .5652289&    .8025451&    .0000314&           1\\
Female              &    .2721257&    .4450552&           0&           0&           1&           0&           1\\
two-year Journal Impact Factor   &    4.190875&    4.214197&    2.084337&    3.038461&    4.621324&           0&    132.6792\\
ln(two-year Journal Impact Factor)         &    1.189344&    .6338853&    .7375989&    1.113365&    1.531683&           0&    4.887935\\
five-year Journal Impact Factor    &    4.552189&    4.447293&    2.287671&    3.355212&           5&           1&     134.672\\
ln(five-year Journal Impact Factor)          &    1.273767&     .631945&    .8275344&    1.210515&    1.609438&           0&    4.902842\\
Two-Step Credit Score  &    .0369705&    .0480954&    .0132538&    .0232558&    .0419287&    .0003139&    .9090909\\
Three-Years Disruption Score                 &   -.0009666&    .0386683&   -.0045317&   -.0006285&           0&          -1&           1\\
five-year Disruption Score                 &   -.0004151&    .0372708&   -.0043263&   -.0006929&    .0001977&          -1&           1\\
Publication Year                &    2009.975&    6.028572&        2005&        2010&        2015&        2000&        2021\\
two-year citations            &    7.029358&    17.41921&           1&           3&           7&           0&        1523\\
Department Size&    1512.716&    2746.692&         140&         586&        1687&           1&       40865\\
Career Age          &    23.73893&    15.14362&          11&          22&          35&           0&          60\\
Open Access            &     .334913&    .4719611&           0&           0&           1&           0&           1\\
Career Cohort       &    4.318967&    1.979498&           3&           5&           6&           1&           7\\
Past citations&    2149.849&    5860.528&          38&         367&        1827&           0&      178215\\
WomenShare (Surprise References)       &    .3081784&    .0963309&    .2341253&    .3006158&    .3766001&    .0761015&    .6301304\\
WomenShare (Prescience References)       &    .3111562&    .0995479&    .2328184&    .3075731&    .3835366&    .0761015&    .6301304\\
WomenShare (Surprise concepts)      &     .362081&    .1065606&    .2815366&    .3617006&    .4400461&    .0714286&    .6301304\\
WomenShare (Prescience concepts)      &    .3566239&    .1079178&    .2783941&    .3567669&    .4350593&    .0761015&    .6301304\\
\bottomrule
\end{tabular}
\end{adjustbox}
\end{table}

\begin{table}[H]\centering
\def\sym#1{\ifmmode^{#1}\else\(^{#1}\)\fi}
\caption{Summary statistics of \emph{all} (gender-homogeneous and mixed) multi-authored papers (4\,085\,543 publications).}\label{tab:descr_stats_ma}
\begin{adjustbox}{width=\textwidth}
\begin{tabular}{l*{1}{ccccccc}}
\toprule

                    &\multicolumn{1}{c}{Mean}&\multicolumn{1}{c}{Standard Deviation}&\multicolumn{1}{c}{25th perc.}&\multicolumn{1}{c}{50th perc.}&\multicolumn{1}{c}{75th perc.}&\multicolumn{1}{c}{Min value}&\multicolumn{1}{c}{Max value}\\

\midrule 
Surprise (References)      &    .6412111&    .2666128&    .4491986&    .6975946&    .8695936&           0&           1\\
Prescience (References)    &    .5821974&    .2787709&    .3576885&    .6131484&     .827352&           0&           1\\
Surprise (concepts)    &    .5673799&    .2859917&    .3346121&    .5972016&    .8191983&           0&           1\\
Prescience (concepts)  &    .6020616&    .2761986&    .3832973&    .6380665&    .8456904&           0&           1\\
\textit{Female Share}       &    .2981608&    .3441279&           0&          .2&          .5&           0&           1\\
two-year Journal Impact Factor   &    5.097368&    4.224422&    2.927757&    4.120163&    5.832447&           0&    467.3684\\
ln(two-year Journal Impact Factor)         &    1.447362&    .5671963&    1.075355&    1.416447&    1.763747&           0&    6.147118\\
five-year Journal Impact Factor   &    5.459966&    4.297991&    3.177496&    4.456612&    6.209934&           1&     265.071\\
ln(five-year Journal Impact Factor)         &    1.518399&    .5645669&    1.156093&    1.494389&     1.82615&           0&    5.579998\\
Two-Step Credit Score  &    .0368914&    .0400584&    .0156028&    .0255755&    .0434132&    .0001316&          .9\\
Three-Years Disruption Score               &   -.0048742&    .0251893&   -.0067568&   -.0017637&   -.0000682&          -1&           1\\
five-year Disruption Score               &   -.0045224&    .0242881&    -.006507&   -.0017301&   -.0001398&          -1&           1\\
Publication Year             &    2012.145&     6.08752&        2007&        2013&        2017&        2000&        2021\\
two-year Citations          &    11.66267&     40.0359&           2&           6&          13&           0&       40291\\
Department Size&    3505.908&    5086.406&         696&        1750&        4033&           1&       49445\\
Career Age          &    22.22036&    11.59788&          14&    21.33333&          29&           0&          60\\
Open Access              &    .5414223&    .4982813&           0&           1&           1&           0&           1\\
Team Size       &    6.140079&    20.43644&           3&           5&           7&           2&        3222\\
Num Gender Unknown        &    3.108408&    20.11316&           1&           2&           4&           0&        3221\\
\bottomrule
\end{tabular}
\end{adjustbox}
\end{table}

\begin{figure}[H]
\includegraphics[width=0.75\linewidth]{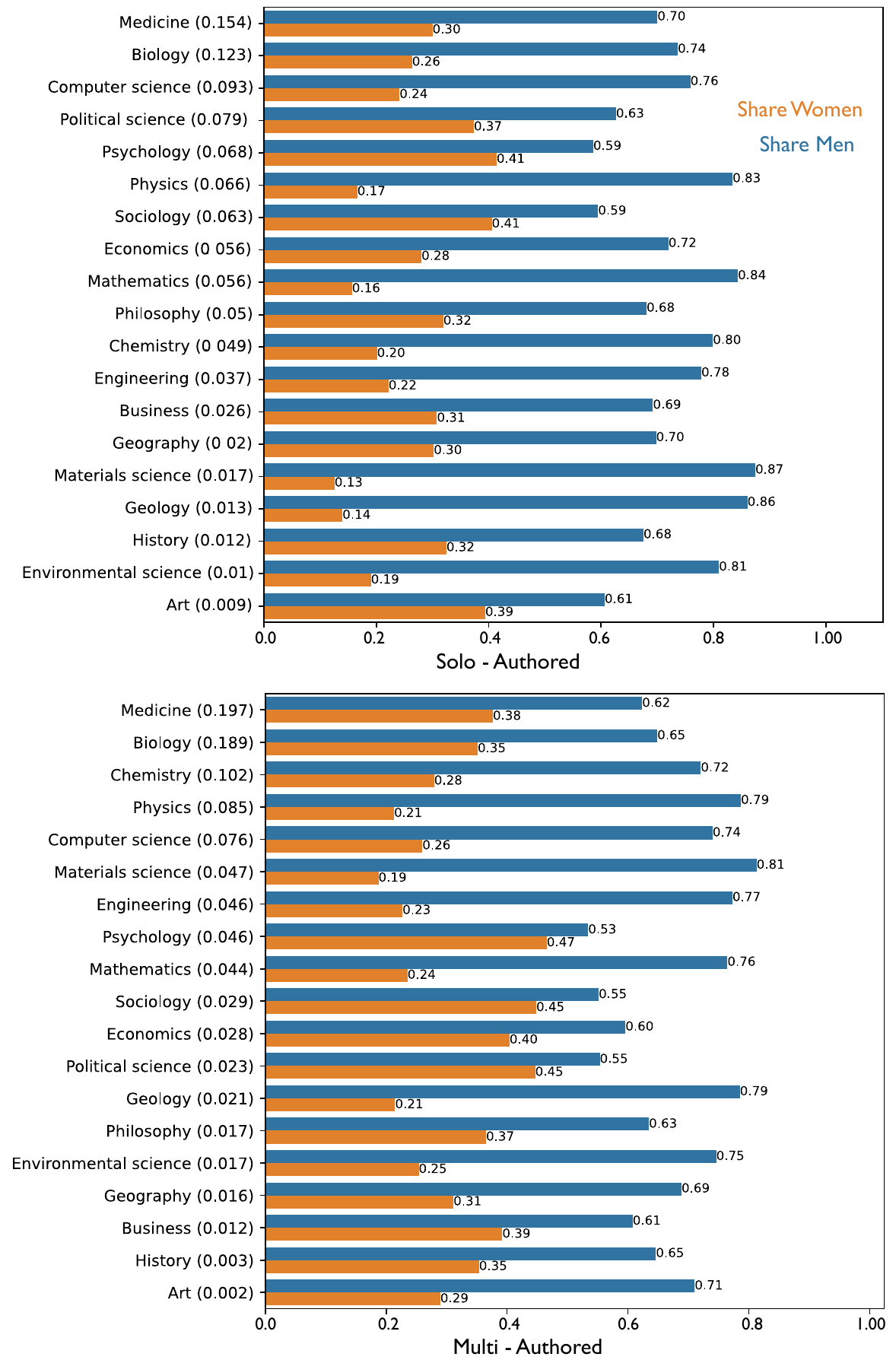}
   \caption{Sample share of papers authored by women (orange) and men (blue) in (a) solo-authored, and (b) multi-authored papers by discipline (level-zero fields). The share is computed as the number of papers in discipline j by women (men) divided by the total number of papers in discipline j. Disciplines are ordered by sample share of solo- (multi-) authored papers in each field (in parenthesis). 
   Multi and solo-authored papers display similar gap in the distribution of male and female authors -- with women publishing less papers than men across all disciplines. 
   }\label{fig:share_field}
\end{figure}

\subsection{Author level statistics}

\begin{table}[H]\centering
\def\sym#1{\ifmmode^{#1}\else\(^{#1}\)\fi}
\caption{Summary statistics of solo-authors (141\,786  unique authors), multi-authors (1878225 unique authors),  all authors (1909424 unique authors), and authors of both solo and multi-authored papers (110587 unique authors). 
Women represent 44\% of al authors in our data collections on solo and multi-authored papers (\emph{All Authors}). In solo-authored papers, men are more prominent authors, publishing roughly 2 solo-papers each (mean 1.99, sd 2.54, min 1, max 206), where women solo-authors have on average 1.5 solo-papers (mean 1.52, sd 1.387, min 1, max 67); 71\% (=31\,988/44\,976) of solo-women publish both as solo-author and in multi-authored papers, while 81\% (=78\,599/96\,810) of solo-men publish as both. Women in our solo-authored sample are publishing less papers than men -- both alone and in research teams. Solo-authors are much more productive authors in number of multi-authored papers published with respect to the baseline of people who usually publish the same kind of paper. The sample features 73\% (=32\,833/44\,976) of women with just one solo-authored paper, and 62\% (60\,023/96\,810) of men with just one solo-authored work.
}\label{table_descr_auth}
\begin{adjustbox}{width=\textwidth}
\begin{tabular}{l*{1}{ccccccc}}
\toprule

                    &\multicolumn{1}{c}{Mean}&\multicolumn{1}{c}{Standard Deviation}&\multicolumn{1}{c}{25th perc.}&\multicolumn{1}{c}{50th perc.}&\multicolumn{1}{c}{75th perc.}&\multicolumn{1}{c}{Min value}&\multicolumn{1}{c}{Max value}\\
\midrule 
\multicolumn{1}{c}{Solo-authors} & & & & & \\
\midrule
Publications as multi-author count &  28.84956 & 45.46335 &           4&          12&          35&           1&         925\\
Women's share as authors   &    .3172104&    .4653918&           0&           0&           1&           0&           1\\
Female solo-authors Publications count  (total authors: 44976)  & 1.52 & 1.38 & 1 & 1 & 2 & 1& 67 \\
Male solo-authors Publications count (total authors: 96810)  & 1.99 & 2.55&1&1&2&1&
 206 \\
\midrule 
\multicolumn{1}{c}{Multi-authors} & & & & & \\
\midrule
Publications as multi-author count   &    6.720795&    17.08841&           1&           2&           5&           1&         925\\
Publications as solo-author count    &    2.005308&    2.488205&           1&           1&           2&           1&         206\\
Women's share as authors      &    .4445192&    .4969125&           0&           0&           1&           0&           1\\
Female multi-authors Publications count (total authors: 834907)
&4.982&11.84&1&2&4&1&740\\
Male multi-authors Publications count (total authors: 1043318) 
& 8.112 & 20.228 & 1 & 2 & 6 & 1 & 925  \\
\midrule 
\multicolumn{1}{c}{All authors} & & & & & \\
\midrule
Number of multi-authored papers  &    6.720795&    17.08841&           1&           2&           5&           1&         925\\
Average share of women in teams &    .3771454&    .2912426&    .1333333&    .3333333&    .5740741&           0&           1\\
Women's share as authors        &     .444058&    .4968608&           0&           0&           1&           0&           1\\
Female multi-authored Publications count  (total authors: 834907) & 4.98&11.83&1&2&4&1&740\\
Female solo-authored Publications count (total authors: 44976) &1.518&1.387&1&1&2&1&67 \\
Male multi-authored Publications count  (total authors: 1043318) &  8.112&20.228&1&2&6&1&925\\
Male solo-authored Publications count (total authors: 96810)  & 1.9953&2.5482&1&1&2&1&206\\
\midrule
\multicolumn{1}{c}{Sample of authors in both solo and multi-authored papers} & & & & & & &\\
\midrule
Number of Publications as multi-author &    28.84956&    45.46335&           4&          12&          35&           1&         925\\
Number of Publications as solo-author  &    2.005308&    2.488205&           1&           1&           2&           1&         206\\
Women's share as authors    &    .2892564&    .4534192&           0&           0&           1&           0&           1\\
Number of multi-authored Publications for women (total authors: 31988)  & 21.504&34.805&3&
8&26&1&680\\
Number of solo-authored Publications for women (total counts: 31988) &1.648&1.5728&1&1&2&1& 67\\
Number of multi-authored Publications for men (total authors: 78599) & 31.838&48.828&4&14&39&1&925 \\
Number of solo-authored Publications for men (total authors: 78599)  & 2.150&2.7624&1&1&2&1& 206 \\
Number of Multi-authored Publications by women with one solo-authored paper  (total authors: 21948)
&17.063&28.0429&2&7&20&1&1578\\
Number of Multi-authored Publications by men with one solo-authored paper  (total authors: 46001)
&24.247&37.587&3& 11&30
&1&745\\
\bottomrule
\end{tabular}
\end{adjustbox}
\end{table}


\begin{figure}[H]
\includegraphics[width=1.0\linewidth]{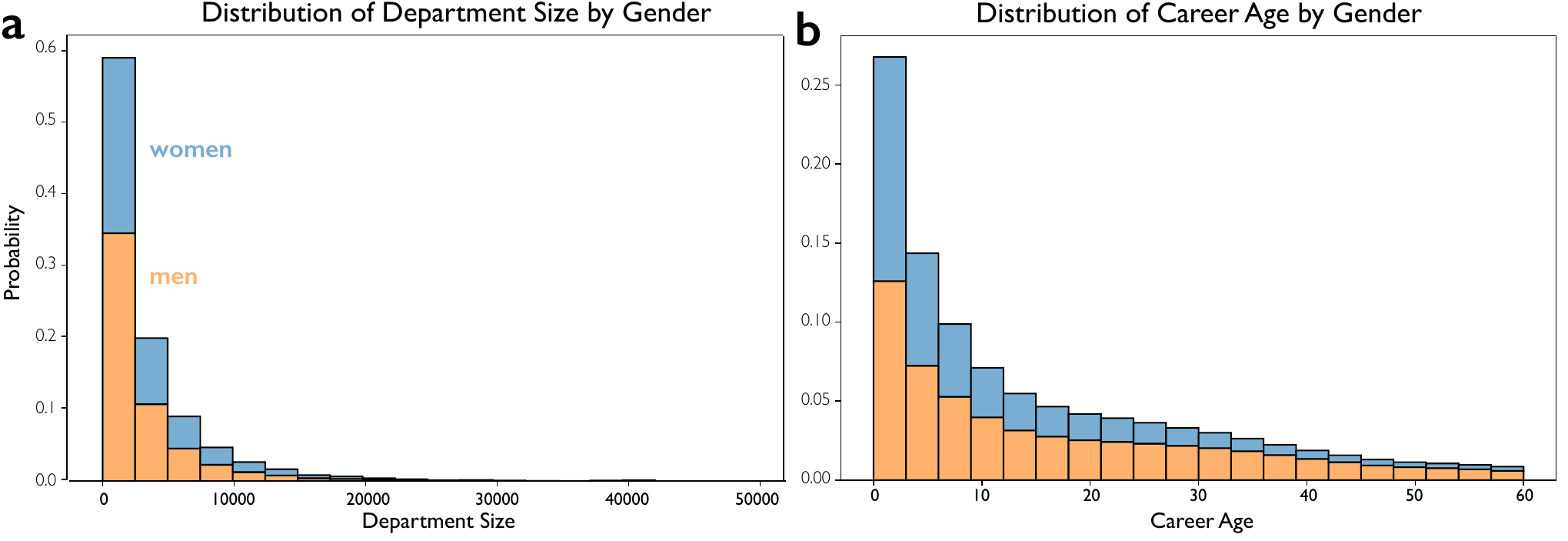}
  \caption{Sample distribution of (a) average author's department size, and (b) average author's career age, of unique US authors in multi-authored papers. }
  \label{fig:fig_auth_dep_car_ma}
\end{figure}

\begin{figure}[H]
\includegraphics[width=1.0\linewidth]{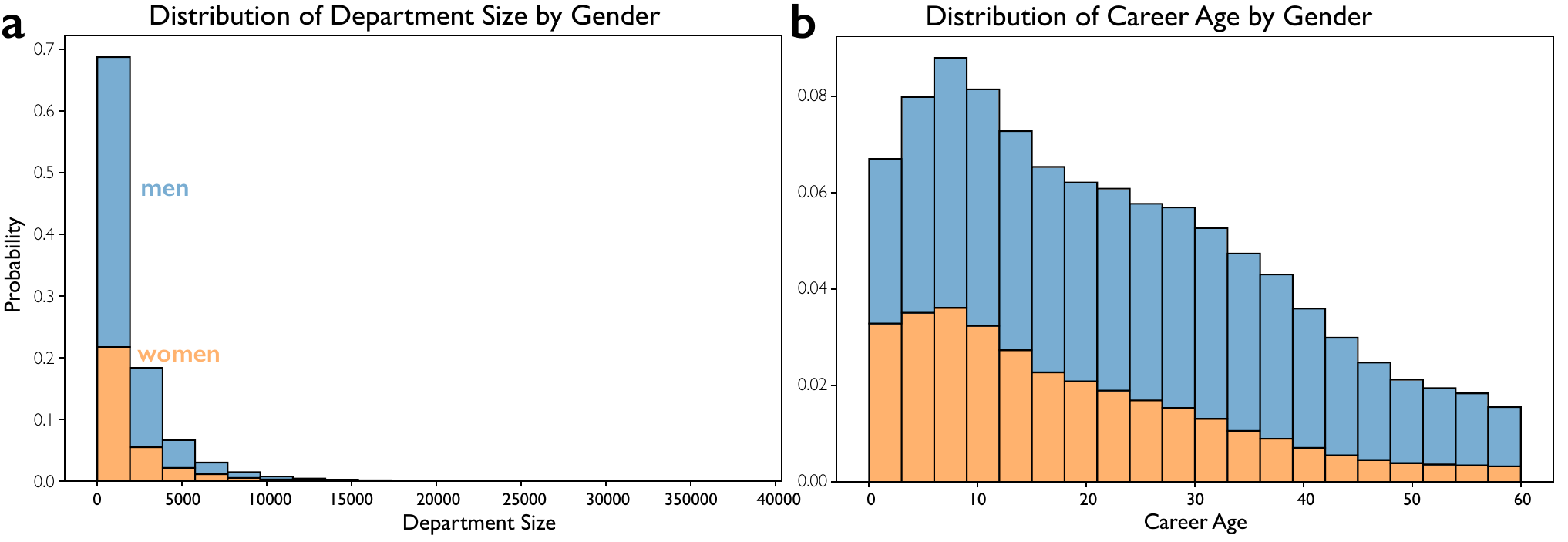}
  \caption{Sample distribution of (a) average author's department size, and (b) average author's career age of unique US authors in solo-authored papers. }
  \label{fig:fig_auth_dep_car_solo}
\end{figure}


\begin{figure}[H]
\includegraphics[width=0.75\linewidth]{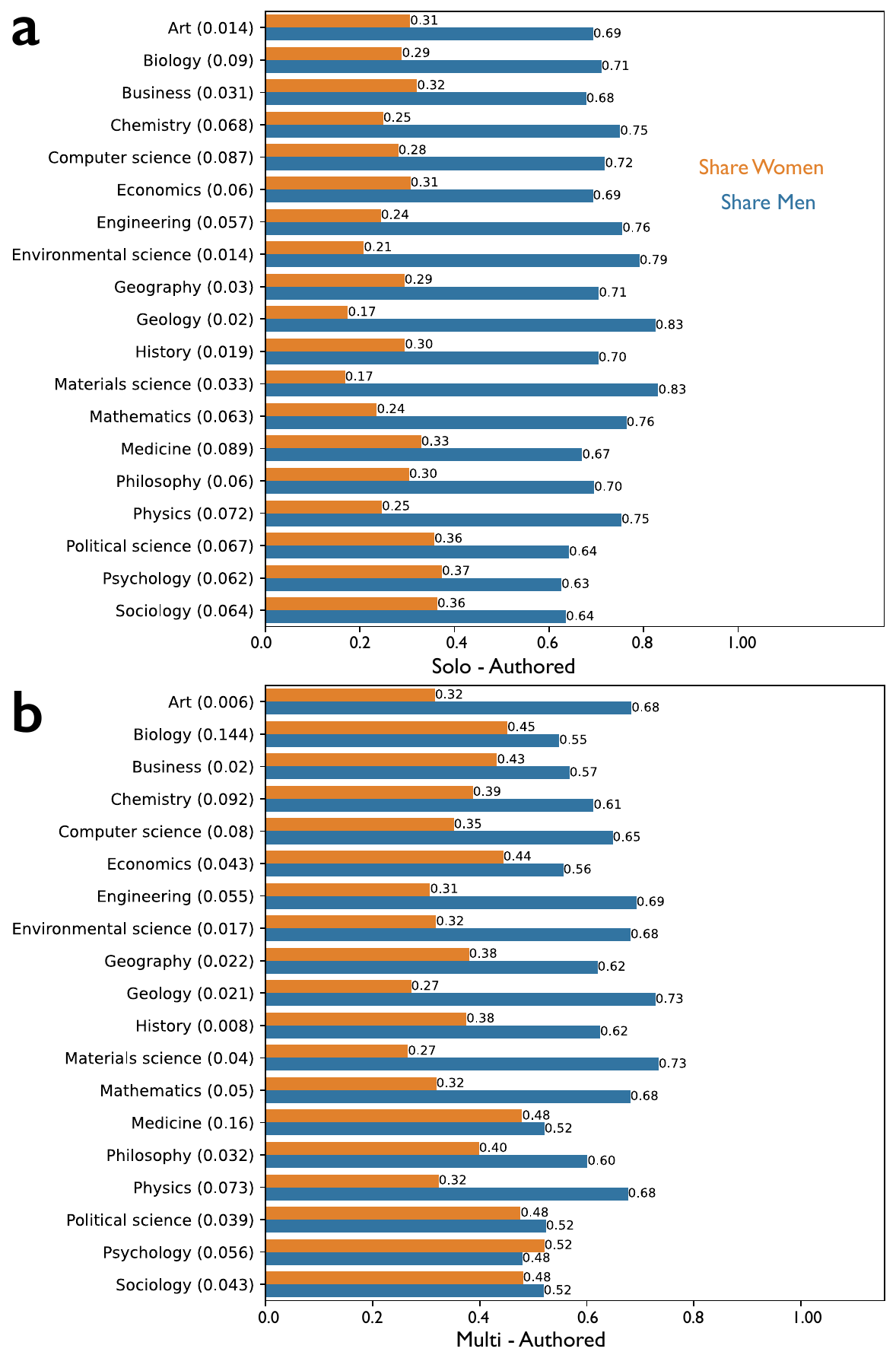}
  \caption{Sample share of authors in disciplines by gender in (a) solo-authored and (b) multi-authored papers. We consider unique authors in our sample. The numerator is number of unique women (men) authors in a given discipline, the denominator is the total number of unique authors in a given discipline. Next to disciplines, we report the sample share of authors in each discipline.}\label{fig:fig_auth_disc}
\end{figure}

\begin{figure}[H]
\includegraphics[width=1.0\linewidth]{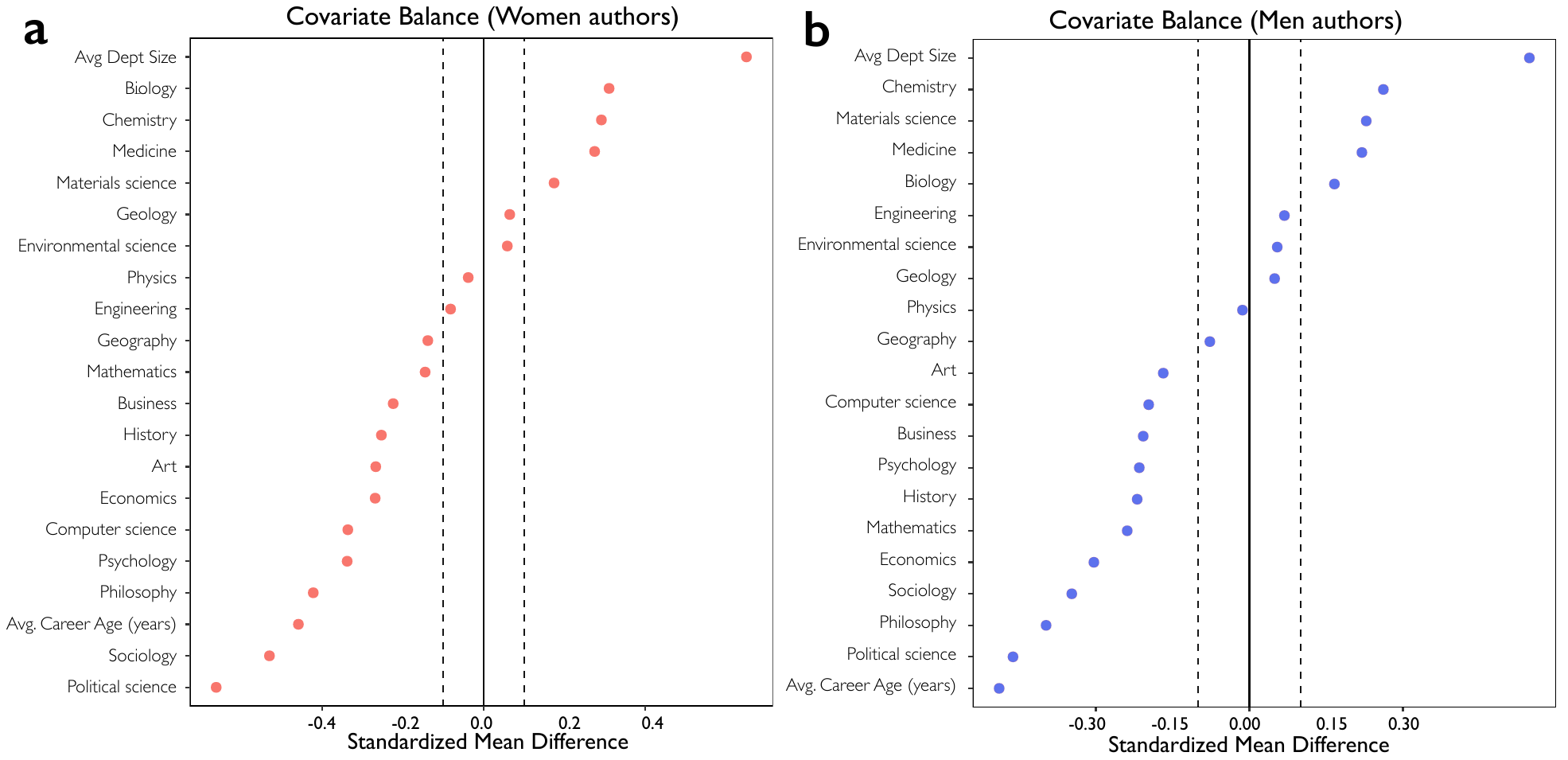}
  \caption{Standardized covariate mean differences between authors of solo-papers and multi-authored papers, considering (a) women US authors and (b) men US authors, alongside a 0.1 threshold.  Discipline variables indicate the average share of publications of across disciplines  -- for an author \emph{j}, the numerator is the number of papers published in a given discipline by author \emph{j}, the denominator is the total number number of publications by author \emph{j}. A positive standardized mean difference indicates a higher sample average among multi-authored papers.}
  \label{fig:fig_auth_loveplot}
\end{figure}

\section{Solo-authored papers: Main Models Regression Tables}
In this section, we report the tables with the linear regression estimates of equation (1) of the main text, as well as those examining the relationship of surprise and prescience with scholarly rewards by gender in solo-authored papers -- corresponding to the model in equation (2) of the main text. The tables are discussed in the main text.

\begin{table}[H]\centering
\def\sym#1{\ifmmode^{#1}\else\(^{#1}\)\fi}
\caption{Linear regression estimates of eq. (1) (main text) for content and context surprise and prescience scores in solo-authored papers by gender, with author-level clustered standard errors. Research field fixed effects and dummies for publishing year are omitted.}\label{tab1_reg1}
\begin{adjustbox}{width=\textwidth}
\begin{tabular}{l*{4}{D{.}{.}{-1}}}
\toprule
                    &\multicolumn{1}{c}{(1)}&\multicolumn{1}{c}{(2)}&\multicolumn{1}{c}{(3)}&\multicolumn{1}{c}{(4)}\\
                    &\multicolumn{1}{c}{Surprise (References)}&\multicolumn{1}{c}{Prescience (References)}&\multicolumn{1}{c}{Surprise (concepts)}&\multicolumn{1}{c}{Prescience (concepts)}\\
\midrule
Female=1            &      0.0258\sym{***}&      0.0241\sym{***}&    -0.00525\sym{**} &    -0.00773\sym{***}\\
                    &     (13.13)         &     (12.66)         &     (-3.24)         &     (-4.65)         \\
\addlinespace
DepSize& -0.00000158\sym{***}&-0.000000757\sym{*}  & 0.000000221         &  0.00000212\sym{***}\\
                    &     (-4.55)         &     (-2.28)         &      (0.80)         &      (7.21)         \\
\addlinespace
CareerAge          &    -0.00131\sym{***}&    -0.00128\sym{***}&   -0.000166\sym{***}&   -0.000349\sym{***}\\
                    &    (-22.44)         &    (-22.66)         &     (-3.47)         &     (-7.06)         \\
\addlinespace
Constant            &       0.547\sym{***}&       0.556\sym{***}&       0.487\sym{***}&       0.547\sym{***}\\
                    &    (140.10)         &    (142.04)         &    (128.83)         &    (148.65)         \\
\midrule
Observations        &      226208         &      226208         &      226208         &      226208         \\
\bottomrule
\multicolumn{5}{l}{\footnotesize \textit{t} statistics in parentheses}\\
\multicolumn{5}{l}{\footnotesize \sym{*} \(p<0.05\), \sym{**} \(p<0.01\), \sym{***} \(p<0.001\)}\\
\end{tabular}
\end{adjustbox}
\end{table}

\begin{table}[H]\centering
\def\sym#1{\ifmmode^{#1}\else\(^{#1}\)\fi}
\caption{Linear regression estimates of eq. (2) (main text) for two-step credit score, five-year disruption scores, and two-year journal impact factor on \emph{surprise score} by gender in solo-authored papers. We compute author-level clustered standard errors. We control for career age, average number of citations from institution of author in year of publication and in same field of paper, publication year with 2020 as baseline year (omitted), open access, and dummy variables for level-one fields (omitted). For disruption, we include additional controls given by the two-year journal impact factor, two-year citations.}\label{tab_reg2_surprise}
\begin{adjustbox}{width=\textwidth}
\begin{tabular}{l*{6}{D{.}{.}{-1}}}
\toprule
                    &\multicolumn{1}{c}{(1)}&\multicolumn{1}{c}{(2)}&\multicolumn{1}{c}{(3)}&\multicolumn{1}{c}{(4)}&\multicolumn{1}{c}{(5)}&\multicolumn{1}{c}{(6)}\\
                    &\multicolumn{1}{c}{TwoStepsCreditScore}&\multicolumn{1}{c}{TwoStepsCreditScore}&\multicolumn{1}{c}{5-year Disruption}&\multicolumn{1}{c}{5-year Disruption}&\multicolumn{1}{c}{\textit{ln(2-year Journal Impact Factor)}}&\multicolumn{1}{c}{\textit{ln(2-year Journal Impact Factor)}}\\
\midrule
Female=1            &    -0.00151         &    -0.00146\sym{*}  &     0.00618\sym{***}&     0.00179\sym{***}&     -0.0665\sym{***}&     -0.0656\sym{***}\\
                    &     (-1.28)         &     (-2.23)         &      (8.35)         &      (4.47)         &     (-7.26)         &    (-10.15)         \\
\addlinespace
\textit{Surprise (References)}    &     -0.0121\sym{***}&                     &     -0.0127\sym{***}&                     &     0.00415         &                     \\
                    &    (-16.85)         &                     &    (-25.62)         &                     &      (0.70)         &                     \\
\addlinespace
Female=1 $\times$ \textit{Surprise (References)}&     0.00102         &                     &    -0.00778\sym{***}&                     &      0.0410\sym{***}&                     \\
                    &      (0.72)         &                     &     (-8.41)         &                     &      (3.42)         &                     \\
\addlinespace
DepSize& -0.00000137\sym{***}& -0.00000137\sym{***}&   -4.85e-08\sym{*}  &   -2.89e-08         &   0.0000261\sym{***}&   0.0000261\sym{***}\\
                    &    (-18.88)         &    (-18.99)         &     (-2.06)         &     (-1.22)         &     (33.48)         &     (33.53)         \\
\addlinespace
CareerAge          &   -0.000127\sym{***}&   -0.000115\sym{***}&   0.0000345\sym{***}&   0.0000534\sym{***}&     0.00226\sym{***}&     0.00225\sym{***}\\
                    &     (-9.76)         &     (-8.85)         &      (5.61)         &      (8.56)         &     (21.37)         &     (21.25)         \\
\addlinespace

\textit{OpenAccess}             &    -0.00328\sym{***}&    -0.00330\sym{***}&   -0.000553\sym{**} &   -0.000415\sym{*}  &       0.147\sym{***}&       0.146\sym{***}\\
                    &    (-10.08)         &    (-10.10)         &     (-3.17)         &     (-2.37)         &     (45.14)         &     (45.08)         \\
\addlinespace

\textit{Surprise (Concepts)}   &                     &    -0.00180\sym{**} &                     &   0.0000814         &                     &     -0.0298\sym{***}\\
                    &                     &     (-3.27)         &                     &      (0.25)         &                     &     (-5.59)         \\
\addlinespace
Female=1 $\times$ \textit{Surprise (Concepts)}&                     &    0.000602         &                     &    -0.00141\sym{*}  &                     &      0.0461\sym{***}\\
                    &                     &      (0.59)         &                     &     (-2.25)         &                     &      (4.76)         \\
\addlinespace
impact\_factor\_2y    &                     &                     &   -0.000167\sym{***}&  -0.0000618\sym{*}  &                     &                     \\
                    &                     &                     &     (-5.86)         &     (-2.20)         &                     &                     \\
\addlinespace
2y\_cites            &                     &                     &  -0.0000128         &  -0.0000585\sym{***}&                     &                     \\
                    &                     &                     &     (-0.86)         &     (-4.00)         &                     &                     \\
\addlinespace
Constant            &      0.0399\sym{***}&      0.0334\sym{***}&     0.00542\sym{***}&    -0.00162\sym{**} &       0.911\sym{***}&       0.928\sym{***}\\
                    &     (46.87)         &     (48.05)         &      (8.03)         &     (-2.68)         &    (108.58)         &    (115.02)         \\
\midrule
Observations        &       96513         &       96513         &      212916         &      212916         &      212402         &      212402         \\
\bottomrule
\multicolumn{7}{l}{\footnotesize \textit{t} statistics in parentheses}\\
\multicolumn{7}{l}{\footnotesize \sym{*} \(p<0.05\), \sym{**} \(p<0.01\), \sym{***} \(p<0.001\)}\\
\end{tabular}
\end{adjustbox}
\end{table}

\begin{table}[H]\centering
\def\sym#1{\ifmmode^{#1}\else\(^{#1}\)\fi}
\caption{Linear regression model estimates of eq. (2) (main text) for two-step credit score, five-year disruption scores on \emph{prescience score} by gender in solo-authored papers. We compute author-level clustered standard errors. We control for career age, average number of citations from institution of author in year of publication and in same field of paper, publication year with 2020 as baseline year (omitted), open access, and dummy variables for level-one fields (omitted). For disruption, we include additional controls given by the two-year journal impact factor, two-year citations.}\label{tab_reg2_prescience}
\begin{adjustbox}{width=\textwidth}
\begin{tabular}{l*{4}{D{.}{.}{-1}}}
\toprule
                    &\multicolumn{1}{c}{(1)}&\multicolumn{1}{c}{(2)}&\multicolumn{1}{c}{(3)}&\multicolumn{1}{c}{(4)}\\
                    &\multicolumn{1}{c}{TwoStepsCreditScore}&\multicolumn{1}{c}{TwoStepsCreditScore}&\multicolumn{1}{c}{5-year Disruption}&\multicolumn{1}{c}{5-year Disruption}\\
\midrule
Female=1            &   -0.000286         &    -0.00243\sym{**} &     0.00568\sym{***}&     0.00142\sym{**} \\
                    &     (-0.26)         &     (-3.15)         &      (8.31)         &      (3.27)         \\
\addlinespace
\textit{Prescience (References)}  &     -0.0114\sym{***}&                     &     -0.0120\sym{***}&                     \\
                    &    (-16.94)         &                     &    (-26.05)         &                     \\
\addlinespace
Female=1 $\times$ \textit{Prescience (References)}&   -0.000783         &                     &    -0.00755\sym{***}&                     \\
                    &     (-0.58)         &                     &     (-8.43)         &                     \\
\addlinespace
DepSize& -0.00000138\sym{***}& -0.00000135\sym{***}&   -3.59e-08         &   -3.05e-08         \\
                    &    (-19.02)         &    (-18.84)         &     (-1.53)         &     (-1.29)         \\
\addlinespace
CareerAge          &   -0.000127\sym{***}&   -0.000119\sym{***}&   0.0000364\sym{***}&   0.0000539\sym{***}\\
                    &     (-9.82)         &     (-9.18)         &      (5.92)         &      (8.63)         \\
\addlinespace
\textit{OpenAccess}             &    -0.00337\sym{***}&    -0.00319\sym{***}&   -0.000520\sym{**} &   -0.000422\sym{*}  \\
                    &    (-10.33)         &     (-9.79)         &     (-2.98)         &     (-2.41)         \\
\addlinespace
\textit{Prescience (Concepts)} &                     &     -0.0105\sym{***}&                     &     0.00126\sym{***}\\
                    &                     &    (-16.56)         &                     &      (3.51)         \\
\addlinespace
Female=1 $\times$ \textit{Prescience (Concepts)}&                     &     0.00213         &                     &   -0.000687         \\
                    &                     &      (1.93)         &                     &     (-1.02)         \\
\addlinespace
impact\_factor\_2y    &                     &                     &   -0.000151\sym{***}&  -0.0000620\sym{*}  \\
                    &                     &                     &     (-5.33)         &     (-2.21)         \\
\addlinespace
2y\_cites            &                     &                     &  -0.0000186         &  -0.0000600\sym{***}\\
                    &                     &                     &     (-1.27)         &     (-4.09)         \\
\addlinespace
Constant            &      0.0395\sym{***}&      0.0384\sym{***}&     0.00519\sym{***}&    -0.00224\sym{***}\\
                    &     (48.02)         &     (51.30)         &      (7.79)         &     (-3.63)         \\
\midrule
Observations        &       96513         &       96513         &      212916         &      212916         \\
\bottomrule
\multicolumn{5}{l}{\footnotesize \textit{t} statistics in parentheses}\\
\multicolumn{5}{l}{\footnotesize \sym{*} \(p<0.05\), \sym{**} \(p<0.01\), \sym{***} \(p<0.001\)}\\
\end{tabular}
\end{adjustbox}
\end{table}

\section{Unconditional model estimates}

We estimate the unconditional model (1) and model (2) from the main text. To account for sample variation, we restrict the estimation to the subsample of solo-authored papers that would be used once controls are included -- that is, we consider only observations without missing values on any control variables. This ensures that unconditional gender differences and conditional gender s are estimated on the same sample, making them directly comparable. We find that unconditional estimates of gender differences are in line with the results of the conditional models, indicating that our set of control variables do not absorb gender differences. 
Differences in significance of estimates between the conditional and unconditional models may reflect just noise, or variance inflation from adding controls. Previous works suggest caution in interpreting aggregate results, as they compare \emph{apples} to \emph{oranges} and may prove misleading \supercite{doi:10.1177/21582440231184847}.

\begin{table}[H]\centering
\def\sym#1{\ifmmode^{#1}\else\(^{#1}\)\fi}
\caption{Linear regression estimates of eq. (1) (main text) for content and context surprise and prescience scores in solo-authored papers by gender, with  author-level clustered standard errors. }
\begin{adjustbox}{width=\textwidth}
\begin{tabular}{l*{4}{D{.}{.}{-1}}}
\toprule
                    &\multicolumn{1}{c}{(1)}&\multicolumn{1}{c}{(2)}&\multicolumn{1}{c}{(3)}&\multicolumn{1}{c}{(4)}\\
                    &\multicolumn{1}{c}{Surprise (References)}&\multicolumn{1}{c}{Prescience (References)}&\multicolumn{1}{c}{Surprise (concepts)}&\multicolumn{1}{c}{Prescience (concepts)}\\
\midrule
Female=1            &      0.0416\sym{***}&      0.0368\sym{***}&    -0.00813\sym{***}&     -0.0176\sym{***}\\
                    &     (20.67)         &     (19.29)         &     (-5.28)         &    (-10.31)         \\
\addlinespace
Constant            &       0.577\sym{***}&       0.540\sym{***}&       0.537\sym{***}&       0.555\sym{***}\\
                    &    (478.56)         &    (483.11)         &    (613.70)         &    (565.93)         \\
\midrule
Observations        &      226208         &      226208         &      226208         &      226208         \\
\bottomrule
\multicolumn{5}{l}{\footnotesize \textit{t} statistics in parentheses}\\
\multicolumn{5}{l}{\footnotesize \sym{*} \(p<0.05\), \sym{**} \(p<0.01\), \sym{***} \(p<0.001\)}\\
\end{tabular}
\end{adjustbox}
\end{table}

\begin{table}[H]\centering
\def\sym#1{\ifmmode^{#1}\else\(^{#1}\)\fi}
\caption{Linear regression estimates of eq. (2) (main text) for log fo two-step credit score, five-year disruption scores, and two-year journal impact factor on \emph{surprise score} by gender in solo-authored papers. We compute  author-level clustered standard errors.}
\begin{adjustbox}{width=\textwidth}
\begin{tabular}{l*{2}{D{.}{.}{-1}}}
\toprule
                    &\multicolumn{1}{c}{(1)}&\multicolumn{1}{c}{(2)}\\
                    &\multicolumn{1}{c}{ln(2-year Journal Impact Factor)}&\multicolumn{1}{c}{ln(2-year Journal Impact Factor)}\\
\midrule
Female=1            &      -0.120\sym{***}&      -0.140\sym{***}\\
                    &    (-10.03)         &    (-18.14)         \\
\addlinespace
\textit{Surprise (References)}    &      0.0327\sym{***}&                     \\
                    &      (4.76)         &                     \\
\addlinespace
Female=1 $\times$ \textit{Surprise (References)}&      0.0237         &                     \\
                    &      (1.51)         &                     \\
\addlinespace
\textit{Surprise (Concepts)}   &                     &     -0.0220\sym{***}\\
                    &                     &     (-3.55)         \\
\addlinespace
Female=1 $\times$ \textit{Surprise (Concepts)}&                     &      0.0670\sym{***}\\
                    &                     &      (5.85)         \\
\addlinespace
Constant            &       1.199\sym{***}&       1.230\sym{***}\\
                    &    (237.48)         &    (291.69)         \\
\midrule
Observations        &      212402         &      212402         \\
\bottomrule
\multicolumn{3}{l}{\footnotesize \textit{t} statistics in parentheses}\\
\multicolumn{3}{l}{\footnotesize \sym{*} \(p<0.05\), \sym{**} \(p<0.01\), \sym{***} \(p<0.001\)}\\
\end{tabular}
\end{adjustbox}
\end{table}

\begin{table}[H]\centering
\def\sym#1{\ifmmode^{#1}\else\(^{#1}\)\fi}
\caption{Linear regression estimates of eq. (2) (main text) for two-step credit score on \emph{surprise score} by gender in solo-authored papers. We compute author-level clustered standard errors.}\label{tab_reg2_surprise}
\begin{adjustbox}{width=\textwidth}
\begin{tabular}{l*{4}{D{.}{.}{-1}}}
\toprule
                    &\multicolumn{1}{c}{(1)}&\multicolumn{1}{c}{(2)}&\multicolumn{1}{c}{(3)}&\multicolumn{1}{c}{(4)}\\
                    &\multicolumn{1}{c}{TwoStepsCreditScore}&\multicolumn{1}{c}{TwoStepsCreditScore}&\multicolumn{1}{c}{TwoStepsCreditScore}&\multicolumn{1}{c}{TwoStepsCreditScore}\\
\midrule
Female=1            &    -0.00370\sym{**} &    -0.00110         &    -0.00140         &    -0.00439\sym{***}\\
                    &     (-2.72)         &     (-0.84)         &     (-1.83)         &     (-4.86)         \\
\addlinespace
\textit{Surprise (References)}    &     -0.0229\sym{***}&                     &                     &                     \\
                    &    (-26.65)         &                     &                     &                     \\
\addlinespace
Female=1 $\times$ \textit{Surprise (References)}&     0.00384\sym{*}  &                     &                     &                     \\
                    &      (2.37)         &                     &                     &                     \\
\addlinespace
\textit{Prescience (References)}  &                     &     -0.0225\sym{***}&                     &                     \\
                    &                     &    (-27.09)         &                     &                     \\
\addlinespace
Female=1 $\times$ \textit{Prescience (References)}&                     &    0.000230         &                     &                     \\
                    &                     &      (0.14)         &                     &                     \\
\addlinespace
\textit{Surprise (Concepts)}   &                     &                     &    -0.00232\sym{***}&                     \\
                    &                     &                     &     (-3.47)         &                     \\
\addlinespace
Female=1 $\times$ \textit{Surprise (Concepts)}&                     &                     &    -0.00116         &                     \\
                    &                     &                     &     (-0.96)         &                     \\
\addlinespace
\textit{Prescience (Concepts)} &                     &                     &                     &     -0.0219\sym{***}\\
                    &                     &                     &                     &    (-29.18)         \\
\addlinespace
Female=1 $\times$ \textit{Prescience (Concepts)}&                     &                     &                     &     0.00362\sym{**} \\
                    &                     &                     &                     &      (2.80)         \\
\addlinespace
Constant            &      0.0529\sym{***}&      0.0518\sym{***}&      0.0388\sym{***}&      0.0502\sym{***}\\
                    &     (74.16)         &     (77.02)         &     (88.07)         &     (91.27)         \\
\midrule
Observations        &       96513         &       96513         &       96513         &       96513         \\
\bottomrule
\multicolumn{5}{l}{\footnotesize \textit{t} statistics in parentheses}\\
\multicolumn{5}{l}{\footnotesize \sym{*} \(p<0.05\), \sym{**} \(p<0.01\), \sym{***} \(p<0.001\)}\\
\end{tabular}
\end{adjustbox}
\end{table}

\begin{table}[H]\centering
\def\sym#1{\ifmmode^{#1}\else\(^{#1}\)\fi}
\caption{Linear regression model estimates of eq. (2) (main text) five-year disruption scores on \emph{prescience score} by gender in solo-authored papers. We compute author-level clustered standard errors.}\label{tab_reg2_prescience}
\begin{adjustbox}{width=\textwidth}
\begin{tabular}{l*{4}{D{.}{.}{-1}}}
\toprule
                    &\multicolumn{1}{c}{(1)}&\multicolumn{1}{c}{(2)}&\multicolumn{1}{c}{(3)}&\multicolumn{1}{c}{(4)}\\
                    &\multicolumn{1}{c}{5-year Disruption}&\multicolumn{1}{c}{5-year Disruption}&\multicolumn{1}{c}{5-year Disruption}&\multicolumn{1}{c}{5-year Disruption}\\
\midrule
Female=1            &     0.00890\sym{***}&     0.00833\sym{***}&     0.00364\sym{***}&     0.00414\sym{***}\\
                    &     (11.75)         &     (11.89)         &      (9.12)         &      (9.65)         \\
\addlinespace
\textit{Surprise (References)}    &     -0.0120\sym{***}&                     &                     &                     \\
                    &    (-26.14)         &                     &                     &                     \\
\addlinespace
Female=1 $\times$ \textit{Surprise (References)}&    -0.00883\sym{***}&                     &                     &                     \\
                    &     (-9.28)         &                     &                     &                     \\
\addlinespace
\textit{Prescience (References)}  &                     &     -0.0112\sym{***}&                     &                     \\
                    &                     &    (-26.35)         &                     &                     \\
\addlinespace
Female=1 $\times$ \textit{Prescience (References)}&                     &    -0.00861\sym{***}&                     &                     \\
                    &                     &     (-9.34)         &                     &                     \\
\addlinespace
\textit{Surprise (Concepts)}   &                     &                     &   -0.000420         &                     \\
                    &                     &                     &     (-1.27)         &                     \\
\addlinespace
Female=1 $\times$ \textit{Surprise (Concepts)}&                     &                     &    -0.00127\sym{*}  &                     \\
                    &                     &                     &     (-2.00)         &                     \\
\addlinespace
\textit{Prescience (Concepts)} &                     &                     &                     &    0.000222         \\
                    &                     &                     &                     &      (0.64)         \\
\addlinespace
Female=1 $\times$ \textit{Prescience (Concepts)}&                     &                     &                     &    -0.00217\sym{**} \\
                    &                     &                     &                     &     (-3.19)         \\
\addlinespace
Constant            &     0.00554\sym{***}&     0.00466\sym{***}&    -0.00119\sym{***}&    -0.00154\sym{***}\\
                    &     (15.43)         &     (14.47)         &     (-5.85)         &     (-6.75)         \\
\midrule
Observations        &      212916         &      212916         &      212916         &      212916         \\
\bottomrule
\multicolumn{5}{l}{\footnotesize \textit{t} statistics in parentheses}\\
\multicolumn{5}{l}{\footnotesize \sym{*} \(p<0.05\), \sym{**} \(p<0.01\), \sym{***} \(p<0.001\)}\\
\end{tabular}
\end{adjustbox}
\end{table}

\section{Assessing the role of individual controls}

To explore how individual covariates relate to observed gender differences, we estimate regression models (1) and (2) from the main text by including one control at a time. 

\subsection{Gender differences in approach to novelty and uptake}

Overall, the results are consistent with the fully specified models (Table \ref{tab1_reg1}).

\begin{table}[H]\centering
\def\sym#1{\ifmmode^{#1}\else\(^{#1}\)\fi}
\caption{Linear regression estimates of gender differences in content and context surprise and prescience scores in solo-authored papers, with author-level clustered standard errors, conditioning on controls separately.}
\begin{adjustbox}{width=\textwidth}
\begin{tabular}{l*{4}{D{.}{.}{-1}}}
\toprule
                    &\multicolumn{1}{c}{(1)}&\multicolumn{1}{c}{(2)}&\multicolumn{1}{c}{(3)}&\multicolumn{1}{c}{(4)}\\
                    &\multicolumn{1}{c}{\textit{Surprise (References)}}&\multicolumn{1}{c}{\textit{Prescience (References)}}&\multicolumn{1}{c}{\textit{Surprise (Concepts)}}&\multicolumn{1}{c}{\textit{Prescience (Concepts)}}\\          
\midrule
&\multicolumn{4}{c}{Department Size}\\
\midrule
Female=1            &      0.0415\sym{***}&      0.0368\sym{***}&    -0.00812\sym{***}&     -0.0176\sym{***}\\
                    &     (20.69)         &     (19.29)         &     (-5.28)         &    (-10.36)         \\
\addlinespace
DepSize& -0.00000388\sym{***}& -0.00000470\sym{***}&    8.20e-08         &  0.00000763\sym{***}\\
                    &    (-10.38)         &    (-14.40)         &      (0.34)         &     (26.06)         \\
\addlinespace
Constant            &       0.582\sym{***}&       0.547\sym{***}&       0.537\sym{***}&       0.543\sym{***}\\
                    &    (482.85)         &    (470.27)         &    (566.32)         &    (522.47)         \\
\midrule
&\multicolumn{4}{c}{Career Age}\\
\midrule
Female=1            &      0.0319\sym{***}&      0.0265\sym{***}&    -0.00984\sym{***}&     -0.0165\sym{***}\\
                    &     (15.94)         &     (13.72)         &     (-6.27)         &     (-9.43)         \\
\addlinespace
CareerAge          &    -0.00143\sym{***}&    -0.00153\sym{***}&   -0.000254\sym{***}&    0.000174\sym{***}\\
                    &    (-21.96)         &    (-25.31)         &     (-5.35)         &      (3.30)         \\
\addlinespace
Constant            &       0.613\sym{***}&       0.579\sym{***}&       0.543\sym{***}&       0.550\sym{***}\\
                    &    (356.46)         &    (342.62)         &    (377.11)         &    (345.65)         \\
\midrule
&\multicolumn{4}{c}{Research Field Dummies (omitted)}\\
\midrule
Female=1            &      0.0305\sym{***}&      0.0261\sym{***}&    -0.00492\sym{**} &    -0.00784\sym{***}\\
                    &     (15.54)         &     (13.73)         &     (-3.11)         &     (-4.82)         \\
\addlinespace
Constant            &       0.490\sym{***}&       0.461\sym{***}&       0.468\sym{***}&       0.505\sym{***}\\
                    &    (213.37)         &    (204.34)         &    (219.92)         &    (237.55)         \\
\midrule
&\multicolumn{4}{c}{Publication Year (omitted)}\\
\midrule
Female=1            &      0.0469\sym{***}&      0.0457\sym{***}&    -0.00701\sym{***}&     -0.0138\sym{***}\\
                    &     (23.31)         &     (24.11)         &     (-4.55)         &     (-8.06)         \\
\addlinespace
Constant            &       0.605\sym{***}&       0.603\sym{***}&       0.553\sym{***}&       0.593\sym{***}\\
                    &    (184.09)         &    (184.52)         &    (175.13)         &    (190.97)         \\
\midrule
Observations        &      226208         &      226208         &      226208         &      226208         \\
\bottomrule
\multicolumn{5}{l}{\footnotesize \textit{t} statistics in parentheses}\\
\multicolumn{5}{l}{\footnotesize \sym{*} \(p<0.05\), \sym{**} \(p<0.01\), \sym{***} \(p<0.001\)}\\
\end{tabular}
\end{adjustbox}
\end{table}

\subsection{Gender differences in returns to innovation}

We include covariates one at a time to show how estimates of gender differences in returns to novelty and uptake change when conditioning on key dimensions individually. Estimates remain broadly consistent with the full models for disruption (Table \ref{tab:cd_controls_1}) and journal impact (Table \ref{tab:jif_controls_1}).

When estimating the effect of novelty on two-step credit scores, the average gender difference shifts from negative to positive once only field dummies are included as covariates, except for equally concept prescience (Table \ref{tab:cred_controls_1}). This suggests that women’s field choices partially account for gender differences in citation outcomes. This result complements existing literature \supercite{doi:10.1177/21582440231184847, 10.1162/qss_a_00117}. Controlling for research fields is particularly important when examining the relationship between novelty and scholarly rewards, as innovation and citation patterns vary substantially across disciplines. At the same time, women may self-select into less recognized fields, affecting citation outcomes.

Nevertheless, research fields do not appear to strongly mediate gender differences in the returns to novelty: the sign and magnitude of estimated gender gap in the marginal effect of novelty on citations remains consistent with the fully specified model. Notably, field dummies absorb much of the gender differences in how reference surprise gets rewarded in downstream citations. The conditional specification of the models still more appropriate, as it shows whether novelty translates differently in rewards by gender for similar people across relevant dimensions -- e.g. at the same career stage, within similar institution in size, in the same field and year of publication. 

\begin{table}[H]\centering
\def\sym#1{\ifmmode^{#1}\else\(^{#1}\)\fi}
\caption{Linear regression estimates of two-year journal impact factor on content and context surprise scores by gender in solo-authored papers, conditioning on controls separately, with clustered standard errors at the author level. The table continues in the next page. }\label{tab:jif_controls_1}
\begin{adjustbox}{width=0.8\textwidth}
\begin{tabular}{l*{2}{D{.}{.}{-1}}}
\toprule
                    &\multicolumn{1}{c}{(1)}&\multicolumn{1}{c}{(2)}\\
                    &\multicolumn{1}{c}{ln(2-year Journal Impact Factor)}&\multicolumn{1}{c}{ln(2-year Journal Impact Factor)}\\
\midrule
&\multicolumn{2}{c}{Department Size}\\
\midrule
Female=1            &      -0.118\sym{***}&      -0.134\sym{***}\\
                    &     (-9.89)         &    (-17.98)         \\
\addlinespace
\textit{Surprise (References)}    &      0.0534\sym{***}&                     \\
                    &      (8.04)         &                     \\
\addlinespace
Female=1 $\times$ \textit{Surprise (References)}&      0.0203         &                     \\
\end{tabular}
\end{adjustbox}
\end{table}

\begin{table}[H]\centering
\def\sym#1{\ifmmode^{#1}\else\(^{#1}\)\fi}
\caption*{Table \ref{tab:jif_controls_1} continued.}
\begin{adjustbox}{width=0.8\textwidth}
\begin{tabular}{l*{2}{D{.}{.}{-1}}}
\toprule
                    &\multicolumn{1}{c}{(1)}&\multicolumn{1}{c}{(2)}\\
                    &\multicolumn{1}{c}{ln(2-year Journal Impact Factor)}&\multicolumn{1}{c}{ln(2-year Journal Impact Factor)}\\
\midrule
&\multicolumn{2}{c}{Publication Year (omitted)}\\
\midrule
Female=1            &      -0.135\sym{***}&      -0.154\sym{***}\\
                    &    (-11.36)         &    (-20.08)         \\
\addlinespace
\textit{Surprise (References)}    &      0.0585\sym{***}&                     \\
                    &      (8.66)         &                     \\
\addlinespace
Female=1 $\times$ \textit{Surprise (References)}&      0.0247         &                     \\
\end{tabular}
\end{adjustbox}
\end{table}

\begin{table}[H]\centering
\def\sym#1{\ifmmode^{#1}\else\(^{#1}\)\fi}
\caption{Linear regression estimates of five-year disruption on content and context surprise and prescience scores by gender in solo-authored papers, conditioning on controls separately. We compute author-level clustered standard errors. The table continues in the next page. }\label{tab:cd_controls_1}
\begin{adjustbox}{width=0.8\textwidth}
\begin{tabular}{l*{4}{D{.}{.}{-1}}}
\toprule
                    &\multicolumn{1}{c}{(1)}&\multicolumn{1}{c}{(2)}&\multicolumn{1}{c}{(3)}&\multicolumn{1}{c}{(4)}\\
                    &\multicolumn{1}{c}{5-year Disruption}&\multicolumn{1}{c}{5-year Disruption}&\multicolumn{1}{c}{5-year Disruption}&\multicolumn{1}{c}{5-year Disruption}\\
\midrule
&\multicolumn{4}{c}{Department Size}\\
\midrule
Female=1            &     0.00894\sym{***}&     0.00836\sym{***}&     0.00380\sym{***}&     0.00431\sym{***}\\
                    &     (11.81)         &     (11.93)         &      (9.47)         &     (10.01)         \\
\addlinespace
\textit{Surprise (References)}    &     -0.0121\sym{***}&                     &                     &                     \\
                    &    (-26.27)         &                     &                     &                     \\
\addlinespace
Female=1 $\times$ \textit{Surprise (References)}&    -0.00882\sym{***}&                     &                     &                     \\
\end{tabular}
\end{adjustbox}
\end{table}

\begin{table}[H]\centering
\def\sym#1{\ifmmode^{#1}\else\(^{#1}\)\fi}
\caption*{Table \ref{tab:cd_controls_1} continued (i).}
\begin{adjustbox}{width=0.9\textwidth}
\begin{tabular}{l*{4}{D{.}{.}{-1}}}
\toprule
                    &\multicolumn{1}{c}{(1)}&\multicolumn{1}{c}{(2)}&\multicolumn{1}{c}{(3)}&\multicolumn{1}{c}{(4)}\\
                     &\multicolumn{1}{c}{5-year Disruption}&\multicolumn{1}{c}{5-year Disruption}&\multicolumn{1}{c}{5-year Disruption}&\multicolumn{1}{c}{5-year Disruption}\\
\midrule
&\multicolumn{4}{c}{Publication Year (omitted)}\\
\midrule
Female=1            &     0.00884\sym{***}&     0.00832\sym{***}&     0.00351\sym{***}&     0.00403\sym{***}\\
                    &     (11.66)         &     (11.86)         &      (8.79)         &      (9.38)         \\
\addlinespace
\textit{Surprise (References)}    &     -0.0119\sym{***}&                     &                     &                     \\
                    &    (-25.83)         &                     &                     &                     \\
\addlinespace
Female=1 $\times$ \textit{Surprise (References)}&    -0.00879\sym{***}&                     &                     &                     \\
\end{tabular}
\end{adjustbox}
\end{table}

\begin{table}[H]\centering
\def\sym#1{\ifmmode^{#1}\else\(^{#1}\)\fi}
\caption*{Table \ref{tab:cd_controls_1} continued (ii).}
\begin{adjustbox}{width=0.9\textwidth}
\begin{tabular}{l*{4}{D{.}{.}{-1}}}
\toprule
                    &\multicolumn{1}{c}{(1)}&\multicolumn{1}{c}{(2)}&\multicolumn{1}{c}{(3)}&\multicolumn{1}{c}{(4)}\\
                    &\multicolumn{1}{c}{5-year Disruption}&\multicolumn{1}{c}{5-year Disruption}&\multicolumn{1}{c}{5-year Disruption}&\multicolumn{1}{c}{5-year Disruption}\\
\midrule
&\multicolumn{4}{c}{Two-year journal impact}\\
\midrule
Female=1            &     0.00881\sym{***}&     0.00820\sym{***}&     0.00349\sym{***}&     0.00401\sym{***}\\
                    &     (11.64)         &     (11.72)         &      (8.74)         &      (9.35)         \\
\addlinespace
\textit{Surprise (References)}    &     -0.0121\sym{***}&                     &                     &                     \\
                    &    (-26.30)         &                     &                     &                     \\
\addlinespace
Female=1 $\times$ \textit{Surprise (References)}&    -0.00888\sym{***}&                     &                     &                     \\
\end{tabular}
\end{adjustbox}
\end{table}

\begin{table}[H]\centering
\def\sym#1{\ifmmode^{#1}\else\(^{#1}\)\fi}
\caption{Linear regression estimates of two-step credit score on content and context surprise and prescience scores by gender in solo-authored papers, conditioning on controls separately. We compute  author-level clustered standard errors. The table continues in the next page. }\label{tab:cred_controls_1}
\begin{adjustbox}{width=0.8\textwidth}
\begin{tabular}{l*{4}{D{.}{.}{-1}}}
\toprule
                    &\multicolumn{1}{c}{(1)}&\multicolumn{1}{c}{(2)}&\multicolumn{1}{c}{(3)}&\multicolumn{1}{c}{(4)}\\
                    &\multicolumn{1}{c}{TwoStepsCreditScore}&\multicolumn{1}{c}{TwoStepsCreditScore}&\multicolumn{1}{c}{TwoStepsCreditScore}&\multicolumn{1}{c}{TwoStepsCreditScore}\\
\midrule
&\multicolumn{4}{c}{Department Size}\\
\midrule
Female=1            &    -0.00369\sym{**} &    -0.00110         &    -0.00145         &    -0.00452\sym{***}\\
                    &     (-2.73)         &     (-0.85)         &     (-1.90)         &     (-5.01)         \\
\addlinespace
\textit{Surprise (References)}    &     -0.0225\sym{***}&                     &                     &                     \\
                    &    (-26.39)         &                     &                     &                     \\
\addlinespace
Female=1 $\times$ \textit{Surprise (References)}&     0.00382\sym{*}  &                     &                     &                     \\
\end{tabular}
\end{adjustbox}
\end{table}

\begin{table}[H]\centering
\def\sym#1{\ifmmode^{#1}\else\(^{#1}\)\fi}
\caption*{Table \ref{tab:cred_controls_1} continued.}
\begin{adjustbox}{width=0.9\textwidth}
\begin{tabular}{l*{4}{D{.}{.}{-1}}}
\toprule
                    &\multicolumn{1}{c}{(1)}&\multicolumn{1}{c}{(2)}&\multicolumn{1}{c}{(3)}&\multicolumn{1}{c}{(4)}\\
                    &\multicolumn{1}{c}{TwoStepsCreditScore}&\multicolumn{1}{c}{TwoStepsCreditScore}&\multicolumn{1}{c}{TwoStepsCreditScore}&\multicolumn{1}{c}{TwoStepsCreditScore}\\
\midrule
&\multicolumn{4}{c}{Publication Year (omitted)}\\
\midrule
Female=1            &    -0.00483\sym{***}&    -0.00272\sym{*}  &    -0.00290\sym{***}&    -0.00505\sym{***}\\
                    &     (-3.67)         &     (-2.18)         &     (-4.00)         &     (-5.98)         \\
\addlinespace
\textit{Surprise (References)}    &     -0.0230\sym{***}&                     &                     &                     \\
                    &    (-27.77)         &                     &                     &                     \\
\addlinespace
Female=1 $\times$ \textit{Surprise (References)}&     0.00353\sym{*}  &                     &                     &                     \\
\end{tabular}
\end{adjustbox}
\end{table}

\section{Heterogeneity of gender difference in novelty and prescience in solo-authored papers}
 We augment model (1) of the main text by interacting our variable of interest, $Female$, with regression covariates, such as career age, department size, and other relevant sources of variations, like women's share in field of publication, disciplines, or past citations of authors.

\subsection{Women's share in field of publication}\label{app:womens_share}

The share of women in a field may influence women’s approaches to novelty, as \emph{critical mass} affects innovation opportunities and the prevalence of gender discrimination \supercite{price_little_1986, Etzkowitz1994}. 
For each paper, we measure women’s share in a field as the proportion of solo-authored publications by women in the scientific fields where the paper attains its highest surprise or prescience, computed separately for concepts and references. These shares are calculated using solo-authored publications in each field up to the year of the paper’s publication.

We model a paper’s novelty -- measured as surprise or prescience in concepts or references —- as a function of the author’s gender, the share of women in the relevant field, and the interaction between the two, while controlling for publication year, career age, department size, and research field. This allows us to examine whether women’s investment in novelty varies with the proportion of women active in the field.

\begin{figure}[H]
\includegraphics[width=1.0\linewidth]{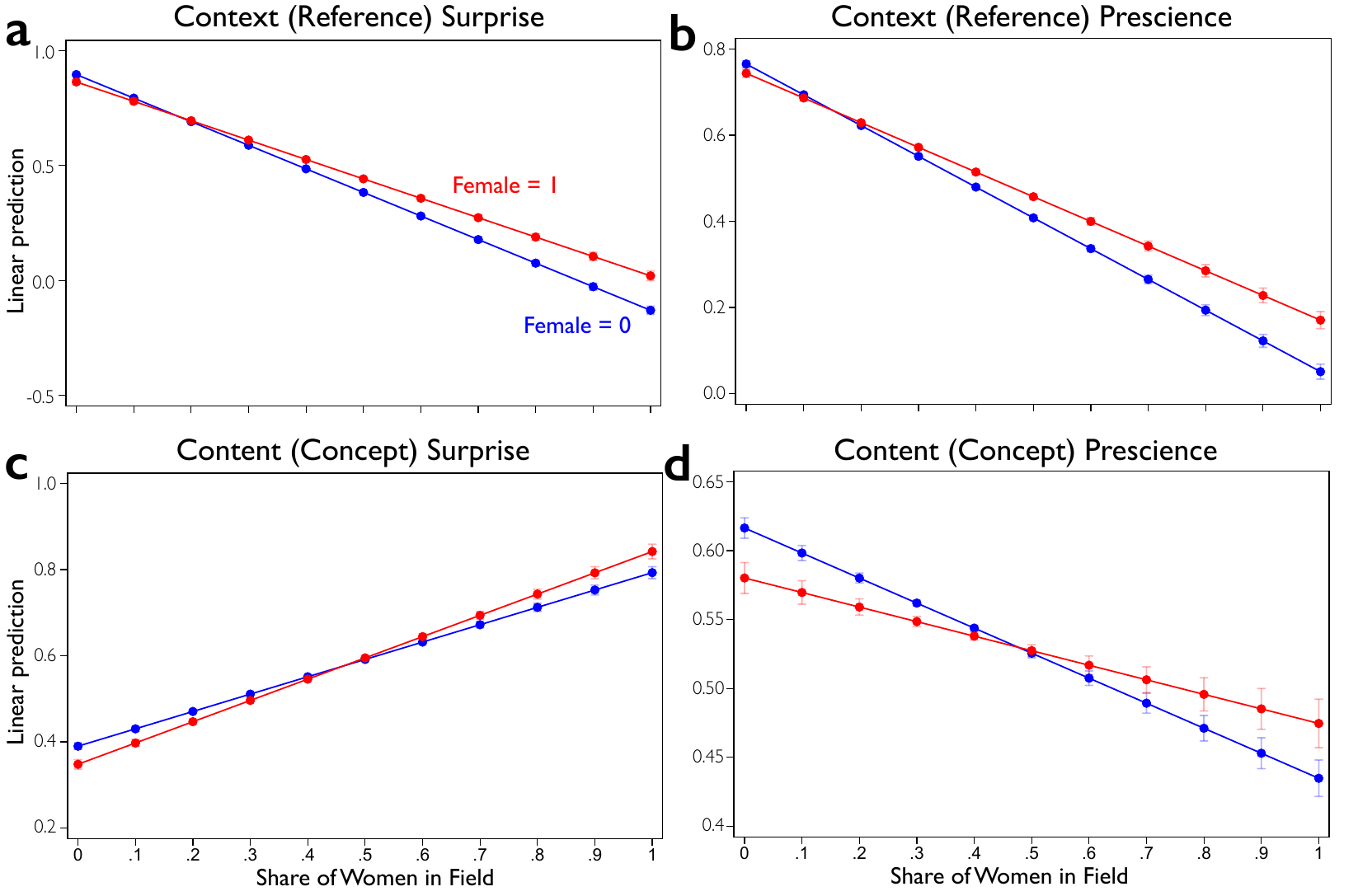}
  \caption{Predicted values of surprise and prescience of science by share of women in the field, for women and men's solo authored papers, marginalized over the observed distribution of covariates (year of publication, department size, level-one field of publication, career age). }
  \label{fig:fig1_w_field}
\end{figure}

Figure \ref{fig:fig1_w_field} shows predicted surprise and prescience by women’s share in the field. Reference surprise and prescience decline as the share of women increases, with men’s novelty decreasing more sharply -- Fig. \ref{fig:fig1_w_field} (a-b). Content surprise rises with women’s presence. Concept prescience also declines with women’s share, but women experience less reduction than men -- Fig. \ref{fig:fig1_w_field} (c-d).


Regression estimates (Table \ref{reg1_ws}, which include indicators for female authorship and women’s field share as both main effects and interactions) show that the association between women’s field share and novelty differs by author gender. In particular, the interaction between female authorship and women’s share of solo-authored papers in a field is positive for both surprise and prescience, indicating that gender differences in novelty vary systematically with women’s representation in the field.

At the same time, columns (1) and (2) show that, holding women’s field share and other covariates constant, women’s papers are on average less surprising or prescient in their combinations of references. This baseline gap is attenuated -- and eventually reversed -- at higher levels of women’s representation among solo authors in the field. This pattern is consistent 
with the idea that women’s papers tend to become more creative in their use of sources as women’s participation in solo-authored research within a field increases.

\begin{table}[H]\centering
\def\sym#1{\ifmmode^{#1}\else\(^{#1}\)\fi}
\caption{Linear Regression model estimates of surprise and prescience score by genders and  women's share of the topic of maximal surprise (prescience) of solo-authored papers, with author-level clustered standard errors. Research field fixed effects and dummies for publishing year are omitted.}\label{reg1_ws}
\begin{adjustbox}{width=\textwidth}
\begin{tabular}{lcccc}
\toprule
                    &\multicolumn{1}{c}{(1)}&\multicolumn{1}{c}{(2)}&\multicolumn{1}{c}{(3)}&\multicolumn{1}{c}{(4)}\\
                    &\multicolumn{1}{c}{\textit{Surprise (References)}}&\multicolumn{1}{c}{\textit{Prescience (References)}}&\multicolumn{1}{c}{\textit{Surprise (concepts)}}&\multicolumn{1}{c}{\textit{Prescience (concepts)}}\\
\midrule
Female=1            &     -0.0321\sym{***}&     -0.0215\sym{***}&     -0.0420\sym{***}&     -0.0362\sym{***}\\
                    &     (-5.87)         &     (-3.98)         &     (-7.31)         &     (-6.24)         \\
\addlinespace
\textit{WomenShare (Surprise References)}        &      -1.026\sym{***}&                     &                     &                     \\
                    &    (-79.87)         &                     &                     &                     \\
\addlinespace
Female=1 $\times$ \textit{WomenShare (Surprise References)}&       0.182\sym{***}&                     &                     &                     \\
                    &     (10.92)         &                     &                     &                     \\
\addlinespace
DepSize& -0.00000105\sym{**} &-0.000000367         &    2.24e-08         &  0.00000220\sym{***}\\
                    &     (-3.05)         &     (-1.10)         &      (0.08)         &      (7.49)         \\
\addlinespace
CareerAge          &    -0.00122\sym{***}&    -0.00123\sym{***}&   -0.000184\sym{***}&   -0.000341\sym{***}\\
                    &    (-21.38)         &    (-22.09)         &     (-3.86)         &     (-6.92)         \\
\addlinespace
\textit{WomenShare (Prescience References)}        &                     &      -0.714\sym{***}&                     &                     \\
                    &                     &    (-56.64)         &                     &                     \\
\addlinespace
Female=1 $\times$ \textit{WomenShare (Prescience References)}&                     &       0.141\sym{***}&                     &                     \\
                    &                     &      (8.73)         &                     &                     \\
\addlinespace
\textit{WomenShare (Surprise Concepts)}       &                     &                     &       0.403\sym{***}&                     \\
                    &                     &                     &     (37.38)         &                     \\
\addlinespace
Female=1 $\times$ \textit{WomenShare (Surprise Concepts)}&                     &                     &      0.0910\sym{***}&                     \\
                    &                     &                     &      (6.50)         &                     \\
\addlinespace
\textit{WomenShare (Prescience Concepts)}       &                     &                     &                     &      -0.182\sym{***}\\
                    &                     &                     &                     &    (-17.53)         \\
\addlinespace
Female=1 $\times$ \textit{WomenShare (Prescience Concepts)}&                     &                     &                     &      0.0760\sym{***}\\
                    &                     &                     &                     &      (5.26)         \\
\addlinespace
Constant            &       0.844\sym{***}&       0.765\sym{***}&       0.368\sym{***}&       0.602\sym{***}\\
                    &    (160.26)         &    (144.22)         &     (73.53)         &    (124.02)         \\
\midrule
Observations        &      226208         &      226208         &      226208         &      226208         \\
\bottomrule
\multicolumn{5}{l}{\footnotesize \textit{t} statistics in parentheses}\\
\multicolumn{5}{l}{\footnotesize \sym{*} \(p<0.05\), \sym{**} \(p<0.01\), \sym{***} \(p<0.001\)}\\
\end{tabular}
\end{adjustbox}
\end{table}

\subsection{Department size}\label{app:inst}

We examine how the size of the author’s institution -- proxied by the number of publications from the institution within the relevant field -- interacts with gender of solo-authors, and the joint trend in surprise and prescience. For papers associated with multiple fields, we use the largest department size. 

\begin{figure}[H]
\includegraphics[width=1.0\linewidth]{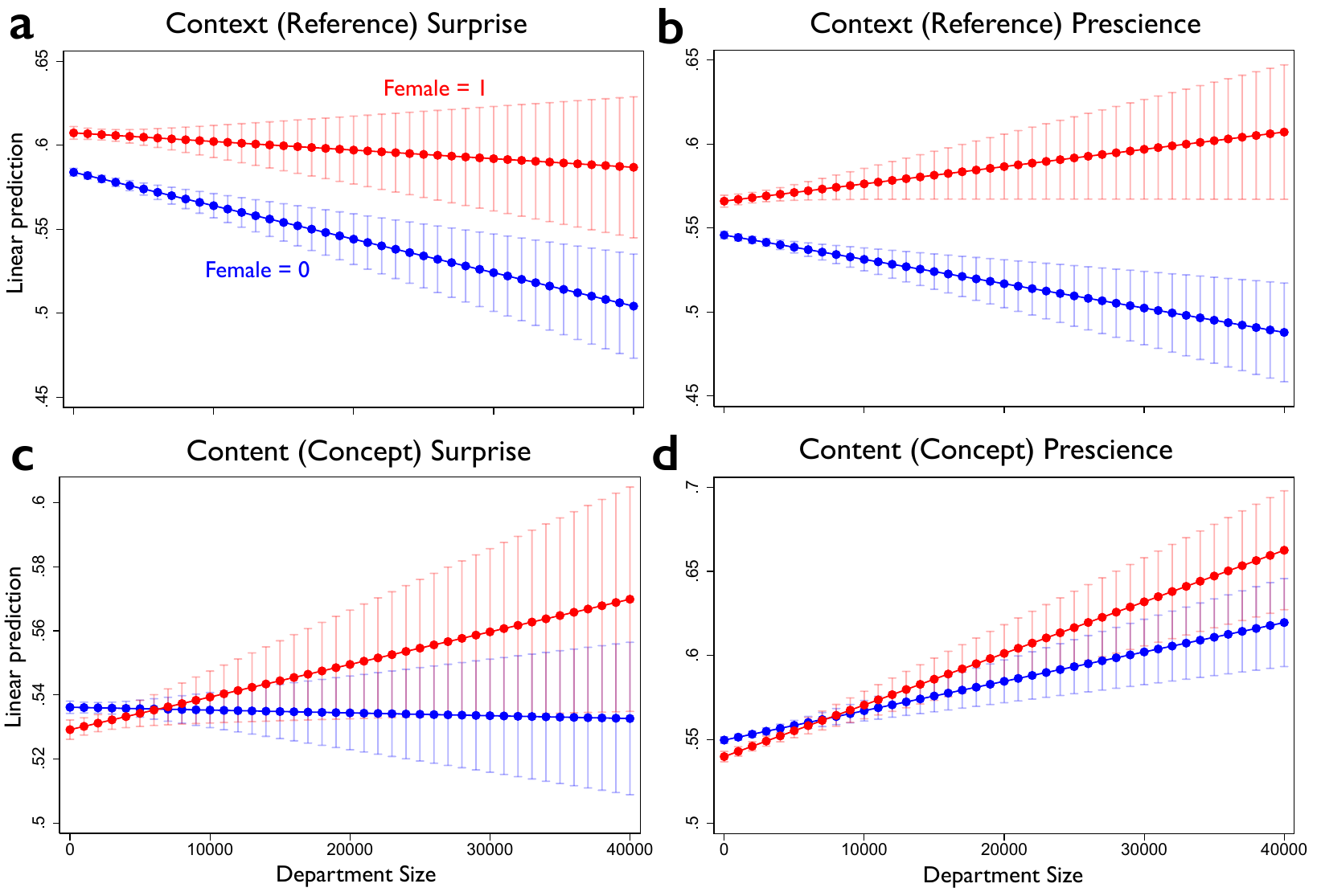}
  \caption{Predicted values of surprise and prescience by department size for women and men's solo authored papers, marginalized over the observed distribution of covariates (year of publication, level-one field of publication, career age).}
  \label{fig:fig1_inst}
\end{figure}

In Figure \ref{fig:fig1_inst}, predicted patterns show that larger departments are generally associated with lower reference-based surprise and prescience. However, women’s solo-authored papers are less affected by these declines: they maintain higher reference surprise and experience increasing reference prescience as department size grows. From Table \ref{tab_inst_1}, column (3), content-based surprise and prescience also tend to rise with department size for women. Overall, women appear to benefit relatively more from larger departments in terms of novelty, while men’s novelty scores show weaker or negative effects with department size.

\begin{table}[H]\centering
\def\sym#1{\ifmmode^{#1}\else\(^{#1}\)\fi}
\caption{Linear regression model estimates for content and context surprise and prescience scores, with author-level clustered standard errors. Research field fixed effects and dummies for publishing year are omitted.} \label{tab_inst_1}
\begin{adjustbox}{width=\textwidth}
\begin{tabular}{lcccc}
\toprule
                    &\multicolumn{1}{c}{(1)}&\multicolumn{1}{c}{(2)}&\multicolumn{1}{c}{(3)}&\multicolumn{1}{c}{(4)}\\
                    &\multicolumn{1}{c}{\textit{Surprise (References)}}&\multicolumn{1}{c}{\textit{Prescience (References)}}&\multicolumn{1}{c}{\textit{Surprise (concepts)}}&\multicolumn{1}{c}{\textit{Prescience (concepts)}}\\
\midrule
Female=1            &      0.0234\sym{***}&      0.0202\sym{***}&    -0.00699\sym{***}&    -0.00982\sym{***}\\
                    &     (10.61)         &      (9.57)         &     (-3.86)         &     (-5.29)         \\
\addlinespace
DepSize& -0.00000199\sym{***}& -0.00000145\sym{***}&   -8.77e-08         &  0.00000175\sym{***}\\
                    &     (-4.89)         &     (-3.73)         &     (-0.28)         &      (5.07)         \\
\addlinespace
Female=1 $\times$ DepSize&  0.00000148\sym{*}  &  0.00000248\sym{***}&  0.00000111\sym{*}  &  0.00000132\sym{*}  \\
                    &      (2.26)         &      (4.01)         &      (2.11)         &      (2.40)         \\
\addlinespace
CareerAge          &    -0.00131\sym{***}&    -0.00128\sym{***}&   -0.000166\sym{***}&   -0.000350\sym{***}\\
                    &    (-22.45)         &    (-22.68)         &     (-3.48)         &     (-7.07)         \\
\addlinespace
Constant            &       0.547\sym{***}&       0.557\sym{***}&       0.487\sym{***}&       0.548\sym{***}\\
                    &    (139.77)         &    (142.04)         &    (128.69)         &    (148.42)         \\
\midrule
Observations        &      226208         &      226208         &      226208         &      226208         \\
\midrule
Observations        &      226208         &      226208         &      226208         &      226208         \\
\bottomrule
\multicolumn{5}{l}{\footnotesize \textit{t} statistics in parentheses}\\
\multicolumn{5}{l}{\footnotesize \sym{*} \(p<0.05\), \sym{**} \(p<0.01\), \sym{***} \(p<0.001\)}\\
\end{tabular}
\end{adjustbox}
\end{table}

\subsection{Career age}\label{app:career_cohort}

We examine how gender differences in novelty vary across career age cohorts for solo-authored papers by interacting gender with career age intervals, ranging from 0–5 up to 40–60 years since first publication.

Predicted patterns show that women's solo-authored papers maintain higher reference-based surprise and prescience than men's across most career stages, with peaks at late stages (>40 years). For content-based novelty, men-authored solo-papers are ahead of women's across all career stages, with the gender gap narrowing as careers progress. Instead, for content-based prescience, women's works starts below men's, and the gender gap in prescience widens mid-career (5–20 years) before virtually disappearing in later stages. 
Regression estimates are reported in Table \ref{tab_cc}.

\begin{table}[htbp]\centering
\def\sym#1{\ifmmode^{#1}\else\(^{#1}\)\fi}
\caption{Linear regression model estimates for content and context surprise and prescience scores, with  author-level clustered standard errors. Research field fixed effects and dummies for publishing year are omitted.}\label{tab_cc}
\begin{adjustbox}{width=\textwidth}
\begin{tabular}{lcccc}
\toprule
                    &\multicolumn{1}{c}{(1)}&\multicolumn{1}{c}{(2)}&\multicolumn{1}{c}{(3)}&\multicolumn{1}{c}{(4)}\\
                    &\multicolumn{1}{c}{Surprise (References)  }&\multicolumn{1}{c}{\textit{Prescience (References)}}&\multicolumn{1}{c}{\textit{Surprise (concepts)}}&\multicolumn{1}{c}{\textit{Prescience (concepts)}}\\
\midrule
Female=1            &      0.0171\sym{***}&      0.0118\sym{**} &     -0.0112\sym{**} &    -0.00903\sym{*}  \\
                    &      (4.46)         &      (3.18)         &     (-3.06)         &     (-2.47)         \\
\addlinespace
career\_cohort=2     &      0.0238\sym{***}&      0.0198\sym{***}&    -0.00134         &     0.00723\sym{*}  \\
                    &      (7.03)         &      (5.99)         &     (-0.41)         &      (2.18)         \\
\addlinespace
career\_cohort=3     &     0.00724\sym{*}  &     0.00405         &    -0.00148         &     0.00917\sym{**} \\
                    &      (2.04)         &      (1.16)         &     (-0.44)         &      (2.71)         \\
\addlinespace
career\_cohort=4     &    -0.00313         &    -0.00568         &    -0.00514         &     0.00593         \\
                    &     (-0.86)         &     (-1.59)         &     (-1.49)         &      (1.73)         \\
\addlinespace
career\_cohort=5     &     -0.0223\sym{***}&     -0.0261\sym{***}&    -0.00724\sym{*}  &    -0.00259         \\
                    &     (-6.83)         &     (-8.16)         &     (-2.38)         &     (-0.84)         \\
\addlinespace
career\_cohort=6     &     -0.0399\sym{***}&     -0.0409\sym{***}&    -0.00972\sym{**} &    -0.00667\sym{*}  \\
                    &    (-11.61)         &    (-12.19)         &     (-3.14)         &     (-2.12)         \\
\addlinespace
career\_cohort=7     &     -0.0500\sym{***}&     -0.0533\sym{***}&    -0.00858\sym{**} &     -0.0142\sym{***}\\
                    &    (-13.55)         &    (-15.21)         &     (-2.69)         &     (-4.29)         \\
\addlinespace
Female=1 $\times$ career\_cohort=2&    -0.00264         &   -0.000708         &     0.00750         &    -0.00715         \\
                    &     (-0.51)         &     (-0.14)         &      (1.48)         &     (-1.42)         \\
\addlinespace
Female=1 $\times$ career\_cohort=3&     0.00521         &     0.00751         &     0.00324         &    -0.00563         \\
                    &      (0.91)         &      (1.35)         &      (0.61)         &     (-1.07)         \\
\addlinespace
Female=1 $\times$ career\_cohort=4&     0.00154         &     0.00346         &     0.00363         &    -0.00278         \\
                    &      (0.26)         &      (0.59)         &      (0.65)         &     (-0.51)         \\
\addlinespace
Female=1 $\times$ career\_cohort=5&     0.00945         &      0.0148\sym{**} &     0.00774         &     0.00435         \\
                    &      (1.80)         &      (2.86)         &      (1.56)         &      (0.88)         \\
\addlinespace
Female=1 $\times$ career\_cohort=6&      0.0170\sym{*}  &      0.0208\sym{***}&     0.00879         &     0.00488         \\
                    &      (2.52)         &      (3.29)         &      (1.62)         &      (0.90)         \\
\addlinespace
Female=1 $\times$ career\_cohort=7&      0.0338\sym{***}&      0.0446\sym{***}&      0.0102         &      0.0175\sym{**} \\
                    &      (4.79)         &      (6.64)         &      (1.74)         &      (2.89)         \\
\addlinespace
DepSize& -0.00000161\sym{***}&-0.000000791\sym{*}  & 0.000000215         &  0.00000209\sym{***}\\
                    &     (-4.63)         &     (-2.38)         &      (0.78)         &      (7.12)         \\
\addlinespace
Constant            &       0.532\sym{***}&       0.545\sym{***}&       0.489\sym{***}&       0.540\sym{***}\\
                    &    (120.87)         &    (124.58)         &    (113.12)         &    (127.73)         \\
\midrule
Observations        &      226208         &      226208         &      226208         &      226208         \\
\bottomrule
\bottomrule
\multicolumn{5}{l}{\footnotesize \textit{t} statistics in parentheses}\\
\multicolumn{5}{l}{\footnotesize \sym{*} \(p<0.05\), \sym{**} \(p<0.01\), \sym{***} \(p<0.001\)}\\
\end{tabular}
\end{adjustbox}
\end{table}

\begin{figure}[H]
\includegraphics[width=1.0\linewidth]{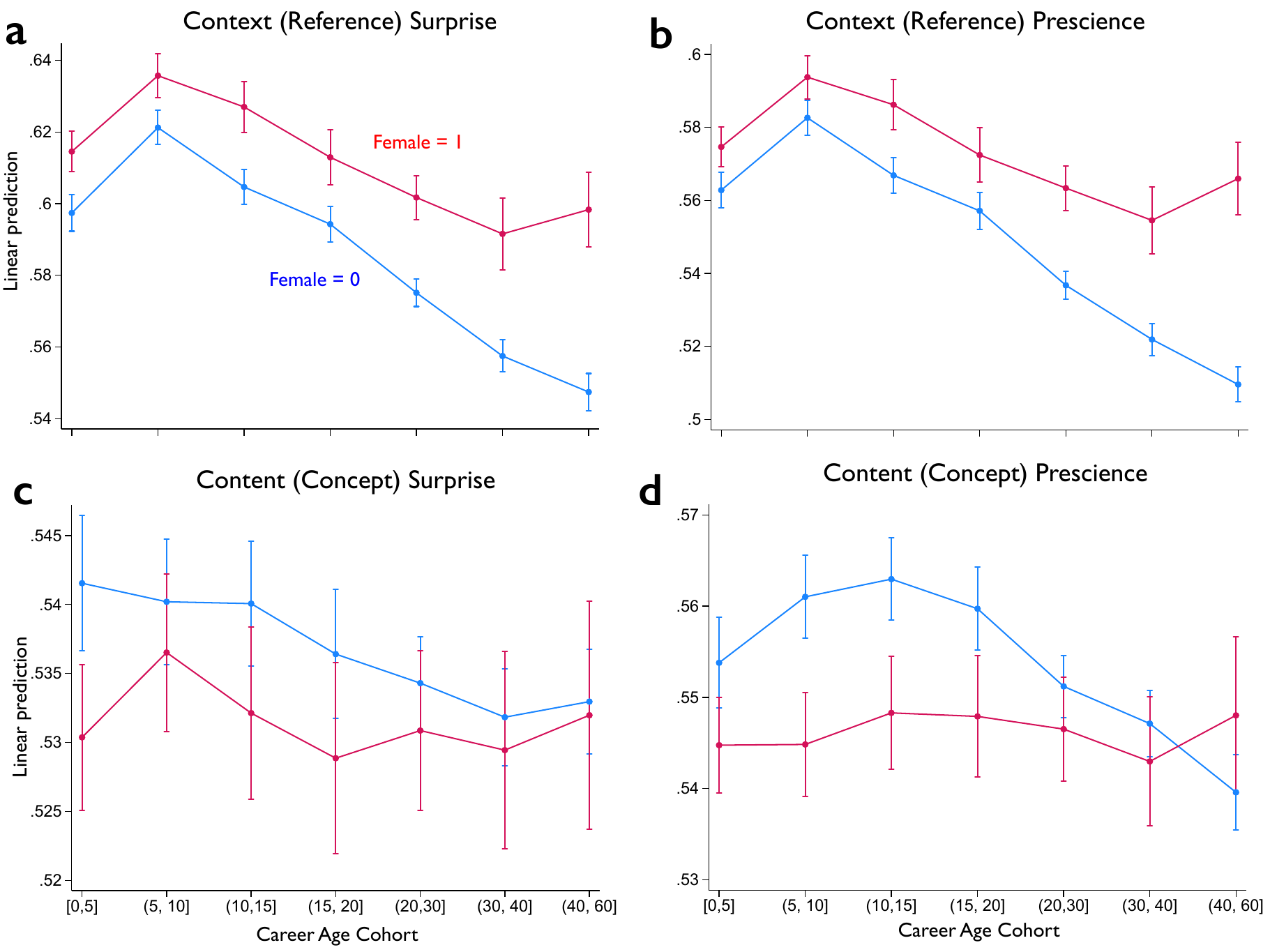}
  \caption{Predicted reference surprise and prescience (row one) and concept surprise and prescience (row two) by gender across career age cohorts, marginalized over the observed distribution of covariates (year of publication, department size, level-one field of publication).}
  \label{fig:fig_reg1_cc_2}
\end{figure}

\subsection{Past citations}\label{app:past_cite}
We check whether the differences in novelty production in solo-authored papers by genders derives from reputation of the author, as more prominent scientists could inherently have higher prescience because of their status, or produce more surprising papers. We proxy status within science with past citations of the solo-authors up until the year of publication of the focal paper.

Higher past citations are associated with lower surprise and prescience in reference combinations, as well as lower concept-based surprise, but with higher prescience in concept combinations—an outcome in which men’s work tends to be more prescient. However, increases in past citations are more strongly associated with uptake of women’s concept combinations than men’s. By contrast, women’s reference-based novelty shows no differential association with prior reputation (Table \ref{tab_past_cites}).  

\begin{figure}[H]
\includegraphics[width=1.0\linewidth]{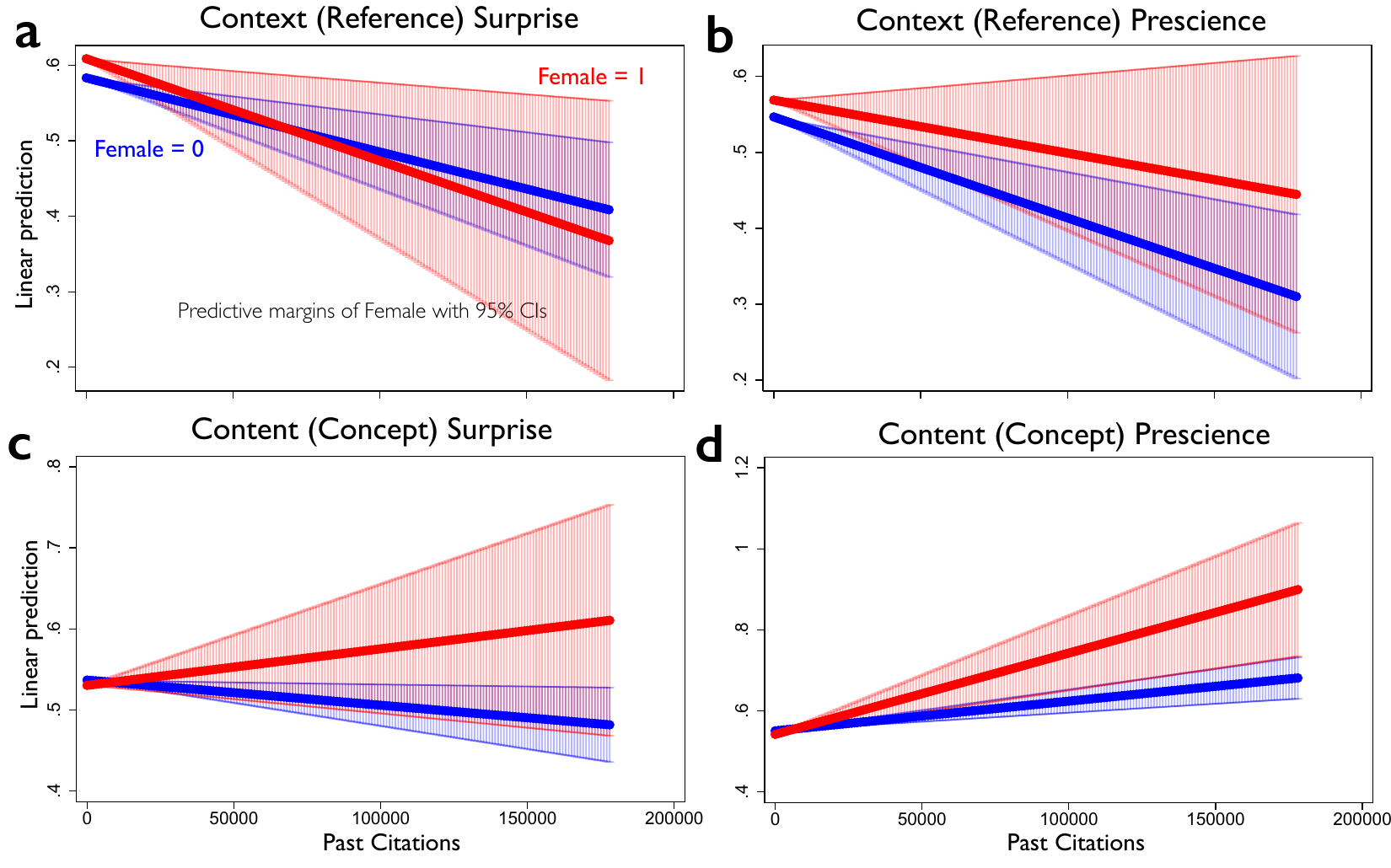}
  \caption{Predicted values of surprise and prescience by past citations of women and men in solo authored papers, marginalized over the observed distribution of covariates (year of publication, department size, level-one field of publication, career age).}
  \label{fig:fig_past_cite}
\end{figure}

\begin{table}[H]\centering
\def\sym#1{\ifmmode^{#1}\else\(^{#1}\)\fi}
\caption{Linear Regression model estimates of surprise and prescience scores by gender and  
past citations of solo-authored papers, with author-level clustered standard errors. Research field fixed effects and dummies for publishing year are omitted.}\label{tab_past_cites}
\begin{adjustbox}{width=\textwidth}
\begin{tabular}{lcccc}
\toprule
                    &\multicolumn{1}{c}{(1)}&\multicolumn{1}{c}{(2)}&\multicolumn{1}{c}{(3)}&\multicolumn{1}{c}{(4)}\\
                    &\multicolumn{1}{c}{\textit{Surprise (References)}}&\multicolumn{1}{c}{\textit{Prescience (References)}}&\multicolumn{1}{c}{\textit{surprise  (concepts)}}&\multicolumn{1}{c}{\textit{Prescience (concepts)}}\\
\midrule
Female=1            &      0.0254\sym{***}&      0.0222\sym{***}&    -0.00644\sym{***}&    -0.00866\sym{***}\\
                    &     (12.37)         &     (11.27)         &     (-3.86)         &     (-5.03)         \\
\addlinespace
PastCitations&-0.000000981\sym{***}& -0.00000133\sym{***}&-0.000000310\sym{*}  & 0.000000733\sym{***}\\
                    &     (-3.79)         &     (-4.25)         &     (-2.33)         &      (4.93)         \\
\addlinespace
Female=1 $\times$ PastCitations&-0.000000372         & 0.000000631         & 0.000000760         &  0.00000127\sym{**} \\
                    &     (-0.65)         &      (1.08)         &      (1.80)         &      (2.63)         \\
\addlinespace
CareerAge          &    -0.00121\sym{***}&    -0.00116\sym{***}&   -0.000147\sym{**} &   -0.000440\sym{***}\\
                    &    (-19.34)         &    (-18.91)         &     (-2.98)         &     (-8.53)         \\
\addlinespace
DepSize& -0.00000123\sym{***}&-0.000000335         & 0.000000291         &  0.00000181\sym{***}\\
                    &     (-3.49)         &     (-1.01)         &      (1.04)         &      (6.03)         \\
\addlinespace
Constant            &       0.546\sym{***}&       0.555\sym{***}&       0.487\sym{***}&       0.548\sym{***}\\
                    &    (139.49)         &    (141.63)         &    (128.61)         &    (148.64)         \\
\midrule
Observations        &      226208         &      226208         &      226208         &      226208         \\
\bottomrule
\multicolumn{5}{l}{\footnotesize \textit{t} statistics in parentheses}\\
\multicolumn{5}{l}{\footnotesize \sym{*} \(p<0.05\), \sym{**} \(p<0.01\), \sym{***} \(p<0.001\)}\\
\end{tabular}
\end{adjustbox}
\end{table}

\subsection{Discipline}\label{app:level0}


We examine whether gender differences in solo-authored novelty vary across disciplines (level-zero concepts). We estimate linear regression models including interactions between the gender indicator and dummy variables for each discipline (for a total of 19 disciplines), which allows the gender gap in surprise and prescience to differ across fields. 


Table \ref{tab1_disc} reports the interactions between female authorship and discipline for solo-authored papers. Positive interaction coefficients indicate that the gender difference in novelty is relatively larger in that discipline, whereas negative coefficients indicate a smaller gender difference. For reference-based novelty, interactions are positive in computer science, medicine, and philosophy, and negative in art, business, and political science. For concept-based novelty, interactions are positive in biology and medicine. 

Estimates for the full multi-authored sample are consistent with 
solo-authored papers: 
increasing women's share in teams is positively and significantly associated to innovation in computer science, economics, environmental science, geology, materials science, and philosophy. Conversely, women’s participated multi-authored contributions in political science, psychology, and sociology are generally less innovative. 


\begin{table}[H]\centering
\def\sym#1{\ifmmode^{#1}\else\(^{#1}\)\fi}
\caption{Linear regression model estimates for content and context surprise and prescience scores by discipline (level zero field) in solo-authored papers. We compute author-level clustered standard errors. Research field fixed effects and dummies for publishing year are omitted.\label{tab1_disc}}
\begin{adjustbox}{width=0.6\textwidth}
\begin{tabular}{lcccc}
\toprule
                    &\multicolumn{1}{c}{(1)}&\multicolumn{1}{c}{(2)}&\multicolumn{1}{c}{(3)}&\multicolumn{1}{c}{(4)}\\
                    &\multicolumn{1}{c}{\textit{Surprise (References)}}&\multicolumn{1}{c}{\textit{Prescience (References)}}&\multicolumn{1}{c}{\textit{Surprise (concepts)}}&\multicolumn{1}{c}{\textit{Prescience (concepts)}}\\
\midrule

Female=1            &      0.0211\sym{***}&      0.0226\sym{***}&     -0.0165\sym{***}&     -0.0168\sym{***}\\
                    &      (4.50)         &      (5.02)         &     (-4.01)         &     (-4.07)         \\
\addlinespace
Art=1            &      0.0287\sym{**} &      0.0336\sym{***}&     -0.0216\sym{*}  &      0.0109         \\
                    &      (3.12)         &      (3.79)         &     (-2.54)         &      (1.27)         \\
\addlinespace
Biology=1        &      0.0196\sym{***}&   -0.000388         &     -0.0239\sym{***}&     -0.0428\sym{***}\\
                    &      (6.86)         &     (-0.14)         &     (-8.99)         &    (-16.78)         \\
\addlinespace
Business=1       &    -0.00728         &    -0.00450         &     -0.0104\sym{*}  &    -0.00463         \\
                    &     (-1.63)         &     (-1.04)         &     (-2.54)         &     (-1.10)         \\
\addlinespace
Chemistry=1      &     -0.0166\sym{***}&     -0.0260\sym{***}&     -0.0170\sym{***}&     -0.0513\sym{***}\\
                    &     (-4.68)         &     (-7.27)         &     (-5.20)         &    (-16.23)         \\
\addlinespace
Computer science=1&      0.0448\sym{***}&      0.0122\sym{***}&     -0.0265\sym{***}&     -0.0125\sym{***}\\
                    &     (16.14)         &      (4.40)         &     (-9.98)         &     (-4.69)         \\
\addlinespace
Economics=1      &      0.0144\sym{***}&      0.0141\sym{***}&    -0.00792\sym{*}  &      0.0134\sym{***}\\
                    &      (3.95)         &      (3.96)         &     (-2.27)         &      (3.86)         \\
\addlinespace
Engineering=1    &      0.0343\sym{***}&      0.0118\sym{*}  &     -0.0111\sym{*}  &    -0.00435         \\
                    &      (6.69)         &      (2.38)         &     (-2.44)         &     (-0.94)         \\
\addlinespace
Environmental science=1&      0.0157\sym{*}  &      0.0408\sym{***}&    -0.00370         &      0.0130\sym{*}  \\
                    &      (2.51)         &      (6.82)         &     (-0.65)         &      (2.31)         \\
\addlinespace
Geography=1      &      0.0137\sym{**} &      0.0178\sym{***}&     -0.0202\sym{***}&     0.00694         \\
                    &      (2.79)         &      (3.72)         &     (-4.46)         &      (1.51)         \\
\addlinespace
Geology=1        &     -0.0259\sym{***}&     -0.0214\sym{**} &     -0.0198\sym{**} &     -0.0478\sym{***}\\
                    &     (-3.59)         &     (-3.17)         &     (-3.15)         &     (-7.76)         \\
\addlinespace
History=1        &     -0.0452\sym{***}&     -0.0347\sym{***}&     -0.0236\sym{***}&     0.00698         \\
                    &     (-7.54)         &     (-5.89)         &     (-4.11)         &      (1.22)         \\
\addlinespace
Materials science=1&     -0.0113         &     0.00108         &     -0.0143\sym{*}  &    -0.00955         \\
                    &     (-1.75)         &      (0.17)         &     (-2.30)         &     (-1.63)         \\
\addlinespace
Mathematics=1    &     0.00818\sym{*}  &      0.0116\sym{**} &    -0.00595         &     0.00644         \\
                    &      (2.17)         &      (3.19)         &     (-1.60)         &      (1.77)         \\
\addlinespace
Medicine=1       &     -0.0490\sym{***}&     -0.0229\sym{***}&     -0.0593\sym{***}&     -0.0327\sym{***}\\
                    &    (-16.29)         &     (-7.74)         &    (-21.56)         &    (-11.94)         \\
\addlinespace
Philosophy=1     &     0.00707         &      0.0105\sym{*}  &      0.0117\sym{*}  &      0.0208\sym{***}\\
                    &      (1.36)         &      (2.09)         &      (2.41)         &      (4.24)         \\
\addlinespace
Physics=1        &      0.0172\sym{***}&      0.0394\sym{***}&     -0.0146\sym{***}&    -0.00853\sym{*}  \\
                    &      (4.53)         &     (10.51)         &     (-3.91)         &     (-2.40)         \\
\addlinespace
Political science=1&     -0.0379\sym{***}&     -0.0313\sym{***}&     -0.0186\sym{**} &      0.0192\sym{**} \\
                    &     (-5.18)         &     (-4.43)         &     (-2.78)         &      (2.72)         \\
\addlinespace
Psychology=1     &      0.0270\sym{***}&      0.0305\sym{***}&     -0.0274\sym{***}&     -0.0238\sym{***}\\
                    &      (8.26)         &      (9.69)         &     (-9.17)         &     (-7.82)         \\
\addlinespace
Sociology=1      &      0.0124\sym{***}&     0.00962\sym{**} &    0.000168         &    -0.00156         \\
                    &      (3.56)         &      (2.84)         &      (0.05)         &     (-0.48)         \\
\addlinespace
Female=1 $\times$ Art=1&     -0.0221\sym{**} &     -0.0265\sym{***}&    -0.00315         &     -0.0143         \\
                    &     (-2.72)         &     (-3.33)         &     (-0.41)         &     (-1.88)         \\
\addlinespace
Female=1 $\times$ Biology=1&    0.000534         &     0.00131         &     0.00799\sym{*}  &     0.00960\sym{**} \\
                    &      (0.14)         &      (0.36)         &      (2.38)         &      (2.91)         \\
\addlinespace
Female=1 $\times$ Business=1&     -0.0192\sym{**} &     -0.0193\sym{**} &     0.00772         &     0.00466         \\
                    &     (-3.09)         &     (-3.14)         &      (1.43)         &      (0.82)         \\
\addlinespace
Female=1 $\times$ Chemistry=1&     -0.0129\sym{**} &    -0.00746         &      0.0118\sym{**} &     0.00808         \\
                    &     (-2.61)         &     (-1.53)         &      (2.62)         &      (1.82)         \\
\addlinespace
Female=1 $\times$ Computer science=1&      0.0135\sym{***}&     0.00192         &     0.00724\sym{*}  &    -0.00786\sym{*}  \\
                    &      (3.77)         &      (0.55)         &      (2.14)         &     (-2.30)         \\
\addlinespace
Female=1 $\times$ Economics=1&     0.00584         &      0.0104\sym{*}  &    -0.00134         &     0.00273         \\
                    &      (1.34)         &      (2.44)         &     (-0.33)         &      (0.67)         \\
\addlinespace
Female=1 $\times$ Engineering=1&     0.00544         &     0.00159         &    -0.00860         &     -0.0205\sym{***}\\
                    &      (1.05)         &      (0.31)         &     (-1.78)         &     (-4.21)         \\
\addlinespace
Female=1 $\times$ Environmental science=1&      0.0126         &      0.0132         &      0.0208\sym{*}  &      0.0219\sym{*}  \\
                    &      (1.07)         &      (1.19)         &      (2.03)         &      (2.15)         \\
\addlinespace
Female=1 $\times$ Geography=1&     0.00959         &     0.00604         &  -0.0000779         &    -0.00886         \\
                    &      (1.51)         &      (0.98)         &     (-0.01)         &     (-1.50)         \\
\addlinespace
Female=1 $\times$ Geology=1&    -0.00473         &     -0.0140         &     0.00581         &     0.00868         \\
                    &     (-0.42)         &     (-1.33)         &      (0.56)         &      (0.90)         \\
\addlinespace
Female=1 $\times$ History=1&    -0.00115         &    -0.00352         &    -0.00347         &     -0.0127         \\
                    &     (-0.15)         &     (-0.49)         &     (-0.48)         &     (-1.77)         \\
\addlinespace
Female=1 $\times$ Materials science=1&     0.00312         &     0.00211         &     0.00837         &      0.0229\sym{**} \\
                    &      (0.33)         &      (0.22)         &      (0.98)         &      (2.64)         \\
\addlinespace
Female=1 $\times$ Mathematics=1&   -0.000606         &    -0.00257         &    -0.00261         &     0.00655         \\
                    &     (-0.12)         &     (-0.54)         &     (-0.55)         &      (1.39)         \\
\addlinespace
Female=1 $\times$ Medicine=1&     0.00814\sym{*}  &     0.00926\sym{*}  &      0.0124\sym{***}&      0.0154\sym{***}\\
                    &      (1.98)         &      (2.37)         &      (3.60)         &      (4.45)         \\
\addlinespace
Female=1 $\times$ Philosophy=1&     0.00872\sym{*}  &      0.0133\sym{***}&     0.00835\sym{*}  &     0.00179         \\
                    &      (2.13)         &      (3.32)         &      (2.13)         &      (0.45)         \\
\addlinespace
Female=1 $\times$ Physics=1&     0.00119         &     0.00323         &    -0.00299         &    -0.00927\sym{*}  \\
                    &      (0.26)         &      (0.73)         &     (-0.69)         &     (-2.14)         \\
\addlinespace
Female=1 $\times$ Political science=1&    -0.00861\sym{*}  &    -0.00658         &     0.00241         &     0.00632         \\
                    &     (-2.14)         &     (-1.69)         &      (0.64)         &      (1.65)         \\
\addlinespace
Female=1 $\times$ Psychology=1&    -0.00182         &    -0.00688         &    -0.00472         &    -0.00878\sym{*}  \\
                    &     (-0.47)         &     (-1.84)         &     (-1.35)         &     (-2.44)         \\
\addlinespace
Female=1 $\times$ Sociology=1&    -0.00166         &    -0.00854\sym{*}  &   -0.000169         &     0.00735         \\
                    &     (-0.40)         &     (-2.11)         &     (-0.04)         &      (1.89)         \\
\addlinespace
DepSize& -0.00000149\sym{***}&-0.000000677\sym{*}  & 0.000000205         &  0.00000212\sym{***}\\
                    &     (-4.31)         &     (-2.04)         &      (0.74)         &      (7.23)         \\
\addlinespace
CareerAge          &    -0.00127\sym{***}&    -0.00125\sym{***}&   -0.000132\sym{**} &   -0.000324\sym{***}\\
                    &    (-21.86)         &    (-22.24)         &     (-2.77)         &     (-6.59)         \\
\addlinespace
Constant            &       0.548\sym{***}&       0.554\sym{***}&       0.514\sym{***}&       0.566\sym{***}\\
                    &    (131.33)         &    (133.00)         &    (128.22)         &    (144.32)         \\
\midrule
Observations        &      226208         &      226208         &      226208         &      226208         \\
\bottomrule
\multicolumn{5}{l}{\footnotesize \textit{t} statistics in parentheses}\\
\multicolumn{5}{l}{\footnotesize \sym{*} \(p<0.05\), \sym{**} \(p<0.01\), \sym{***} \(p<0.001\)}\\
\end{tabular}
\end{adjustbox}
\end{table}

\begin{table}[H]\centering
\def\sym#1{\ifmmode^{#1}\else\(^{#1}\)\fi}
\caption{Linear regression model estimates for gender differences on content and context surprise and prescience scores by discipline (level zero field) in multi-authored papers. We compute White-robust standard errors. Research field fixed effects and dummies for publishing year are omitted.}\label{tab1_disc_ma}
\begin{adjustbox}{width=0.55\textwidth}
\begin{tabular}{lcccc}
\toprule
                    &\multicolumn{1}{c}{(1)}&\multicolumn{1}{c}{(2)}&\multicolumn{1}{c}{(3)}&\multicolumn{1}{c}{(4)}\\
                    &\multicolumn{1}{c}{\textit{Surprise (References)}}&\multicolumn{1}{c}{\textit{Prescience (References)}}&\multicolumn{1}{c}{\textit{Surprise (concepts)}}&\multicolumn{1}{c}{\textit{Prescience (concepts)}}\\
\midrule
\textit{Female Share}       &      0.0189\sym{***}&      0.0102\sym{***}&    -0.00849\sym{***}&     -0.0144\sym{***}\\
                    &     (18.64)         &      (9.47)         &     (-7.47)         &    (-13.37)         \\
\addlinespace
Art=1            &      0.0373\sym{***}&      0.0512\sym{***}&      0.0135\sym{*}  &      0.0711\sym{***}\\
                    &      (5.60)         &      (8.20)         &      (1.98)         &     (10.36)         \\
\addlinespace
Biology=1        &      0.0438\sym{***}&     0.00401\sym{***}&     0.00267\sym{***}&     -0.0198\sym{***}\\
                    &     (83.39)         &      (7.13)         &      (4.53)         &    (-35.47)         \\
\addlinespace
Business=1       &     -0.0205\sym{***}&    -0.00840\sym{***}&     -0.0162\sym{***}&    -0.00306         \\
                    &    (-12.19)         &     (-5.09)         &     (-9.47)         &     (-1.75)         \\
\addlinespace
Chemistry=1      &      0.0182\sym{***}&    -0.00364\sym{***}&     -0.0216\sym{***}&     -0.0255\sym{***}\\
                    &     (34.97)         &     (-6.20)         &    (-35.51)         &    (-44.11)         \\
\addlinespace
Computer science=1&      0.0434\sym{***}&      0.0168\sym{***}&     -0.0191\sym{***}&    0.000157         \\
                    &     (62.83)         &     (23.54)         &    (-25.47)         &      (0.22)         \\
\addlinespace
Economics=1      &     0.00146         &      0.0109\sym{***}&     -0.0235\sym{***}&      0.0309\sym{***}\\
                    &      (1.14)         &      (8.48)         &    (-17.27)         &     (23.13)         \\
\addlinespace
Engineering=1    &      0.0294\sym{***}&     0.00675\sym{***}&     -0.0115\sym{***}&     0.00299\sym{**} \\
                    &     (27.94)         &      (6.41)         &    (-10.54)         &      (2.86)         \\
\addlinespace

Environmental science=1&      0.0289\sym{***}&      0.0517\sym{***}&    -0.00130         &     -0.0106\sym{***}\\
                    &     (24.63)         &     (44.21)         &     (-1.02)         &     (-8.55)         \\
\addlinespace
Geography=1      &      0.0183\sym{***}&      0.0269\sym{***}&     -0.0161\sym{***}&      0.0262\sym{***}\\
                    &     (14.85)         &     (21.81)         &    (-11.79)         &     (19.88)         \\
\addlinespace
Geology=1        &     0.00123         &    -0.00698\sym{***}&    -0.00904\sym{***}&     -0.0345\sym{***}\\
                    &      (1.00)         &     (-5.65)         &     (-6.70)         &    (-26.68)         \\
\addlinespace
History=1        &     -0.0150\sym{***}&     0.00650\sym{**} &   -0.000927         &      0.0605\sym{***}\\
                    &     (-5.97)         &      (2.64)         &     (-0.34)         &     (23.24)         \\
\addlinespace
Materials science=1&    -0.00382\sym{***}&      0.0189\sym{***}&     -0.0204\sym{***}&     -0.0197\sym{***}\\
                    &     (-4.65)         &     (21.62)         &    (-22.18)         &    (-22.71)         \\
\addlinespace
Mathematics=1    &     0.00776\sym{***}&      0.0136\sym{***}&     -0.0197\sym{***}&      0.0102\sym{***}\\
                    &      (7.89)         &     (13.78)         &    (-18.68)         &     (10.08)         \\
\addlinespace
Medicine=1       &     -0.0186\sym{***}&     0.00819\sym{***}&     -0.0325\sym{***}&     -0.0230\sym{***}\\
                    &    (-32.81)         &     (13.59)         &    (-51.61)         &    (-38.42)         \\
\addlinespace
Philosophy=1     &      0.0190\sym{***}&      0.0231\sym{***}&      0.0157\sym{***}&      0.0667\sym{***}\\
                    &      (5.73)         &      (7.14)         &      (4.59)         &     (19.17)         \\
\addlinespace
Physics=1        &      0.0152\sym{***}&      0.0347\sym{***}&     -0.0132\sym{***}&    -0.00325\sym{***}\\
                    &     (20.88)         &     (45.68)         &    (-16.43)         &     (-4.25)         \\
\addlinespace
Political science=1&    -0.00592         &     0.00853\sym{**} &     -0.0279\sym{***}&      0.0417\sym{***}\\
                    &     (-1.93)         &      (2.85)         &     (-9.30)         &     (13.32)         \\
\addlinespace
Psychology=1     &      0.0472\sym{***}&      0.0464\sym{***}&     0.00807\sym{***}&      0.0216\sym{***}\\
                    &     (58.90)         &     (56.03)         &      (8.83)         &     (25.17)         \\
\addlinespace
Sociology=1      &      0.0260\sym{***}&      0.0256\sym{***}&    -0.00450\sym{*}  &     0.00937\sym{***}\\
                    &     (15.80)         &     (15.72)         &     (-2.54)         &      (5.34)         \\
\addlinespace
Art=1 $\times$ \textit{Female Share}&     0.00661         &    0.000918         &     -0.0149\sym{**} &     -0.0402\sym{***}\\
                    &      (1.42)         &      (0.21)         &     (-3.01)         &     (-8.36)         \\
\addlinespace
Biology=1 $\times$ \textit{Female Share}&    -0.00651\sym{***}&    -0.00257\sym{**} &     0.00517\sym{***}&   0.0000634         \\
                    &     (-7.69)         &     (-2.89)         &      (5.48)         &      (0.07)         \\
\addlinespace
Business=1 $\times$ \textit{Female Share}&     -0.0409\sym{***}&     -0.0268\sym{***}&     0.00508\sym{*}  &     0.00676\sym{**} \\
                    &    (-17.79)         &    (-11.91)         &      (2.13)         &      (2.79)         \\
\addlinespace
Chemistry=1 $\times$ \textit{Female Share}&    -0.00358\sym{***}&    -0.00638\sym{***}&     0.00308\sym{**} &     0.00800\sym{***}\\
                    &     (-3.92)         &     (-6.34)         &      (2.93)         &      (7.98)         \\
\addlinespace
Computer science=1 $\times$ \textit{Female Share}&      0.0258\sym{***}&      0.0120\sym{***}&     0.00135         &    -0.00780\sym{***}\\
                    &     (23.94)         &     (10.80)         &      (1.13)         &     (-6.74)         \\
\addlinespace
Economics=1 $\times$ \textit{Female Share}&     0.00452\sym{**} &      0.0108\sym{***}&     0.00651\sym{***}&     0.00533\sym{***}\\
                    &      (3.02)         &      (7.21)         &      (3.99)         &      (3.41)         \\
\addlinespace
Engineering=1 $\times$ \textit{Female Share}&      0.0186\sym{***}&     0.00379\sym{**} &    -0.00900\sym{***}&     -0.0144\sym{***}\\
                    &     (13.59)         &      (2.69)         &     (-6.04)         &     (-9.96)         \\
\addlinespace
Environmental science=1 $\times$ \textit{Female Share}&      0.0196\sym{***}&      0.0111\sym{***}&     0.00152         &      0.0252\sym{***}\\
                    &      (8.59)         &      (4.83)         &      (0.60)         &     (10.17)         \\
\addlinespace
Geography=1 $\times$ \textit{Female Share}&    0.000114         &     0.00241         &     0.00648\sym{**} &    0.000649         \\
                    &      (0.06)         &      (1.19)         &      (2.90)         &      (0.30)         \\
\addlinespace
Geology=1 $\times$ \textit{Female Share}&     0.00989\sym{***}&     0.00575\sym{**} &     0.00492\sym{*}  &     0.00605\sym{**} \\
                    &      (4.86)         &      (2.76)         &      (2.11)         &      (2.74)         \\
\addlinespace
History=1 $\times$ \textit{Female Share}&      0.0130\sym{**} &      0.0267\sym{***}&     -0.0278\sym{***}&     -0.0326\sym{***}\\
                    &      (3.17)         &      (6.83)         &     (-6.16)         &     (-7.56)         \\
\addlinespace
Materials science=1 $\times$ \textit{Female Share}&     0.00811\sym{***}&      0.0250\sym{***}&     0.00927\sym{***}&      0.0176\sym{***}\\
                    &      (5.51)         &     (15.96)         &      (5.55)         &     (11.01)         \\
\addlinespace
Mathematics=1 $\times$ \textit{Female Share}&     0.00182         &     0.00157         &   -0.000425         &      0.0102\sym{***}\\
                    &      (1.34)         &      (1.14)         &     (-0.28)         &      (7.14)         \\
\addlinespace
Medicine=1 $\times$ \textit{Female Share}&    -0.00592\sym{***}&    -0.00346\sym{***}&     0.00143         &      0.0115\sym{***}\\
                    &     (-7.07)         &     (-3.84)         &      (1.51)         &     (12.73)         \\
\addlinespace
Philosophy=1 $\times$ \textit{Female Share}&      0.0146\sym{***}&      0.0150\sym{***}&      0.0117\sym{***}&     -0.0135\sym{***}\\
                    &      (8.52)         &      (8.64)         &      (6.27)         &     (-7.34)         \\
\addlinespace
Physics=1 $\times$ \textit{Female Share}&   -0.000536         &     0.00564\sym{***}&    -0.00356\sym{**} &    -0.00353\sym{**} \\
                    &     (-0.48)         &      (4.92)         &     (-2.86)         &     (-2.99)         \\
\addlinespace
Political science=1 $\times$ \textit{Female Share}&     -0.0115\sym{***}&    -0.00671\sym{***}&    -0.00552\sym{**} &     0.00868\sym{***}\\
                    &     (-6.97)         &     (-4.16)         &     (-3.23)         &      (5.03)         \\
\addlinespace
Psychology=1 $\times$ \textit{Female Share}&     0.00714\sym{***}&     0.00296\sym{*}  &    -0.00990\sym{***}&     -0.0141\sym{***}\\
                    &      (6.22)         &      (2.55)         &     (-7.54)         &    (-11.43)         \\
\addlinespace
Sociology=1 $\times$ \textit{Female Share}&    -0.00482\sym{***}&    -0.00304\sym{*}  &     -0.0138\sym{***}&     -0.0136\sym{***}\\
                    &     (-3.55)         &     (-2.22)         &     (-9.01)         &     (-9.56)         \\
\addlinespace
DepartmentSize&  0.00000159\sym{***}&  0.00000124\sym{***}& 0.000000326\sym{***}&  0.00000192\sym{***}\\
                    &     (54.42)         &     (38.99)         &      (9.85)         &     (60.43)         \\
\addlinespace
CareerAge         &   -0.000174\sym{***}&   -0.000554\sym{***}&   -0.000139\sym{***}&   -0.000450\sym{***}\\
                    &    (-15.60)         &    (-46.56)         &    (-11.24)         &    (-38.42)         \\
\addlinespace
TeamSize        &     0.00139\sym{***}&    0.000694\sym{***}&     0.00102\sym{***}&     0.00273\sym{***}\\
                    &     (25.73)         &     (13.01)         &     (17.99)         &     (34.73)         \\
\addlinespace
$TeamSize^{2}$ &-0.000000421\sym{***}&-0.000000380\sym{***}&-0.000000162\sym{***}&-0.000000433\sym{***}\\
                    &    (-32.45)         &    (-28.76)         &    (-16.00)         &    (-28.38)         \\
\addlinespace
\textit{Num Gender Unknown}         &   -0.000308\sym{***}&    0.000265\sym{***}&   -0.000589\sym{***}&    -0.00159\sym{***}\\
                    &     (-4.80)         &      (4.17)         &     (-9.25)         &    (-17.65)         \\
\addlinespace
Constant            &       0.527\sym{***}&       0.529\sym{***}&       0.536\sym{***}&       0.591\sym{***}\\
                    &    (531.52)         &    (509.80)         &    (493.17)         &    (573.93)         \\
\midrule
Observations        &     4085543         &     4085543         &     4085543         &     4085543         \\
\bottomrule
\multicolumn{5}{l}{\footnotesize \textit{t} statistics in parentheses}\\
\multicolumn{5}{l}{\footnotesize \sym{*} \(p<0.05\), \sym{**} \(p<0.01\), \sym{***} \(p<0.001\)}\\
\end{tabular}
\end{adjustbox}
\end{table}

\section{Prescience conditional on initial surprise}\label{app:5ranks}

We examine how prescience varies with initial surprise of solo-authored papers across genders. We consider increasing thresholds of initial surprise (0.1 to 0.9, in 0.2 increments), and re-estimate the model on each subsample. 


For reference combinations (Table \ref{app:tab1_rank5}), the gender gap in uptake decreases as initial surprise increases. For concept combinations (Table \ref{app:tab2_rank5}), women’s papers initially receive less uptake, but the gap is no longer statistically distinguishable from zero for papers above the 0.9 percentile. Table \ref{app:tab3_rank5} shows no significant gender difference in two-step credit score for equally prescient reference combinations, no matter the initial level of surprise. Instead, Table \ref{app:tab4_rank5} reports significant gender differences in downstream credit for concept combinations with mid-to-low initial surprise (0.1–0.5), with the gap becoming statistically indistinguishable from zero at higher surprise levels (>0.7).

For equally prescient works, women’s are on average more disruptive than men’s, with gender differences diminishing with increasing initial level of surprise, until they become insignificant at higher initial surprise values (>0.5 for concepts, >0.7 for references) -- see Tables \ref{app:tab5_rank5} and \ref{app:tab6_rank5}. 

Overall, we find that gender is no longer significantly impacting recognition when reference or concept combinations are highly unexpected or surprise at their time of publication -- i.e., with highest surprise scores in our sample.

\begin{table}[htbp]\centering
\def\sym#1{\ifmmode^{#1}\else\(^{#1}\)\fi}
\caption{Linear regression model estimates for prescience in reference combinations conditional on increasing thresholds of initial surprise score in solo-authored papers. We include author-level clustered standard errors. Research field fixed effects and dummies for publication year are omitted.} \label{app:tab1_rank5}
\begin{adjustbox}{width=\textwidth}
\begin{tabular}{l*{5}{D{.}{.}{-1}}}
\toprule
                    &\multicolumn{5}{c}{\textbf{Prescience (References)}}\\
                    \multicolumn{1}{c}{\textit{Surprise (References) Score: }}&\multicolumn{1}{c}{\textit{>=0.1}}&\multicolumn{1}{c}{\textit{>=0.3}}&\multicolumn{1}{c}{\textit{>=0.5}}&\multicolumn{1}{c}{\textit{>=0.7}}&\multicolumn{1}{c}{\textit{>=0.9}}\\
\midrule
Female=1            &      0.0201\sym{***}&      0.0145\sym{***}&     0.00909\sym{***}&     0.00536\sym{***}&     0.00274\sym{*}  \\
                    &     (11.32)         &      (8.84)         &      (5.95)         &      (3.84)         &      (2.24)         \\
\addlinespace
DepSize&-0.000000369         & 0.000000863\sym{*}  & 0.000000976\sym{**} &  0.00000123\sym{***}& 0.000000809\sym{**} \\
                    &     (-1.10)         &      (2.50)         &      (2.80)         &      (3.81)         &      (2.71)         \\
\addlinespace
CareerAge          &   -0.000975\sym{***}&   -0.000536\sym{***}&   -0.000212\sym{***}&  -0.0000367         &   0.0000984\sym{*}  \\
                    &    (-17.79)         &    (-10.38)         &     (-4.33)         &     (-0.79)         &      (2.40)         \\
\addlinespace
Constant            &       0.585\sym{***}&       0.665\sym{***}&       0.742\sym{***}&       0.826\sym{***}&       0.925\sym{***}\\
                    &    (150.19)         &    (181.29)         &    (213.16)         &    (258.72)         &    (332.68)         \\
\midrule
Observations        &      208503         &      173387         &      138600         &       99842         &       47982         \\
\bottomrule
\bottomrule
\multicolumn{6}{l}{\footnotesize \textit{t} statistics in parentheses}\\
\multicolumn{6}{l}{\footnotesize \sym{*} \(p<0.05\), \sym{**} \(p<0.01\), \sym{***} \(p<0.001\)}\\
\end{tabular}
\end{adjustbox}
\end{table}

\begin{table}[htbp]\centering
\def\sym#1{\ifmmode^{#1}\else\(^{#1}\)\fi}
\caption{Linear regression model estimates for prescience in concept combinations conditional on increasing thresholds of initial surprise score (higher than 0.1, 0.3, 0.5, 0.7, 0.9) in solo-authored papers. We include  author-level clustered standard errors. Research field fixed effects and dummies for publication year are omitted.} \label{app:tab2_rank5}
\begin{adjustbox}{width=\textwidth}
\begin{tabular}{l*{5}{D{.}{.}{-1}}}
\toprule
                    &\multicolumn{5}{c}{\textbf{Prescience (concepts)}}\\
                    \multicolumn{1}{c}{\textit{Surprise (concepts) Score:} }&\multicolumn{1}{c}{\textit{>=0.1}}&\multicolumn{1}{c}{\textit{>=0.3}}&\multicolumn{1}{c}{\textit{>=0.5}}&\multicolumn{1}{c}{\textit{>=0.7}}&\multicolumn{1}{c}{\textit{>=0.9}}\\
\midrule
Female=1            &    -0.00755\sym{***}&    -0.00773\sym{***}&    -0.00536\sym{**} &    -0.00477\sym{*}  &    -0.00167         \\
                    &     (-4.28)         &     (-4.29)         &     (-2.86)         &     (-2.27)         &     (-0.60)         \\
\addlinespace
DepSize&  0.00000275\sym{***}&  0.00000230\sym{***}&  0.00000182\sym{***}&  0.00000147\sym{***}&  0.00000140\sym{**} \\
                    &      (8.91)         &      (7.67)         &      (5.97)         &      (4.38)         &      (2.91)         \\
\addlinespace
CareerAge          &   -0.000310\sym{***}&   -0.000353\sym{***}&   -0.000364\sym{***}&   -0.000334\sym{***}&   -0.000307\sym{***}\\
                    &     (-5.92)         &     (-6.66)         &     (-6.55)         &     (-5.44)         &     (-3.76)         \\
\addlinespace
Constant            &       0.557\sym{***}&       0.663\sym{***}&       0.734\sym{***}&       0.807\sym{***}&       0.889\sym{***}\\
                    &    (140.62)         &    (168.57)         &    (180.71)         &    (177.39)         &    (145.95)         \\
\midrule
Observations        &      209293         &      166188         &      122332         &       78118         &       29309         \\
\bottomrule
\bottomrule
\bottomrule
\multicolumn{6}{l}{\footnotesize \textit{t} statistics in parentheses}\\
\multicolumn{6}{l}{\footnotesize \sym{*} \(p<0.05\), \sym{**} \(p<0.01\), \sym{***} \(p<0.001\)}\\
\end{tabular}
\end{adjustbox}
\end{table}

\begin{table}[H]\centering
\def\sym#1{\ifmmode^{#1}\else\(^{#1}\)\fi}
\caption{Linear regression model estimates for Two-step credit score in solo-authored papers in prescience in reference combinations conditional on increasing thresholds of initial surprise score (higher than 0.1, 0.3, 0.5, 0.7, 0.9). We include  author-level clustered standard errors. Research field fixed effects and dummies for publication year are omitted.}\label{app:tab3_rank5}
\begin{adjustbox}{width=\textwidth}
\begin{tabular}{l*{5}{D{.}{.}{-1}}}
\toprule
                    &\multicolumn{5}{c}{\textbf{Two-step Credit Score}}\\
                 \multicolumn{1}{c}{\textit{Surprise (References) Score:}}&\multicolumn{1}{c}{\textit{>=0.1}}&\multicolumn{1}{c}{\textit{>=0.3}}&\multicolumn{1}{c}{\textit{>=0.5}}&\multicolumn{1}{c}{\textit{>=0.7}}&\multicolumn{1}{c}{\textit{>=0.9}}\\
\midrule
Female=1            &   -0.000215         &     0.00115         &     0.00196         &     0.00228         &     0.00831         \\
                    &     (-0.20)         &      (0.95)         &      (1.37)         &      (1.17)         &      (1.52)         \\
\addlinespace
\textit{Prescience (References)}&     -0.0110\sym{***}&     -0.0103\sym{***}&     -0.0102\sym{***}&     -0.0101\sym{***}&     -0.0144\sym{***}\\
                    &    (-16.37)         &    (-14.26)         &    (-11.94)         &     (-8.75)         &     (-5.10)         \\
\addlinespace
Female=1 $\times$ \textit{Prescience (References)}&   -0.000874         &    -0.00259         &    -0.00346\sym{*}  &    -0.00376         &    -0.00975         \\
                    &     (-0.67)         &     (-1.76)         &     (-2.05)         &     (-1.70)         &     (-1.65)         \\
\addlinespace
DepSize& -0.00000132\sym{***}& -0.00000119\sym{***}& -0.00000108\sym{***}&-0.000000916\sym{***}&-0.000000698\sym{***}\\
                    &    (-18.48)         &    (-16.49)         &    (-14.64)         &    (-11.31)         &     (-7.33)         \\
\addlinespace
CareerAge          &   -0.000127\sym{***}&   -0.000119\sym{***}&   -0.000111\sym{***}&   -0.000104\sym{***}&  -0.0000643\sym{***}\\
                    &     (-9.96)         &     (-9.48)         &     (-8.67)         &     (-7.85)         &     (-4.22)         \\
\addlinespace
\textit{OpenAccess}             &    -0.00324\sym{***}&    -0.00271\sym{***}&    -0.00245\sym{***}&    -0.00189\sym{***}&    -0.00159\sym{***}\\
                    &    (-10.11)         &     (-8.61)         &     (-7.76)         &     (-5.76)         &     (-4.09)         \\
\addlinespace
Constant            &      0.0388\sym{***}&      0.0375\sym{***}&      0.0362\sym{***}&      0.0346\sym{***}&      0.0366\sym{***}\\
                    &     (47.39)         &     (44.98)         &     (38.80)         &     (28.85)         &     (13.63)         \\
\midrule
Observations        &       93182         &       83988         &       71914         &       55443         &       29378         \\
\bottomrule
\multicolumn{6}{l}{\footnotesize \textit{t} statistics in parentheses}\\
\multicolumn{6}{l}{\footnotesize \sym{*} \(p<0.05\), \sym{**} \(p<0.01\), \sym{***} \(p<0.001\)}\\
\end{tabular}
\end{adjustbox}
\end{table}

\begin{table}[H]\centering
\def\sym#1{\ifmmode^{#1}\else\(^{#1}\)\fi}
\caption{Linear regression model estimates for Two-step Credit Score in solo-authored papers in prescience of concept combinations conditional on increasing thresholds of initial surprise score (higher than 0.1, 0.3, 0.5, 0.7, 0.9). We include author-level clustered standard errors. Research field fixed effects and dummies for publication year are omitted.}\label{app:tab4_rank5}
\begin{adjustbox}{width=\textwidth}
\begin{tabular}{l*{5}{D{.}{.}{-1}}}
\toprule
                    &\multicolumn{5}{c}{\textbf{Two-step Credit Score}}\\
                 \multicolumn{1}{c}{\textit{Surprise (concepts) Score: }}&\multicolumn{1}{c}{\textit{>=0.1}}&\multicolumn{1}{c}{\textit{>=0.3}}&\multicolumn{1}{c}{\textit{>=0.5}}&\multicolumn{1}{c}{\textit{>=0.7}}&\multicolumn{1}{c}{\textit{>=0.9}}\\
\midrule
Female=1            &    -0.00225\sym{**} &    -0.00305\sym{**} &    -0.00329\sym{*}  &    -0.00327         &     0.00409         \\
                    &     (-2.91)         &     (-2.97)         &     (-2.38)         &     (-1.54)         &      (0.71)         \\
\addlinespace
\textit{Prescience (Concepts)}&     -0.0106\sym{***}&     -0.0124\sym{***}&     -0.0139\sym{***}&     -0.0146\sym{***}&     -0.0136\sym{***}\\
                    &    (-16.41)         &    (-14.80)         &    (-13.29)         &    (-10.10)         &     (-4.84)         \\
\addlinespace
Female=1 $\times$ \textit{Prescience (Concepts)}&     0.00191         &     0.00295\sym{*}  &     0.00335         &     0.00315         &    -0.00494         \\
                    &      (1.73)         &      (2.13)         &      (1.90)         &      (1.24)         &     (-0.75)         \\
\addlinespace
DepSize& -0.00000136\sym{***}& -0.00000129\sym{***}& -0.00000127\sym{***}& -0.00000127\sym{***}& -0.00000110\sym{***}\\
                    &    (-18.20)         &    (-15.84)         &    (-14.58)         &    (-12.64)         &     (-6.59)         \\
\addlinespace
CareerAge          &   -0.000121\sym{***}&   -0.000122\sym{***}&   -0.000116\sym{***}&   -0.000130\sym{***}&   -0.000154\sym{***}\\
                    &     (-8.96)         &     (-8.39)         &     (-7.40)         &     (-6.69)         &     (-5.20)         \\
\addlinespace
\textit{OpenAccess}             &    -0.00325\sym{***}&    -0.00355\sym{***}&    -0.00325\sym{***}&    -0.00272\sym{***}&    -0.00319\sym{***}\\
                    &     (-9.51)         &     (-9.60)         &     (-7.95)         &     (-5.28)         &     (-3.89)         \\
\addlinespace
Constant            &      0.0388\sym{***}&      0.0404\sym{***}&      0.0423\sym{***}&      0.0435\sym{***}&      0.0445\sym{***}\\
                    &     (50.19)         &     (42.17)         &     (36.64)         &     (27.29)         &     (15.07)         \\
\midrule
Observations        &       89747         &       72941         &       55085         &       35954         &       13760         \\
\bottomrule
\multicolumn{6}{l}{\footnotesize \textit{t} statistics in parentheses}\\
\multicolumn{6}{l}{\footnotesize \sym{*} \(p<0.05\), \sym{**} \(p<0.01\), \sym{***} \(p<0.001\)}\\
\end{tabular}
\end{adjustbox}
\end{table}

\begin{table}[H]\centering
\def\sym#1{\ifmmode^{#1}\else\(^{#1}\)\fi}
\caption{Linear regression model estimates for five-year disruption score in solo-authored papers in prescience reference combinations, conditional on increasing thresholds of initial surprise score (higher than 0.1, 0.3, 0.5, 0.7, 0.9). We include author-level clustered standard errors. Research field fixed effects and dummies for publication year are omitted.}\label{app:tab5_rank5}
\begin{adjustbox}{width=\textwidth}
\begin{tabular}{l*{5}{D{.}{.}{-1}}}
\toprule
&\multicolumn{5}{c}{\textbf{5-year Disruption}}\\
                 \multicolumn{1}{c}{\textit{Surprise (References) Score: }}&\multicolumn{1}{c}{\textit{>=0.1}}&\multicolumn{1}{c}{\textit{>=0.3}}&\multicolumn{1}{c}{\textit{>=0.5}}&\multicolumn{1}{c}{\textit{>=0.7}}&\multicolumn{1}{c}{\textit{>=0.9}}\\
\midrule
Female=1            &     0.00443\sym{***}&     0.00266\sym{***}&     0.00119\sym{***}&    0.000240         &    0.000429         \\
                    &      (7.54)         &      (5.19)         &      (3.30)         &      (0.60)         &      (0.61)         \\
\addlinespace
\textit{Prescience (References)}&    -0.00716\sym{***}&    -0.00244\sym{***}&   -0.000669\sym{*}  &    0.000187         &     0.00161\sym{***}\\
                    &    (-18.20)         &     (-8.58)         &     (-2.54)         &      (0.64)         &      (3.62)         \\
\addlinespace
Female=1 $\times$ \textit{Prescience (References)}&    -0.00581\sym{***}&    -0.00325\sym{***}&    -0.00133\sym{**} &   -0.000212         &   -0.000586         \\
                    &     (-7.53)         &     (-4.98)         &     (-3.03)         &     (-0.45)         &     (-0.75)         \\
\addlinespace
DepSize&   -3.36e-09         &    1.60e-08         &    3.32e-09         &   -1.30e-08         &   -1.75e-08         \\
                    &     (-0.22)         &      (1.20)         &      (0.25)         &     (-0.91)         &     (-1.01)         \\
\addlinespace
CareerAge          &   0.0000222\sym{***}&   0.0000182\sym{***}&   0.0000123\sym{***}&  0.00000844\sym{**} & -0.00000258         \\
                    &      (4.43)         &      (4.80)         &      (4.14)         &      (3.04)         &     (-0.84)         \\
\addlinespace
\textit{OpenAccess}             &   -0.000582\sym{***}&   -0.000553\sym{***}&   -0.000641\sym{***}&   -0.000607\sym{***}&   -0.000460\sym{***}\\
                    &     (-4.39)         &     (-5.13)         &     (-7.28)         &     (-7.08)         &     (-4.74)         \\
\addlinespace
impact\_factor\_2y    &   -0.000143\sym{***}&   -0.000126\sym{***}&   -0.000117\sym{***}&   -0.000100\sym{***}&  -0.0000849\sym{*}  \\
                    &     (-8.47)         &     (-6.92)         &     (-5.39)         &     (-3.82)         &     (-2.38)         \\
\addlinespace
2y\_cites            &  -0.0000680\sym{***}&   -0.000102\sym{***}&   -0.000107\sym{***}&   -0.000115\sym{***}&   -0.000118\sym{***}\\
                    &     (-7.98)         &    (-11.88)         &    (-11.60)         &    (-10.84)         &     (-8.94)         \\
\addlinespace
Constant            &     0.00211\sym{***}&    -0.00220\sym{***}&    -0.00372\sym{***}&    -0.00486\sym{***}&    -0.00614\sym{***}\\
                    &      (3.58)         &     (-4.95)         &     (-8.98)         &    (-13.45)         &    (-11.45)         \\
\midrule
Observations        &      196587         &      164114         &      131648         &       95175         &       45725         \\
\bottomrule
\multicolumn{6}{l}{\footnotesize \textit{t} statistics in parentheses}\\
\multicolumn{6}{l}{\footnotesize \sym{*} \(p<0.05\), \sym{**} \(p<0.01\), \sym{***} \(p<0.001\)}\\
\end{tabular}
\end{adjustbox}
\end{table}

\begin{table}[H]\centering
\def\sym#1{\ifmmode^{#1}\else\(^{#1}\)\fi}
\caption{Linear regression model estimates for five-year disruption score in solo-authored papers in prescience in concepts conditional on increasing thresholds of initial surprise score (higher than 0.1, 0.3, 0.5, 0.7, 0.9). We include  author-level clustered standard errors. Research field fixed effects and dummies for publication year are omitted.}\label{app:tab6_rank5}
\begin{adjustbox}{width=\textwidth}
\begin{tabular}{l*{5}{D{.}{.}{-1}}}
\toprule
&\multicolumn{5}{c}{\textbf{5-year Disruption}}\\
                 \multicolumn{1}{c}{\textit{Surprise (concepts) Score: }}&\multicolumn{1}{c}{\textit{>=0.1}}&\multicolumn{1}{c}{\textit{>=0.3}}&\multicolumn{1}{c}{\textit{>=0.5}}&\multicolumn{1}{c}{\textit{>=0.7}}&\multicolumn{1}{c}{\textit{>=0.9}}\\
\midrule
Female=1            &     0.00140\sym{***}&     0.00149\sym{**} &   0.0000178         &  -0.0000154         &    -0.00101         \\
                    &      (3.31)         &      (2.61)         &      (0.02)         &     (-0.01)         &     (-0.43)         \\
\addlinespace
\textit{Prescience (Concepts)}&     0.00134\sym{***}&     0.00171\sym{***}&     0.00164\sym{**} &     0.00163\sym{*}  &    0.000701         \\
                    &      (3.62)         &      (3.83)         &      (2.88)         &      (2.27)         &      (0.45)         \\
\addlinespace
Female=1 $\times$ \textit{Prescience (Concepts)}&   -0.000689         &   -0.000881         &    0.000981         &    0.000688         &     0.00185         \\
                    &     (-1.04)         &     (-1.06)         &      (0.98)         &      (0.52)         &      (0.67)         \\
\addlinespace
DepSize&   -3.59e-08         &   -4.42e-08         &   -3.13e-08         &   -1.83e-08         &    5.25e-08         \\
                    &     (-1.41)         &     (-1.56)         &     (-0.99)         &     (-0.50)         &      (0.79)         \\
\addlinespace
CareerAge          &   0.0000562\sym{***}&   0.0000593\sym{***}&   0.0000549\sym{***}&   0.0000555\sym{***}&   0.0000712\sym{***}\\
                    &      (8.66)         &      (8.10)         &      (6.70)         &      (5.50)         &      (4.48)         \\
\addlinespace
\textit{OpenAccess}             &   -0.000288         &   -0.000269         &   -0.000388         &   -0.000670\sym{*}  &    -0.00127\sym{**} \\
                    &     (-1.57)         &     (-1.33)         &     (-1.71)         &     (-2.41)         &     (-2.67)         \\
\addlinespace
impact\_factor\_2y    &  -0.0000733\sym{*}  &  -0.0000656         &  -0.0000621         &  -0.0000385         &  -0.0000667         \\
                    &     (-2.38)         &     (-1.87)         &     (-1.50)         &     (-0.66)         &     (-0.89)         \\
\addlinespace
2y\_cites            &  -0.0000542\sym{***}&  -0.0000522\sym{**} &  -0.0000597\sym{**} &  -0.0000653\sym{**} &  -0.0000668\sym{**} \\
                    &     (-3.52)         &     (-3.12)         &     (-3.14)         &     (-2.81)         &     (-3.21)         \\
\addlinespace
Constant            &    -0.00259\sym{***}&    -0.00389\sym{***}&    -0.00387\sym{***}&    -0.00364\sym{***}&    -0.00339         \\
                    &     (-4.15)         &     (-6.13)         &     (-5.01)         &     (-3.62)         &     (-1.80)         \\
\midrule
Observations        &      196922         &      156476         &      114957         &       73406         &       27446         \\
\bottomrule
\multicolumn{6}{l}{\footnotesize \textit{t} statistics in parentheses}\\
\multicolumn{6}{l}{\footnotesize \sym{*} \(p<0.05\), \sym{**} \(p<0.01\), \sym{***} \(p<0.001\)}\\
\end{tabular}
\end{adjustbox}
\end{table}

\section{Outside subject citation share}

Table \ref{tab1_outside_cite_share} reports the model coefficient estimates behind Figure 4 of the main text.

\begin{table}[H]\centering
\def\sym#1{\ifmmode^{#1}\else\(^{#1}\)\fi}
\caption{Linear Regression model estimates of outside- and inside-subject citation share on surprise and prescience score by authors' gender in solo-authored papers. Outside citation share is equal to $1-InsideSubjectCitationShare$.  Research field fixed effects and dummies for publishing year are omitted. Standard errors are clustered at the author level. \label{tab1_outside_cite_share}}
\begin{tabular}{l*{2}{D{.}{.}{-1}}}
\toprule
                    &\multicolumn{1}{c}{(1)}&\multicolumn{1}{c}{(2)}\\
                    &\multicolumn{1}{c}{outside\_subject\_share}&\multicolumn{1}{c}{inside\_subject\_share}\\
\midrule
Female=1            &      0.0304\sym{***}&     -0.0304\sym{***}\\
                    &      (5.78)         &     (-5.78)         \\
\addlinespace
\textit{Surprise (References)}  &      0.0750\sym{***}&     -0.0750\sym{***}\\
                    &     (23.66)         &    (-23.66)         \\
\addlinespace
Female=1 $\times$ \textit{Surprise (References)}&     -0.0232\sym{***}&      0.0232\sym{***}\\
                    &     (-3.89)         &      (3.89)         \\
\addlinespace
 \textit{Surprise (Concepts)}  &     -0.0196\sym{***}&      0.0196\sym{***}\\
                    &     (-6.75)         &      (6.75)         \\
\addlinespace
Female=1 $\times$  \textit{Surprise (Concepts)}&    -0.00236         &     0.00236         \\
                    &     (-0.43)         &      (0.43)         \\
\addlinespace
DepSize &  0.00000163\sym{***}& -0.00000163\sym{***}\\
                    &      (4.83)         &     (-4.83)         \\
\addlinespace
CareerAge          &    0.000555\sym{***}&   -0.000555\sym{***}\\
                    &      (9.69)         &     (-9.69)         \\
\addlinespace
OpenAccess=1             &      0.0146\sym{***}&     -0.0146\sym{***}\\
                    &      (8.65)         &     (-8.65)         \\
\addlinespace
Constant            &       0.344\sym{***}&       0.656\sym{***}\\
                    &     (74.31)         &    (141.91)         \\
\midrule
Observations        &      189701         &      189701         \\
\bottomrule
\multicolumn{3}{l}{\footnotesize \textit{t} statistics in parentheses}\\
\multicolumn{3}{l}{\footnotesize \sym{*} \(p<0.05\), \sym{**} \(p<0.01\), \sym{***} \(p<0.001\)}\\
\end{tabular}
\end{table}
\section{Multi-authored papers}\label{app:mixed_ma}
We examine how the share of women among authors in multi-authored papers relates to novelty. We estimate models (1) and (2) of the main text with a continuous variable for female share within teams, $FemaleShare_i$, measuring the proportion of clearly-coded women (U.S., white and Hispanic) among the authors of a paper. Controls include average department size of team members (matched with the paper’s field), year of publication, fine-grained research area dummies (level 1), average career age of team members, team size and squared team size, and the number of authors with unknown gender (\emph{Gender Num Unknown}). 

\begin{figure}[H]
\includegraphics[width=1.0\linewidth]{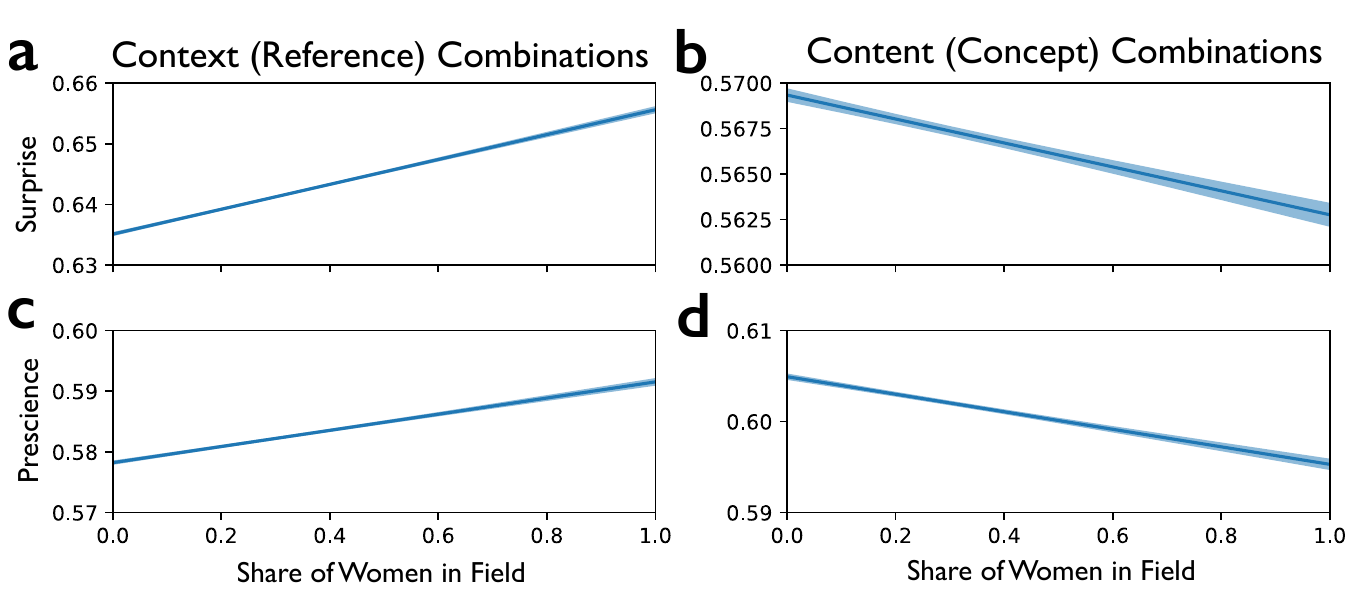}
  \caption{Predicted values of surprise and prescience by women share in multi-authored papers, alongside 95\% confidence intervals, marginalized over observed distribution of covariates.}
  \label{fig:fig_3}
\end{figure}

Table \ref{tab1_ma} reports the regression coefficients, and Figure \ref{fig:fig_3} shows predicted novelty outcomes by women share, with 95\% confidence intervals. Predicted outcomes indicate that increasing the share of women among authors is associated with higher reference surprise and reference prescience, but lower concept surprise and prescience. These patterns are broadly consistent with results for solo-authored papers.

\begin{figure}[H]
\includegraphics[width=1.0\linewidth]{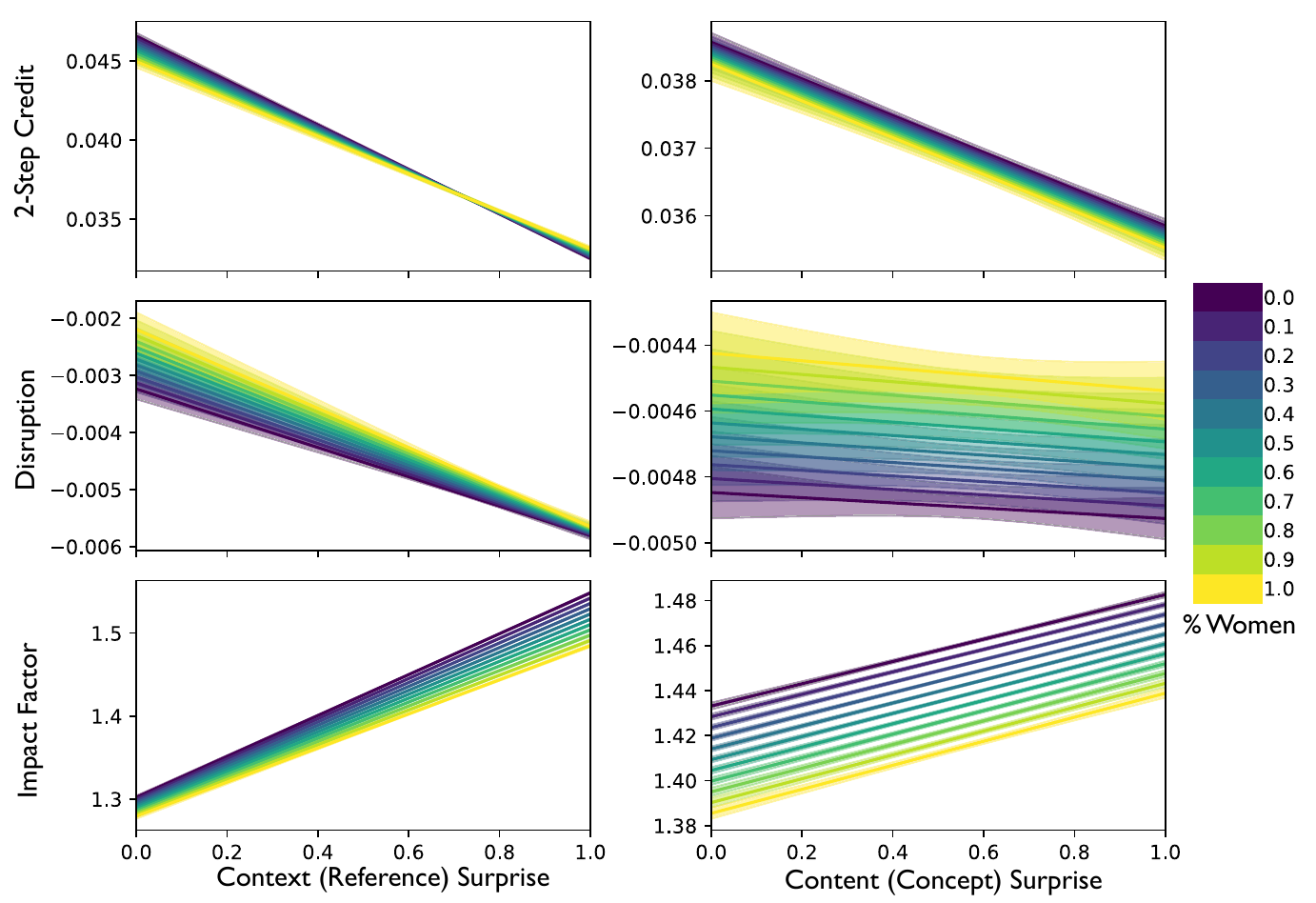}
  \caption{Rewards to \emph{Surprise} by women share in multi-authored papers, alongside 95\% confidence intervals. The solid lines and shaded areas show the estimated margins and their confidence intervals for each level of the share of women in the team, marginalized over observed distribution of covariates.
  }
  \label{fig:fig_5a}
\end{figure}

\begin{figure}[H]
\includegraphics[width=1.0\linewidth]{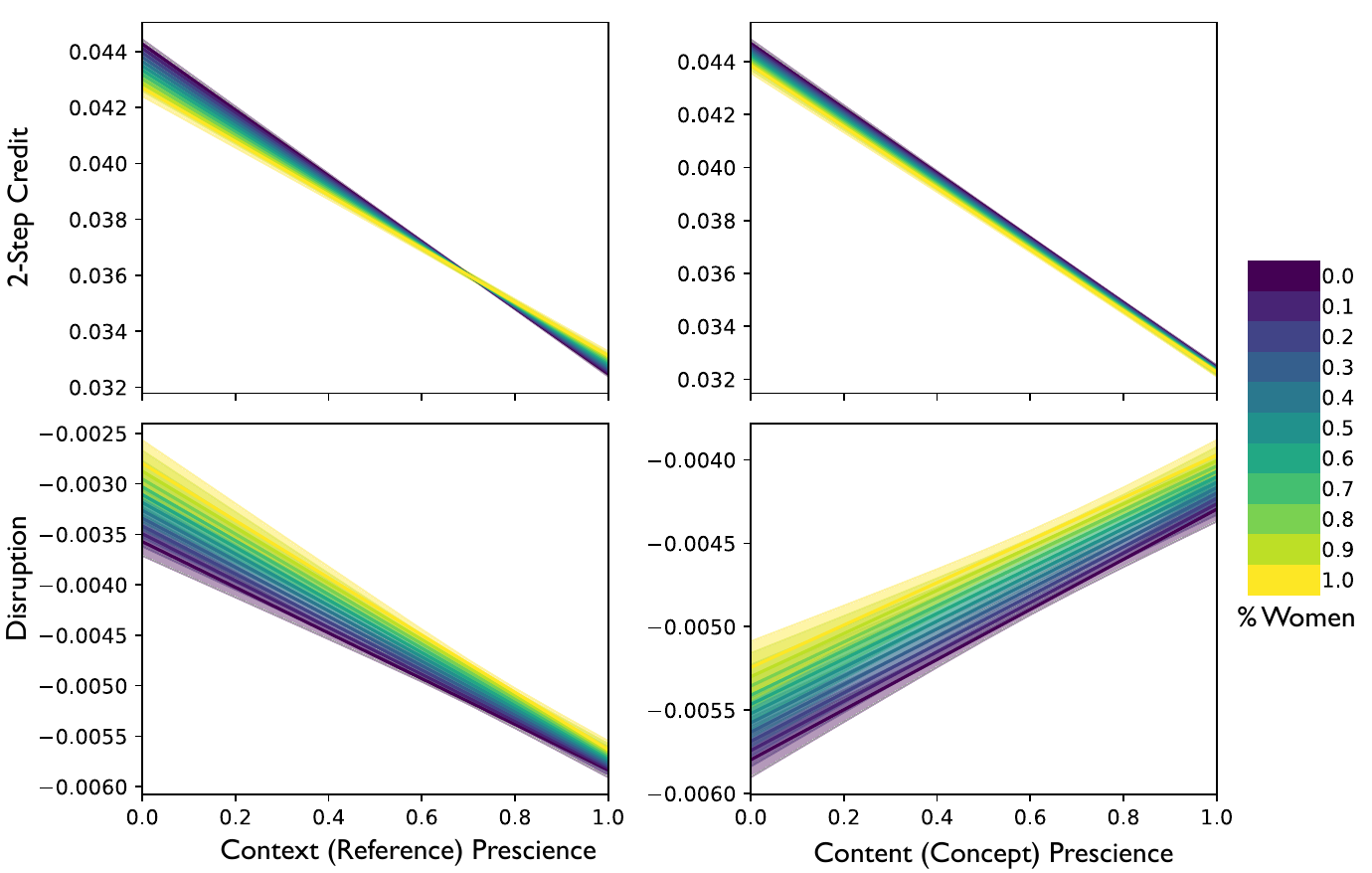}
  \caption{Rewards to \emph{Prescience} by women share in multi-authored papers, alongside 95\% confidence intervals. The solid lines and shaded areas show the estimated margins and their confidence intervals for each level of the share of women in the team, marginalized over observed distribution of covariates.
  }
  \label{fig:fig_5b}
\end{figure}

We further examine how women’s share in teams relates to scholarly rewards (two-step credit, disruption, and journal impact factor). Regression results are reported in Tables \ref{tab3_cred}–\ref{tab3_jif}, and predicted outcomes in Figures \ref{fig:fig_5a} and \ref{fig:fig_5b}. For equally surprising or prescient science, teams with a higher share of clearly coded women tend to receive lower two-step citation credit and are placed in lower-impact journals, but produce more disruptive contributions -- consistent with patterns observed for solo-authored work.

Teams with a higher share of women experience a smaller penalty in downstream citation credit as prescience increases, for both concept and reference combinations (Table \ref{tab3_cred}).

Reference surprise is negatively associated with disruption, and this decline is steeper for teams with more women, again reflecting patterns observed in solo-authored papers, while gender gaps in disruption remain stable as novelty increases (Table \ref{tab3_cd}). Teams with a higher share of women also face lower marginal returns to journal prestige as surprise increases (Table \ref{tab3_jif}). This contrasts with the solo-authored case, where women exhibit a steeper positive relationship between novelty and journal impact factor. As discussed in the main text, this difference disappears when we model the non-linear association between gender composition and journal placement, comparing all-women and all-men teams (Section \ref{app:non_linear_ma}).

\begin{table}[H]\centering
\def\sym#1{\ifmmode^{#1}\else\(^{#1}\)\fi}
\caption{Linear regression model estimates of content and context surprise and prescience scores by gender shares in multi-authored papers. We include White standard errors. Research field fixed effects and dummies for publishing year are omitted.\label{tab1_ma}}
\begin{adjustbox}{width=\textwidth}
\begin{tabular}{lcccc}
\toprule
                    &\multicolumn{1}{c}{(1)}&\multicolumn{1}{c}{(2)}&\multicolumn{1}{c}{(3)}&\multicolumn{1}{c}{(4)}\\
                    &\multicolumn{1}{c}{Surprise (References)}&\multicolumn{1}{c}{Prescience (References)}&\multicolumn{1}{c}{Surprise (concepts)}&\multicolumn{1}{c}{Prescience (concepts)}\\
\midrule
\textit{Female Share}       &      0.0205\sym{***}&      0.0133\sym{***}&    -0.00658\sym{***}&    -0.00963\sym{***}\\
                    &     (52.48)         &     (31.97)         &    (-15.08)         &    (-23.25)         \\
\addlinespace
DepartmentSize&  0.00000154\sym{***}&  0.00000117\sym{***}& 0.000000394\sym{***}&  0.00000196\sym{***}\\
                    &     (52.64)         &     (37.03)         &     (11.91)         &     (61.88)         \\
\addlinespace
CareerAge         &   -0.000160\sym{***}&   -0.000548\sym{***}&   -0.000153\sym{***}&   -0.000468\sym{***}\\
                    &    (-14.24)         &    (-45.96)         &    (-12.37)         &    (-39.95)         \\
\addlinespace
TeamSize        &     0.00142\sym{***}&    0.000716\sym{***}&    0.000993\sym{***}&     0.00267\sym{***}\\
                    &     (26.12)         &     (13.41)         &     (17.56)         &     (34.39)         \\
\addlinespace
$TeamSize^{2}$&-0.000000427\sym{***}&-0.000000374\sym{***}&-0.000000161\sym{***}&-0.000000423\sym{***}\\
                    &    (-32.51)         &    (-28.53)         &    (-15.98)         &    (-28.22)         \\
\addlinespace
\textit{Num Gender Unknown}         &   -0.000326\sym{***}&    0.000223\sym{***}&   -0.000557\sym{***}&    -0.00155\sym{***}\\
                    &     (-5.02)         &      (3.51)         &     (-8.79)         &    (-17.42)         \\
\addlinespace
Constant            &       0.542\sym{***}&       0.539\sym{***}&       0.523\sym{***}&       0.578\sym{***}\\
                    &    (579.17)         &    (551.90)         &    (510.95)         &    (596.49)         \\
\midrule
Observations        &     4085543         &     4085543         &     4085543         &     4085543         \\
\bottomrule
\multicolumn{5}{l}{\footnotesize \textit{t} statistics in parentheses}\\
\multicolumn{5}{l}{\footnotesize \sym{*} \(p<0.05\), \sym{**} \(p<0.01\), \sym{***} \(p<0.001\)}\\
\end{tabular}
\end{adjustbox}
\end{table}

 \begin{table}[H]\centering
\def\sym#1{\ifmmode^{#1}\else\(^{#1}\)\fi}
\caption{Linear regression model estimates of 2-steps credit score for surprise and prescience scores in multi-authored papers. We include White standard errors. Research field fixed effects and dummies for publishing year are omitted.}\label{tab3_cred}
\begin{adjustbox}{width=\textwidth}
\begin{tabular}{lcccc}
\toprule
 &\multicolumn{1}{c}{(1)}&\multicolumn{1}{c}{(2)}&\multicolumn{1}{c}{(3)}&\multicolumn{1}{c}{(4)}\\
                    &\multicolumn{1}{c}{TwoStepsCreditScore}&\multicolumn{1}{c}{TwoStepsCreditScore}&\multicolumn{1}{c}{TwoStepsCreditScore}&\multicolumn{1}{c}{TwoStepsCreditScore}\\
\midrule
\textit{Female Share}       &    -0.00161\sym{***}&    -0.00157\sym{***}&   -0.000325\sym{*}  &   -0.000820\sym{***}\\
                    &     (-6.32)         &     (-7.92)         &     (-2.01)         &     (-4.30)         \\
\addlinespace
\textit{Surprise (References)}    &     -0.0141\sym{***}&                     &                     &                     \\
                    &    (-99.01)         &                     &                     &                     \\
\addlinespace
\textit{Female Share}$\times$ \textit{Surprise (References)}&     0.00222\sym{***}&                     &                     &                     \\
                    &      (6.94)         &                     &                     &                     \\
\addlinespace
DepartmentSize&-0.000000420\sym{***}&-0.000000425\sym{***}&-0.000000424\sym{***}&-0.000000405\sym{***}\\
                    &    (-53.59)         &    (-54.29)         &    (-54.12)         &    (-51.96)         \\
\addlinespace
CareerAge         &  -0.0000140\sym{***}&  -0.0000172\sym{***}&  -0.0000113\sym{***}&  -0.0000169\sym{***}\\
                    &     (-6.62)         &     (-8.10)         &     (-5.33)         &     (-7.96)         \\
\addlinespace
OpenAccess            &    -0.00240\sym{***}&    -0.00278\sym{***}&    -0.00287\sym{***}&    -0.00270\sym{***}\\
                    &    (-47.86)         &    (-55.36)         &    (-56.80)         &    (-53.68)         \\
\addlinespace
TeamSize        &   -0.000285\sym{***}&   -0.000283\sym{***}&   -0.000268\sym{***}&   -0.000244\sym{***}\\
                    &    (-22.53)         &    (-22.42)         &    (-22.14)         &    (-21.19)         \\
\addlinespace
$TeamSize^{2}$ &    1.73e-08\sym{***}&    1.80e-08\sym{***}&    1.92e-08\sym{***}&    1.60e-08\sym{***}\\
                    &     (11.64)         &     (12.08)         &     (12.88)         &     (11.08)         \\
\addlinespace
\textit{Num Gender Unknown}         &    0.000236\sym{***}&    0.000232\sym{***}&    0.000214\sym{***}&    0.000199\sym{***}\\
                    &     (17.22)         &     (16.96)         &     (16.29)         &     (15.89)         \\
\addlinespace
\textit{Prescience (References)}  &                     &     -0.0118\sym{***}&                     &                     \\
                    &                     &    (-98.59)         &                     &                     \\
\addlinespace
\textit{Female Share}$\times$ \textit{Prescience (References)}&                     &     0.00222\sym{***}&                     &                     \\
                    &                     &      (8.34)         &                     &                     \\
\addlinespace
\textit{Surprise (concepts)}   &                     &                     &    -0.00272\sym{***}&                     \\
                    &                     &                     &    (-25.49)         &                     \\
\addlinespace
\textit{Female Share}$\times$ \textit{Surprise (concepts)}&                     &                     & -0.00000244         &                     \\
                    &                     &                     &     (-0.01)         &                     \\
\addlinespace
\textit{Prescience (concepts)} &                     &                     &                     &     -0.0122\sym{***}\\
                    &                     &                     &                     &   (-104.79)         \\
\addlinespace
\textit{Female Share}$\times$ \textit{Prescience (concepts)}&                     &                     &                     &    0.000642\sym{*}  \\
                    &                     &                     &                     &      (2.47)         \\
\addlinespace
Constant            &      0.0354\sym{***}&      0.0339\sym{***}&      0.0287\sym{***}&      0.0345\sym{***}\\
                    &    (239.56)         &    (249.38)         &    (224.02)         &    (255.70)         \\
\midrule
Observations        &     2451029         &     2451029         &     2451029         &     2451029         \\
\multicolumn{5}{l}{\footnotesize \textit{t} statistics in parentheses}\\
\multicolumn{5}{l}{\footnotesize \sym{*} \(p<0.05\), \sym{**} \(p<0.01\), \sym{***} \(p<0.001\)}\\
\end{tabular}
\end{adjustbox}
\end{table}

\begin{table}[H]\centering
\def\sym#1{\ifmmode^{#1}\else\(^{#1}\)\fi}
\caption{Linear regression model estimates  of disruption score of publication on surprise and prescience scores by gender in multi-authored papers, with White standard errors. Research field fixed effects and dummies for publishing year are omitted.}\label{tab3_cd}
\begin{adjustbox}{width=\textwidth}
\begin{tabular}{l*{8}{D{.}{.}{-1}}}
\toprule
                    &\multicolumn{1}{c}{(1)}&\multicolumn{1}{c}{(2)}&\multicolumn{1}{c}{(3)}&\multicolumn{1}{c}{(4)}&\multicolumn{1}{c}{(5)}&\multicolumn{1}{c}{(6)}&\multicolumn{1}{c}{(7)}&\multicolumn{1}{c}{(8)}\\
                    &\multicolumn{1}{c}{3-years Disruption}&\multicolumn{1}{c}{3-years Disruption}&\multicolumn{1}{c}{3-years Disruption}&\multicolumn{1}{c}{3-years Disruption}&\multicolumn{1}{c}{5-year Disruption}&\multicolumn{1}{c}{5-year Disruption}&\multicolumn{1}{c}{5-year Disruption}&\multicolumn{1}{c}{5-year Disruption}\\
\midrule
\textit{Female Share}       &     0.00124\sym{***}&    0.000915\sym{***}&    0.000496\sym{***}&    0.000600\sym{***}&     0.00104\sym{***}&    0.000775\sym{***}&    0.000423\sym{***}&    0.000553\sym{***}\\
                    &      (6.24)         &      (6.12)         &      (5.85)         &      (6.07)         &      (5.39)         &      (5.36)         &      (5.13)         &      (5.76)         \\
\addlinespace
\textit{Surprise (References)}    &    -0.00127\sym{***}&                     &                     &                     &    -0.00257\sym{***}&                     &                     &                     \\
                    &     (-9.55)         &                     &                     &                     &    (-20.53)         &                     &                     &                     \\
\addlinespace
\textit{Female Share}$\times$ \textit{Surprise (References)}&    -0.00113\sym{***}&                     &                     &                     &   -0.000886\sym{***}&                     &                     &                     \\
                    &     (-4.45)         &                     &                     &                     &     (-3.59)         &                     &                     &                     \\
\addlinespace
DepartmentSize&   -2.40e-08\sym{***}&   -2.41e-08\sym{***}&   -2.51e-08\sym{***}&   -2.72e-08\sym{***}&   -2.01e-08\sym{***}&   -2.06e-08\sym{***}&   -2.22e-08\sym{***}&   -2.44e-08\sym{***}\\
                    &     (-8.04)         &     (-8.10)         &     (-8.32)         &     (-9.13)         &     (-7.20)         &     (-7.39)         &     (-7.85)         &     (-8.72)         \\
\addlinespace
CareerAge         & -0.00000769\sym{***}& -0.00000819\sym{***}& -0.00000728\sym{***}& -0.00000662\sym{***}& -0.00000628\sym{***}& -0.00000698\sym{***}& -0.00000550\sym{***}& -0.00000480\sym{***}\\
                    &     (-6.52)         &     (-6.92)         &     (-6.18)         &     (-5.60)         &     (-5.56)         &     (-6.16)         &     (-4.86)         &     (-4.24)         \\
\addlinespace
OpenAccess            &   -0.000606\sym{***}&   -0.000634\sym{***}&   -0.000666\sym{***}&   -0.000693\sym{***}&   -0.000552\sym{***}&   -0.000608\sym{***}&   -0.000658\sym{***}&   -0.000687\sym{***}\\
                    &    (-19.83)         &    (-20.62)         &    (-21.35)         &    (-22.43)         &    (-18.92)         &    (-20.76)         &    (-22.17)         &    (-23.30)         \\
\addlinespace
TeamSize        &   -0.000141\sym{***}&   -0.000142\sym{***}&   -0.000141\sym{***}&   -0.000144\sym{***}&   -0.000129\sym{***}&   -0.000129\sym{***}&   -0.000128\sym{***}&   -0.000131\sym{***}\\
                    &    (-13.97)         &    (-14.01)         &    (-13.89)         &    (-14.10)         &    (-13.99)         &    (-14.05)         &    (-13.84)         &    (-14.07)         \\
\addlinespace
$TeamSize^{2}$ &    1.12e-08\sym{***}&    1.12e-08\sym{***}&    1.16e-08\sym{***}&    1.21e-08\sym{***}&    9.39e-09\sym{***}&    9.56e-09\sym{***}&    1.03e-08\sym{***}&    1.07e-08\sym{***}\\
                    &      (7.66)         &      (7.70)         &      (7.88)         &      (8.18)         &      (7.02)         &      (7.16)         &      (7.54)         &      (7.86)         \\
\addlinespace
\textit{2-year Journal Impact Factor}    &   -0.000270\sym{***}&   -0.000273\sym{***}&   -0.000274\sym{***}&   -0.000277\sym{***}&   -0.000268\sym{***}&   -0.000273\sym{***}&   -0.000276\sym{***}&   -0.000279\sym{***}\\
                    &    (-12.95)         &    (-13.08)         &    (-12.93)         &    (-12.94)         &    (-15.13)         &    (-15.43)         &    (-15.12)         &    (-15.09)         \\
\addlinespace
\textit{two-year citations}           &  -0.0000156         &  -0.0000155         &  -0.0000159         &  -0.0000164         &  -0.0000108         &  -0.0000107         &  -0.0000114         &  -0.0000118         \\
                    &     (-1.83)         &     (-1.82)         &     (-1.85)         &     (-1.87)         &     (-1.50)         &     (-1.49)         &     (-1.55)         &     (-1.58)         \\
\addlinespace
\textit{Num Gender Unknown}         &    0.000115\sym{***}&    0.000116\sym{***}&    0.000114\sym{***}&    0.000115\sym{***}&    0.000107\sym{***}&    0.000107\sym{***}&    0.000105\sym{***}&    0.000106\sym{***}\\
                    &     (12.98)         &     (13.00)         &     (12.83)         &     (12.93)         &     (12.87)         &     (12.87)         &     (12.55)         &     (12.66)         \\
\addlinespace
\textit{Prescience (References)}  &                     &    -0.00133\sym{***}&                     &                     &                     &    -0.00226\sym{***}&                     &                     \\
                    &                     &    (-11.69)         &                     &                     &                     &    (-21.52)         &                     &                     \\
\addlinespace
\textit{Female Share}$\times$ \textit{Prescience (References)}&                     &   -0.000722\sym{***}&                     &                     &                     &   -0.000573\sym{**} &                     &                     \\
                    &                     &     (-3.60)         &                     &                     &                     &     (-2.95)         &                     &                     \\
\addlinespace
\textit{Surprise (concepts)}   &                     &                     &  -0.0000978         &                     &                     &                     &  -0.0000791         &                     \\
                    &                     &                     &     (-1.44)         &                     &                     &                     &     (-1.23)         &                     \\
\addlinespace
\textit{Female Share}$\times$ \textit{Surprise (concepts)}&                     &                     &  -0.0000564         &                     &                     &                     &  -0.0000346         &                     \\
                    &                     &                     &     (-0.45)         &                     &                     &                     &     (-0.28)         &                     \\
\addlinespace
\textit{Prescience (concepts)} &                     &                     &                     &     0.00142\sym{***}&                     &                     &                     &     0.00150\sym{***}\\
                    &                     &                     &                     &     (15.27)         &                     &                     &                     &     (17.61)         \\
\addlinespace
\textit{Female Share}$\times$ \textit{Prescience (concepts)}&                     &                     &                     &   -0.000206         &                     &                     &                     &   -0.000230         \\
                    &                     &                     &                     &     (-1.46)         &                     &                     &                     &     (-1.67)         \\
\addlinespace
Constant            &    -0.00626\sym{***}&    -0.00620\sym{***}&    -0.00684\sym{***}&    -0.00769\sym{***}&    -0.00501\sym{***}&    -0.00515\sym{***}&    -0.00630\sym{***}&    -0.00718\sym{***}\\
                    &    (-38.07)         &    (-39.57)         &    (-49.53)         &    (-52.49)         &    (-32.63)         &    (-35.33)         &    (-48.69)         &    (-52.64)         \\
\midrule
Observations        &     3864378         &     3864378         &     3864378         &     3864378         &     3864455         &     3864455         &     3864455         &     3864455         \\

\bottomrule
\multicolumn{9}{l}{\footnotesize \textit{t} statistics in parentheses}\\
\multicolumn{9}{l}{\footnotesize \sym{*} \(p<0.05\), \sym{**} \(p<0.01\), \sym{***} \(p<0.001\)}\\
\end{tabular}
\end{adjustbox}
\end{table}

\begin{table}[H]\centering
\def\sym#1{\ifmmode^{#1}\else\(^{#1}\)\fi}
\caption{Linear regression model estimates of log of journal impact factor on surprise on multi-authored papers, with White standard errors. Research field fixed effects and dummies for publishing year are omitted.}\label{tab3_jif}
\begin{adjustbox}{width=\textwidth}
\begin{tabular}{lcccc}
\toprule
                    &\multicolumn{1}{c}{(1)}&\multicolumn{1}{c}{(2)}&\multicolumn{1}{c}{(3)}&\multicolumn{1}{c}{(4)}\\
                    &\multicolumn{1}{c}{\textit{ln(2-year Journal Impact Factor)}}&\multicolumn{1}{c}{\textit{ln(2-year Journal Impact Factor)}}&\multicolumn{1}{c}{\textit{ln(5y Impact Factor)}}&\multicolumn{1}{c}{\textit{ln(5y Impact Factor)}}\\
\midrule
\textit{Female Share}       &     -0.0228\sym{***}&     -0.0476\sym{***}&     -0.0227\sym{***}&     -0.0408\sym{***}\\
                    &    (-10.86)         &    (-28.85)         &    (-10.75)         &    (-24.48)         \\
\addlinespace
\textit{Surprise (References)}    &       0.246\sym{***}&                     &       0.263\sym{***}&                     \\
                    &    (176.43)         &                     &    (186.42)         &                     \\
\addlinespace
\textit{Female Share}$\times$ \textit{Surprise (References)}&     -0.0408\sym{***}&                     &     -0.0341\sym{***}&                     \\
                    &    (-14.12)         &                     &    (-11.72)         &                     \\
\addlinespace
DepartmentSize&   0.0000117\sym{***}&   0.0000120\sym{***}&   0.0000120\sym{***}&   0.0000122\sym{***}\\
                    &    (166.48)         &    (169.24)         &    (170.81)         &    (173.70)         \\
\addlinespace
CareerAge         &     0.00101\sym{***}&    0.000950\sym{***}&     0.00105\sym{***}&    0.000989\sym{***}\\
                    &     (43.81)         &     (40.94)         &     (45.31)         &     (42.20)         \\
\addlinespace
OpenAccess            &       0.184\sym{***}&       0.193\sym{***}&       0.178\sym{***}&       0.187\sym{***}\\
                    &    (302.79)         &    (316.54)         &    (290.15)         &    (304.83)         \\
\addlinespace
TeamSize        &      0.0178\sym{***}&      0.0179\sym{***}&      0.0182\sym{***}&      0.0183\sym{***}\\
                    &     (39.46)         &     (39.60)         &     (39.44)         &     (39.58)         \\
\addlinespace
$TeamSize^{2}$ & -0.00000157\sym{***}& -0.00000165\sym{***}& -0.00000151\sym{***}& -0.00000160\sym{***}\\
                    &    (-31.31)         &    (-31.79)         &    (-31.15)         &    (-31.70)         \\
\addlinespace
\textit{Num Gender Unknown}         &     -0.0137\sym{***}&     -0.0136\sym{***}&     -0.0143\sym{***}&     -0.0142\sym{***}\\
                    &    (-28.24)         &    (-27.88)         &    (-28.90)         &    (-28.53)         \\
\addlinespace
\textit{Surprise (concepts)}   &                     &      0.0495\sym{***}&                     &      0.0501\sym{***}\\
                    &                     &     (40.85)         &                     &     (40.97)         \\
\addlinespace
\textit{Female Share}$\times$ \textit{Surprise (concepts)}&                     &     0.00387         &                     &    0.000319         \\
                    &                     &      (1.53)         &                     &      (0.13)         \\
\addlinespace
Constant            &       0.966\sym{***}&       1.075\sym{***}&       1.002\sym{***}&       1.119\sym{***}\\
                    &    (426.11)         &    (480.65)         &    (439.84)         &    (498.09)         \\
\midrule
Observations        &     3861357         &     3861357         &     3864495         &     3864495         \\
\bottomrule
\multicolumn{5}{l}{\footnotesize \textit{t} statistics in parentheses}\\
\multicolumn{5}{l}{\footnotesize \sym{*} \(p<0.05\), \sym{**} \(p<0.01\), \sym{***} \(p<0.001\)}\\
\end{tabular}
\end{adjustbox}
\end{table}

\section{Heterogeneity between solo and mixed multi-authored papers}\label{app:solo_vs_mixed_ma}


Tables \ref{app:tab2_solo_vs_ma}–\ref{app:tab4_solo_vs_ma} report estimates of how gender gaps in the rewards to innovation differ between solo-authored and multi-authored papers (including both mixed-gender and gender-homogeneous teams). We find that gender gaps in disruption and journal impact factor narrow more rapidly with increasing surprise or prescience for solo-authored papers -- particularly for work that combines references —- than for multi-authored papers.

\begin{table}[H]\centering
\def\sym#1{\ifmmode^{#1}\else\(^{#1}\)\fi}
\caption{Linear regression model estimates for content and context surprise and prescience scores comparing gender differences between solo- and multi-authored papers, with White robust standard errors. Research field fixed effects and dummies for publishing year are omitted.}\label{app:tab1_solo_vs_ma}
\begin{adjustbox}{width=\textwidth}
\begin{tabular}{lcccc}
\toprule
                    &\multicolumn{1}{c}{(1)}&\multicolumn{1}{c}{(2)}&\multicolumn{1}{c}{(3)}&\multicolumn{1}{c}{(4)}\\
                    &\multicolumn{1}{c}{Surprise (References)}&\multicolumn{1}{c}{Prescience (References)}&\multicolumn{1}{c}{Surprise (concepts)}&\multicolumn{1}{c}{Prescience (concepts)}\\
\midrule
\textit{Female Share}       &      0.0205\sym{***}&      0.0137\sym{***}&    -0.00634\sym{***}&    -0.00920\sym{***}\\
                    &     (52.63)         &     (32.96)         &    (-14.60)         &    (-22.30)         \\
\addlinespace
DummySolo=1         &     -0.0523\sym{***}&     -0.0475\sym{***}&     -0.0200\sym{***}&     -0.0267\sym{***}\\
                    &    (-67.10)         &    (-60.87)         &    (-26.27)         &    (-36.56)         \\
\addlinespace
DummySolo=1 $\times$ \textit{Female Share}&     0.00607\sym{***}&     0.00548\sym{***}&    -0.00182         &    -0.00485\sym{***}\\
                    &      (4.22)         &      (3.83)         &     (-1.29)         &     (-3.57)         \\
\addlinespace
DepartmentSize&  0.00000154\sym{***}&  0.00000120\sym{***}& 0.000000389\sym{***}&  0.00000194\sym{***}\\
                    &     (52.90)         &     (38.04)         &     (11.88)         &     (61.87)         \\
\addlinespace
CareerAge         &   -0.000261\sym{***}&   -0.000604\sym{***}&   -0.000164\sym{***}&   -0.000449\sym{***}\\
                    &    (-24.00)         &    (-52.56)         &    (-13.91)         &    (-39.91)         \\
\addlinespace
TeamSize        &     0.00146\sym{***}&    0.000716\sym{***}&     0.00105\sym{***}&     0.00263\sym{***}\\
                    &     (26.63)         &     (13.42)         &     (18.47)         &     (34.29)         \\
\addlinespace
$TeamSize^{2}$ &-0.000000436\sym{***}&-0.000000377\sym{***}&-0.000000167\sym{***}&-0.000000418\sym{***}\\
                    &    (-32.71)         &    (-28.70)         &    (-16.39)         &    (-28.13)         \\
\addlinespace
\textit{Num Gender Unknown}         &   -0.000339\sym{***}&    0.000235\sym{***}&   -0.000600\sym{***}&    -0.00152\sym{***}\\
                    &     (-5.19)         &      (3.69)         &     (-9.42)         &    (-17.32)         \\
\addlinespace
Constant            &       0.547\sym{***}&       0.544\sym{***}&       0.522\sym{***}&       0.578\sym{***}\\
                    &    (599.22)         &    (573.48)         &    (528.65)         &    (616.09)         \\
\midrule
Observations        &     4311751         &     4311751         &     4311751         &     4311751         \\
\bottomrule
\bottomrule
\multicolumn{5}{l}{\footnotesize \textit{t} statistics in parentheses}\\
\multicolumn{5}{l}{\footnotesize \sym{*} \(p<0.05\), \sym{**} \(p<0.01\), \sym{***} \(p<0.001\)}\\
\end{tabular}
\end{adjustbox}
\end{table}

\begin{table}[H]\centering
\def\sym#1{\ifmmode^{#1}\else\(^{#1}\)\fi}
\caption{Linear regression model estimates for log of journal impact factor on differential between solo and multi-authored papers in content and context surprise scores by gender, with White standard errors. Research field fixed effects and dummies for publishing year are omitted.}\label{app:tab2_solo_vs_ma}
\begin{adjustbox}{width=\textwidth}

\begin{tabular}{lcccc}
\toprule
                    &\multicolumn{1}{c}{(1)}&\multicolumn{1}{c}{(2)}&\multicolumn{1}{c}{(3)}&\multicolumn{1}{c}{(4)}\\
                    &\multicolumn{1}{c}{\textit{ln(2y Impact Factor)}}&\multicolumn{1}{c}{\textit{ln(2y Impact Factor)}}&\multicolumn{1}{c}{\textit{ln(5y Impact Factor)}}&\multicolumn{1}{c}{\textit{ln(5y Impact Factor)}}\\
\midrule
\textit{Female Share}       &     -0.0239\sym{***}&     -0.0471\sym{***}&     -0.0238\sym{***}&     -0.0402\sym{***}\\
                    &    (-11.39)         &    (-28.58)         &    (-11.25)         &    (-24.17)         \\
\addlinespace
DummySolo=1         &      0.0944\sym{***}&     -0.0107\sym{**} &      0.0954\sym{***}&     -0.0115\sym{***}\\
                    &     (26.10)         &     (-3.25)         &     (26.19)         &     (-3.45)         \\
\addlinespace
DummySolo=1 $\times$ \textit{Female Share}&     -0.0467\sym{***}&     -0.0280\sym{***}&     -0.0491\sym{***}&     -0.0282\sym{***}\\
                    &     (-6.57)         &     (-4.70)         &     (-6.86)         &     (-4.66)         \\
\addlinespace
\textit{Surprise (References)}    &       0.245\sym{***}&                     &       0.262\sym{***}&                     \\
                    &    (175.51)         &                     &    (185.79)         &                     \\
\addlinespace
\textit{Female Share}$\times$ \textit{Surprise (References)}&     -0.0392\sym{***}&                     &     -0.0325\sym{***}&                     \\
                    &    (-13.58)         &                     &    (-11.18)         &                     \\
\addlinespace
DummySolo=1 $\times$ \textit{Surprise (References)}&      -0.228\sym{***}&                     &      -0.235\sym{***}&                     \\
                    &    (-43.75)         &                     &    (-44.66)         &                     \\
\addlinespace
DummySolo=1 $\times$ \textit{Female Share}$\times$ \textit{Surprise (References)}&      0.0822\sym{***}&                     &      0.0903\sym{***}&                     \\
                    &      (8.21)         &                     &      (8.92)         &                     \\
\addlinespace
DepartmentSize&   0.0000118\sym{***}&   0.0000121\sym{***}&   0.0000121\sym{***}&   0.0000123\sym{***}\\
                    &    (169.27)         &    (171.91)         &    (173.42)         &    (176.21)         \\
\addlinespace
CareerAge         &     0.00118\sym{***}&     0.00112\sym{***}&     0.00120\sym{***}&     0.00114\sym{***}\\
                    &     (52.94)         &     (50.06)         &     (53.66)         &     (50.49)         \\
\addlinespace
OpenAccess            &       0.182\sym{***}&       0.190\sym{***}&       0.175\sym{***}&       0.185\sym{***}\\
                    &    (305.72)         &    (319.17)         &    (292.41)         &    (306.77)         \\
\addlinespace
TeamSize        &      0.0175\sym{***}&      0.0176\sym{***}&      0.0179\sym{***}&      0.0180\sym{***}\\
                    &     (39.57)         &     (39.71)         &     (39.55)         &     (39.69)         \\
\addlinespace
$TeamSize^{2}$ & -0.00000154\sym{***}& -0.00000162\sym{***}& -0.00000148\sym{***}& -0.00000157\sym{***}\\
                    &    (-31.29)         &    (-31.77)         &    (-31.11)         &    (-31.66)         \\
\addlinespace
\textit{Num Gender Unknown}         &     -0.0135\sym{***}&     -0.0133\sym{***}&     -0.0141\sym{***}&     -0.0140\sym{***}\\
                    &    (-28.29)         &    (-27.93)         &    (-28.99)         &    (-28.60)         \\
\addlinespace
\textit{Surprise (concepts)}   &                     &      0.0489\sym{***}&                     &      0.0495\sym{***}\\
                    &                     &     (40.38)         &                     &     (40.56)         \\
\addlinespace
\textit{Female Share}$\times$ \textit{Surprise (concepts)}&                     &     0.00323         &                     &   -0.000333         \\
                    &                     &      (1.28)         &                     &     (-0.13)         \\
\addlinespace
DummySolo=1 $\times$ \textit{Surprise (concepts)}&                     &     -0.0747\sym{***}&                     &     -0.0811\sym{***}\\
                    &                     &    (-14.40)         &                     &    (-15.44)         \\
\addlinespace
DummySolo=1 $\times$ \textit{Female Share}$\times$ \textit{Surprise (concepts)}&                     &      0.0523\sym{***}&                     &      0.0564\sym{***}\\
                    &                     &      (5.38)         &                     &      (5.72)         \\
\addlinespace
Constant            &       0.955\sym{***}&       1.064\sym{***}&       0.992\sym{***}&       1.110\sym{***}\\
                    &    (430.81)         &    (487.76)         &    (445.04)         &    (505.92)         \\
\midrule
Observations        &     4073759         &     4073759         &     4077428         &     4077428         \\

\bottomrule
\multicolumn{5}{l}{\footnotesize \textit{t} statistics in parentheses}\\
\multicolumn{5}{l}{\footnotesize \sym{*} \(p<0.05\), \sym{**} \(p<0.01\), \sym{***} \(p<0.001\)}\\
\end{tabular}
\end{adjustbox}
\end{table}

 \begin{table}[htbp]\centering
\def\sym#1{\ifmmode^{#1}\else\(^{#1}\)\fi}
\caption{Linear regression model estimates for 2-steps out credit on differential between solo and multi-authored papers in content and context surprise and prescience scores by gender. We include White standard errors. Research field fixed effects and dummies for publishing year are omitted.}\label{app:tab3_solo_vs_ma}
\begin{adjustbox}{width=\textwidth}
\begin{tabular}{lcccc}
\toprule
 &\multicolumn{1}{c}{(1)}&\multicolumn{1}{c}{(2)}&\multicolumn{1}{c}{(3)}&\multicolumn{1}{c}{(4)}\\
                    &\multicolumn{1}{c}{TwoStepsCreditScore}&\multicolumn{1}{c}{TwoStepsCreditScore}&\multicolumn{1}{c}{TwoStepsCreditScore}&\multicolumn{1}{c}{TwoStepsCreditScore}\\
\midrule

\textit{Female Share}       &    -0.00155\sym{***}&    -0.00154\sym{***}&   -0.000313         &   -0.000797\sym{***}\\
                    &     (-6.10)         &     (-7.76)         &     (-1.94)         &     (-4.18)         \\
\addlinespace
DummySolo=1         &    0.000336         &     0.00160\sym{**} &    0.000228         &     0.00103\sym{*}  \\
                    &      (0.61)         &      (3.24)         &      (0.65)         &      (2.43)         \\
\addlinespace
DummySolo=1 $\times$ \textit{Female Share}&   0.0000495         &     0.00139         &   -0.000922         &    -0.00158\sym{*}  \\
                    &      (0.04)         &      (1.30)         &     (-1.43)         &     (-2.08)         \\
\addlinespace
\textit{Surprise (References)}    &     -0.0140\sym{***}&                     &                     &                     \\
                    &    (-98.65)         &                     &                     &                     \\
\addlinespace
\textit{Female Share}$\times$ \textit{Surprise (References)}&     0.00214\sym{***}&                     &                     &                     \\
                    &      (6.70)         &                     &                     &                     \\
\addlinespace
DummySolo=1 $\times$ \textit{Surprise (References)}&   0.0000868         &                     &                     &                     \\
                    &      (0.13)         &                     &                     &                     \\
\addlinespace
DummySolo=1 $\times$ \textit{Female Share}$\times$ \textit{Surprise (References)}&    -0.00105         &                     &                     &                     \\
                    &     (-0.75)         &                     &                     &                     \\
\addlinespace
DepartmentSize&-0.000000431\sym{***}&-0.000000436\sym{***}&-0.000000435\sym{***}&-0.000000416\sym{***}\\
                    &    (-55.38)         &    (-56.06)         &    (-55.90)         &    (-53.76)         \\
\addlinespace
CareerAge         &  -0.0000223\sym{***}&  -0.0000253\sym{***}&  -0.0000192\sym{***}&  -0.0000245\sym{***}\\
                    &    (-10.69)         &    (-12.11)         &     (-9.15)         &    (-11.74)         \\
\addlinespace
OpenAccess            &    -0.00243\sym{***}&    -0.00280\sym{***}&    -0.00287\sym{***}&    -0.00270\sym{***}\\
                    &    (-48.82)         &    (-56.18)         &    (-57.46)         &    (-54.34)         \\
\addlinespace
TeamSize        &   -0.000276\sym{***}&   -0.000273\sym{***}&   -0.000258\sym{***}&   -0.000235\sym{***}\\
                    &    (-22.28)         &    (-22.17)         &    (-21.83)         &    (-20.83)         \\
\addlinespace
$TeamSize^{2}$ &    1.69e-08\sym{***}&    1.76e-08\sym{***}&    1.88e-08\sym{***}&    1.57e-08\sym{***}\\
                    &     (11.43)         &     (11.86)         &     (12.65)         &     (10.89)         \\
\addlinespace
\textit{Num Gender Unknown}         &    0.000227\sym{***}&    0.000223\sym{***}&    0.000205\sym{***}&    0.000190\sym{***}\\
                    &     (16.96)         &     (16.70)         &     (15.97)         &     (15.53)         \\
\addlinespace
\textit{Prescience (References)}  &                     &     -0.0117\sym{***}&                     &                     \\
                    &                     &    (-98.21)         &                     &                     \\
\addlinespace
\textit{Female Share}$\times$ \textit{Prescience (References)}&                     &     0.00218\sym{***}&                     &                     \\
                    &                     &      (8.17)         &                     &                     \\
\addlinespace
DummySolo=1 $\times$ \textit{Prescience (References)}&                     &    -0.00181\sym{**} &                     &                     \\
                    &                     &     (-2.84)         &                     &                     \\
\addlinespace
DummySolo=1 $\times$ \textit{Female Share}$\times$ \textit{Prescience (References)}&                     &    -0.00306\sym{*}  &                     &                     \\
                    &                     &     (-2.30)         &                     &                     \\
\addlinespace
\textit{Surprise (concepts)}   &                     &                     &    -0.00268\sym{***}&                     \\
                    &                     &                     &    (-25.11)         &                     \\
\addlinespace
\textit{Female Share}$\times$ \textit{Surprise (concepts)}&                     &                     & -0.00000786         &                     \\
                    &                     &                     &     (-0.03)         &                     \\
\addlinespace
DummySolo=1 $\times$ \textit{Surprise (concepts)}&                     &                     &    0.000394         &                     \\
                    &                     &                     &      (0.72)         &                     \\
\addlinespace
DummySolo=1 $\times$ \textit{Female Share}$\times$ \textit{Surprise (concepts)}&                     &                     &    0.000126         &                     \\
                    &                     &                     &      (0.12)         &                     \\
\addlinespace
\textit{Prescience (concepts)} &                     &                     &                     &     -0.0121\sym{***}\\
                    &                     &                     &                     &   (-103.90)         \\
\addlinespace
\textit{Female Share}$\times$ \textit{Prescience (concepts)}&                     &                     &                     &    0.000624\sym{*}  \\
                    &                     &                     &                     &      (2.40)         \\
\addlinespace
DummySolo=1 $\times$ \textit{Prescience (concepts)}&                     &                     &                     &    -0.00143\sym{*}  \\
                    &                     &                     &                     &     (-2.37)         \\
\addlinespace
DummySolo=1 $\times$ \textit{Female Share}$\times$ \textit{Prescience (concepts)}&                     &                     &                     &     0.00118         \\
                    &                     &                     &                     &      (1.07)         \\
\addlinespace
Constant            &      0.0355\sym{***}&      0.0341\sym{***}&      0.0288\sym{***}&      0.0346\sym{***}\\
                    &    (241.26)         &    (251.70)         &    (226.48)         &    (257.62)         \\
\midrule
Observations        &     2547542         &     2547542         &     2547542         &     2547542         \\
\multicolumn{5}{l}{\footnotesize \textit{t} statistics in parentheses}\\
\multicolumn{5}{l}{\footnotesize \sym{*} \(p<0.05\), \sym{**} \(p<0.01\), \sym{***} \(p<0.001\)}\\
\end{tabular}
\end{adjustbox}
\end{table}

\begin{table}[H]\centering
\def\sym#1{\ifmmode^{#1}\else\(^{#1}\)\fi}
\caption{Linear regression model estimates for five-year disruption score of publication on for differential between solo and multi-authored papers in content and context surprise and prescience scores by gender, with White standard errors. Research field fixed effects and dummies for publishing year are omitted.}\label{app:tab4_solo_vs_ma}
\begin{adjustbox}{width=0.8\textwidth}
\begin{tabular}{lcccc}
\toprule
                    &\multicolumn{1}{c}{(1)}&\multicolumn{1}{c}{(2)}&\multicolumn{1}{c}{(3)}&\multicolumn{1}{c}{(4)}\\
                    &\multicolumn{1}{c}{5y Disruption}&\multicolumn{1}{c}{5y Disruption}&\multicolumn{1}{c}{5y Disruption}&\multicolumn{1}{c}{5y Disruption}\\
\midrule
\textit{Female Share}       &     0.00105\sym{***}&    0.000791\sym{***}&    0.000441\sym{***}&    0.000552\sym{***}\\
                    &      (5.46)         &      (5.46)         &      (5.34)         &      (5.74)         \\
\addlinespace
DummySolo=1         &     0.00821\sym{***}&     0.00739\sym{***}&     0.00273\sym{***}&     0.00284\sym{***}\\
                    &     (23.96)         &     (24.23)         &     (13.55)         &     (12.72)         \\
\addlinespace
DummySolo=1 $\times$ \textit{Female Share}&     0.00520\sym{***}&     0.00486\sym{***}&     0.00114\sym{**} &    0.000912\sym{*}  \\
                    &      (6.98)         &      (7.12)         &      (2.84)         &      (2.13)         \\
\addlinespace
\textit{Surprise (References)}    &    -0.00257\sym{***}&                     &                     &                     \\
                    &    (-20.58)         &                     &                     &                     \\
\addlinespace
\textit{Female Share}$\times$ \textit{Surprise (References)}&   -0.000887\sym{***}&                     &                     &                     \\
                    &     (-3.60)         &                     &                     &                     \\
\addlinespace
DummySolo=1 $\times$ \textit{Surprise (References)}&    -0.00982\sym{***}&                     &                     &                     \\
                    &    (-22.19)         &                     &                     &                     \\
\addlinespace
DummySolo=1 $\times$ \textit{Female Share}$\times$ \textit{Surprise (References)}&    -0.00708\sym{***}&                     &                     &                     \\
                    &     (-7.52)         &                     &                     &                     \\
\addlinespace
DepartmentSize&   -1.70e-08\sym{***}&   -1.71e-08\sym{***}&   -2.02e-08\sym{***}&   -2.23e-08\sym{***}\\
                    &     (-6.09)         &     (-6.13)         &     (-7.11)         &     (-7.95)         \\
\addlinespace
CareerAge         & -0.00000307\sym{**} & -0.00000354\sym{**} &-0.000000793         &-0.000000108         \\
                    &     (-2.69)         &     (-3.10)         &     (-0.69)         &     (-0.09)         \\
\addlinespace
OpenAccess            &   -0.000541\sym{***}&   -0.000590\sym{***}&   -0.000640\sym{***}&   -0.000667\sym{***}\\
                    &    (-18.55)         &    (-20.16)         &    (-21.53)         &    (-22.62)         \\
\addlinespace
TeamSize        &   -0.000127\sym{***}&   -0.000129\sym{***}&   -0.000129\sym{***}&   -0.000131\sym{***}\\
                    &    (-13.89)         &    (-14.00)         &    (-13.76)         &    (-14.00)         \\
\addlinespace
$TeamSize^{2}$ &    9.13e-09\sym{***}&    9.30e-09\sym{***}&    1.02e-08\sym{***}&    1.07e-08\sym{***}\\
                    &      (6.84)         &      (6.97)         &      (7.44)         &      (7.76)         \\
\addlinespace
\textit{Num Gender Unknown}         &    0.000107\sym{***}&    0.000107\sym{***}&    0.000105\sym{***}&    0.000107\sym{***}\\
                    &     (12.85)         &     (12.91)         &     (12.56)         &     (12.67)         \\
\addlinespace
two-year Journal Impact Factor   &   -0.000263\sym{***}&   -0.000267\sym{***}&   -0.000266\sym{***}&   -0.000269\sym{***}\\
                    &    (-15.42)         &    (-15.63)         &    (-14.92)         &    (-14.89)         \\
\addlinespace
2y\_cites            &  -0.0000109         &  -0.0000109         &  -0.0000121         &  -0.0000125         \\
                    &     (-1.53)         &     (-1.52)         &     (-1.62)         &     (-1.65)         \\
\addlinespace
\textit{Prescience (References)}  &                     &    -0.00230\sym{***}&                     &                     \\
                    &                     &    (-21.88)         &                     &                     \\
\addlinespace
\textit{Female Share}$\times$ \textit{Prescience (References)}&                     &   -0.000573\sym{**} &                     &                     \\
                    &                     &     (-2.95)         &                     &                     \\
\addlinespace
DummySolo=1 $\times$ \textit{Prescience (References)}&                     &    -0.00895\sym{***}&                     &                     \\
                    &                     &    (-21.89)         &                     &                     \\
\addlinespace
DummySolo=1 $\times$ \textit{Female Share}$\times$ \textit{Prescience (References)}&                     &    -0.00707\sym{***}&                     &                     \\
                    &                     &     (-7.82)         &                     &                     \\
\addlinespace
\textit{Surprise (concepts)}   &                     &                     &  -0.0000748         &                     \\
                    &                     &                     &     (-1.16)         &                     \\
\addlinespace
\textit{Female Share}$\times$ \textit{Surprise (concepts)}&                     &                     &  -0.0000252         &                     \\
                    &                     &                     &     (-0.21)         &                     \\
\addlinespace
DummySolo=1 $\times$ \textit{Surprise (concepts)}&                     &                     &   -0.000144         &                     \\
                    &                     &                     &     (-0.45)         &                     \\
\addlinespace
DummySolo=1 $\times$ \textit{Female Share}$\times$ \textit{Surprise (concepts)}&                     &                     &    -0.00127\sym{*}  &                     \\
                    &                     &                     &     (-2.01)         &                     \\
\addlinespace
\textit{Prescience (concepts)} &                     &                     &                     &     0.00149\sym{***}\\
                    &                     &                     &                     &     (17.35)         \\
\addlinespace
\textit{Female Share}$\times$ \textit{Prescience (concepts)}&                     &                     &                     &   -0.000189         \\
                    &                     &                     &                     &     (-1.37)         \\
\addlinespace
DummySolo=1 $\times$ \textit{Prescience (concepts)}&                     &                     &                     &   -0.000279         \\
                    &                     &                     &                     &     (-0.82)         \\
\addlinespace
DummySolo=1 $\times$ \textit{Female Share}$\times$ \textit{Prescience (concepts)}&                     &                     &                     &   -0.000843         \\
                    &                     &                     &                     &     (-1.24)         \\
\addlinespace
Constant            &    -0.00482\sym{***}&    -0.00491\sym{***}&    -0.00613\sym{***}&    -0.00701\sym{***}\\
                    &    (-31.72)         &    (-34.07)         &    (-48.02)         &    (-51.95)         \\
\midrule
Observations        &     4077371         &     4077371         &     4077371         &     4077371         \\
\bottomrule
\multicolumn{5}{l}{\footnotesize \textit{t} statistics in parentheses}\\
\multicolumn{5}{l}{\footnotesize \sym{*} \(p<0.05\), \sym{**} \(p<0.01\), \sym{***} \(p<0.001\)}\\
\end{tabular}
\end{adjustbox}
\end{table}

\section{Three-year disruption in solo-authored papers}\label{app:3ycd}

Table \ref{tab1_cd} reports the coefficient estimates of model in eq. (2) of the main file on disruption in solo-authored papers, changing the reference period from five-year (as in the main analysis) to the three- years disruption score -- measuring how well papers are able to distance themselves from their literature three-years after publication. We see that the results are in line with the model estimates of the main model in Section C.

\begin{table}[H]\centering
\def\sym#1{\ifmmode^{#1}\else\(^{#1}\)\fi}
\caption{Linear regression model estimates for three-years disruption scores of publication on surprise and prescience scores by gender in solo-authored papers, with  author-level clustered standard errors. Research field dummies and year of publication are omitted.}\label{tab1_cd}
\begin{adjustbox}{width=\textwidth}
\begin{tabular}{l*{4}{D{.}{.}{-1}}}
\toprule
                    &\multicolumn{1}{c}{(1)}&\multicolumn{1}{c}{(2)}&\multicolumn{1}{c}{(3)}&\multicolumn{1}{c}{(4)}\\
                    &\multicolumn{1}{c}{3-years Disruption}&\multicolumn{1}{c}{3-years Disruption}&\multicolumn{1}{c}{3-years Disruption}&\multicolumn{1}{c}{3-years Disruption}\\
\midrule
Female=1            &     0.00595\sym{***}&     0.00181\sym{***}&     0.00552\sym{***}&     0.00155\sym{***}\\
                    &      (7.67)         &      (4.41)         &      (7.69)         &      (3.38)         \\
\addlinespace
\textit{Surprise (References)}    &     -0.0109\sym{***}&                     &                     &                     \\
                    &    (-21.39)         &                     &                     &                     \\
\addlinespace
Female=1 $\times$ \textit{Surprise (References)}&    -0.00739\sym{***}&                     &                     &                     \\
                    &     (-7.61)         &                     &                     &                     \\
\addlinespace
\textit{DepartmentSize}&   -4.79e-08\sym{*}  &   -3.09e-08         &   -3.69e-08         &   -3.25e-08         \\
                    &     (-1.97)         &     (-1.27)         &     (-1.52)         &     (-1.33)         \\
\addlinespace
CareerAge         &   0.0000264\sym{***}&   0.0000429\sym{***}&   0.0000280\sym{***}&   0.0000434\sym{***}\\
                    &      (4.16)         &      (6.66)         &      (4.40)         &      (6.73)         \\
\addlinespace
OpenAccess            &   -0.000640\sym{***}&   -0.000520\sym{**} &   -0.000612\sym{***}&   -0.000527\sym{**} \\
                    &     (-3.62)         &     (-2.93)         &     (-3.46)         &     (-2.98)         \\
\addlinespace
\addlinespace
impact\_factor\_2y    &   -0.000158\sym{***}&  -0.0000657\sym{*}  &   -0.000144\sym{***}&  -0.0000660\sym{*}  \\
                    &     (-5.35)         &     (-2.25)         &     (-4.92)         &     (-2.27)         \\
\addlinespace
2y\_cites            &  -0.0000303\sym{*}  &  -0.0000702\sym{***}&  -0.0000351\sym{*}  &  -0.0000717\sym{***}\\
                    &     (-2.04)         &     (-4.71)         &     (-2.38)         &     (-4.79)         \\
\addlinespace
\textit{Surprise (Concepts)}   &                     &    0.000145         &                     &                     \\
                    &                     &      (0.43)         &                     &                     \\
\addlinespace
Female=1 $\times$ \textit{Surprise (Concepts)}&                     &    -0.00133\sym{*}  &                     &                     \\
                    &                     &     (-2.05)         &                     &                     \\
\addlinespace
\textit{Prescience (References)}  &                     &                     &     -0.0104\sym{***}&                     \\
                    &                     &                     &    (-21.91)         &                     \\
\addlinespace
Female=1 $\times$ \textit{Prescience (References)}&                     &                     &    -0.00722\sym{***}&                     \\
                    &                     &                     &     (-7.67)         &                     \\
\addlinespace
\textit{Prescience (Concepts)} &                     &                     &                     &     0.00135\sym{***}\\
                    &                     &                     &                     &      (3.69)         \\
\addlinespace
Female=1 $\times$ \textit{Prescience (Concepts)}&                     &                     &                     &   -0.000800         \\
                    &                     &                     &                     &     (-1.12)         \\
\addlinespace
Constant            &     0.00377\sym{***}&    -0.00233\sym{***}&     0.00362\sym{***}&    -0.00296\sym{***}\\
                    &      (5.62)         &     (-3.91)         &      (5.46)         &     (-4.87)         \\
\midrule
Observations        &      212883         &      212883         &      212883         &      212883         \\
\bottomrule
\multicolumn{5}{l}{\footnotesize \textit{t} statistics in parentheses}\\
\multicolumn{5}{l}{\footnotesize \sym{*} \(p<0.05\), \sym{**} \(p<0.01\), \sym{***} \(p<0.001\)}\\
\end{tabular}
\end{adjustbox}
\end{table}

\section{Five-year Journal Impact Factor in solo-authored papers}

We report regression estimates from Eq. (2) in the main text, using the natural logarithm of the five-year journal impact factor as the outcome instead of the two-year measure in Table \ref{tab_5yjif}. Results are in line with the model estimates of the main model in Section C. 

\begin{table}[H]\centering
\def\sym#1{\ifmmode^{#1}\else\(^{#1}\)\fi}
\caption{Linear regression model estimates of log of five-year journal impact factor on solo-authored papers' surprise score, with  author-level clustered standard errors. We control for career age, average number of citations from institution of author in year of publication and in same field of paper (Department Size), publication year with 2020 as baseline year (omitted), open access, and dummy variables for field of publication (omitted).}\label{tab_5yjif}
\begin{adjustbox}{width=0.8\textwidth}
\begin{tabular}{lcc}
\toprule
                    &\multicolumn{1}{c}{\textit{ln(2-year Journal Impact Factor)}}&\multicolumn{1}{c}{\textit{ln(5y Impact Factor)}}\\
\midrule
Female=1            &     -0.0685\sym{***}&     -0.0592\sym{***}\\
                    &     (-7.31)         &     (-8.98)         \\
\addlinespace
\textit{Surprise (References)}    &      0.0183\sym{**} &                     \\
                    &      (3.03)         &                     \\
\addlinespace
Female=1 $\times$ \textit{Surprise (References)}&      0.0557\sym{***}&                     \\
                    &      (4.51)         &                     \\
\addlinespace
\textit{DepartmentSize}&   0.0000260\sym{***}&   0.0000259\sym{***}\\
                    &     (33.20)         &     (33.27)         \\
\addlinespace
CareerAge         &     0.00211\sym{***}&     0.00208\sym{***}\\
                    &     (19.66)         &     (19.33)         \\
\addlinespace
OpenAccess            &       0.138\sym{***}&       0.138\sym{***}\\
                    &     (41.75)         &     (41.67)         \\
\addlinespace
\textit{Surprise (Concepts)}   &                     &     -0.0353\sym{***}\\
                    &                     &     (-6.53)         \\
\addlinespace
Female=1 $\times$ \textit{Surprise (Concepts)}&                     &      0.0478\sym{***}\\
                    &                     &      (4.86)         \\
\addlinespace
Constant            &       0.962\sym{***}&       0.990\sym{***}\\
                    &    (114.68)         &    (123.02)         \\
\midrule
Observations        &      212933         &      212933         \\
\bottomrule
\multicolumn{3}{l}{\footnotesize \textit{t} statistics in parentheses}\\
\multicolumn{3}{l}{\footnotesize \sym{*} \(p<0.05\), \sym{**} \(p<0.01\), \sym{***} \(p<0.001\)}\\
\end{tabular}
\end{adjustbox}
\end{table}

\section{Robustness Check: Returns to Novelty and Prescience Controlling for Grants, and Prior Citations}\label{app3:reg2_newgrants_pc}

We re-estimate the baseline regression models on solo-authored papers from the main text, while conditioning on author visibility and prestige, proxied by (i) the number of grants acknowledged in a paper and (ii) authors’ prior citations. Variation in the number of grants or in authors’ past citations does not account for the average gender differences we observe. 
Estimates of the baseline gender coefficient and of the returns to novelty and prescience remain consistent with those reported in the main model (Section C). This indicates that our results are not solely capturing compositional differences in prestige or status. 


\begin{table}[H]\centering
\def\sym#1{\ifmmode^{#1}\else\(^{#1}\)\fi}
\caption{Linear regression model estimates for two-step citation score on solo-authored papers in content and context surprise and prescience scores by gender, controlling for number of awarded grants listed in a paper. We compute author-level clustered standard errors. Research field fixed effects and dummies for publishing year are omitted.}
\begin{adjustbox}{width=\textwidth}
\begin{tabular}{l*{4}{D{.}{.}{-1}}}
\toprule
 &\multicolumn{1}{c}{(1)}&\multicolumn{1}{c}{(2)}&\multicolumn{1}{c}{(3)}&\multicolumn{1}{c}{(4)}\\
                 &\multicolumn{1}{c}{TwoStepsCreditScore}&\multicolumn{1}{c}{TwoStepsCreditScore}&\multicolumn{1}{c}{TwoStepsCreditScore}&\multicolumn{1}{c}{TwoStepsCreditScore}\\
\midrule
Female=1            &    -0.00148         &   -0.000264         &    -0.00143\sym{*}  &    -0.00241\sym{**} \\
                    &     (-1.26)         &     (-0.24)         &     (-2.20)         &     (-3.13)         \\
\addlinespace
\textit{Surprise (References)}    &     -0.0120\sym{***}&                     &                     &                     \\
                    &    (-16.70)         &                     &                     &                     \\
\addlinespace
Female=1 $\times$ \textit{Surprise (References)}&    0.000991         &                     &                     &                     \\
                    &      (0.70)         &                     &                     &                     \\
\end{tabular}
\end{adjustbox}
\end{table}

\begin{table}[H]\centering
\def\sym#1{\ifmmode^{#1}\else\(^{#1}\)\fi}
\caption{Linear regression model estimates for five-year disruption score on solo-authored papers in content and context surprise and prescience scores by gender, controlling for number of awarded grants listed in a paper. We computte author-level clustered standard errors. Research field fixed effects and dummies for publishing year are omitted.}
\begin{adjustbox}{width=\textwidth}
\begin{tabular}{l*{4}{D{.}{.}{-1}}}
\toprule
 &\multicolumn{1}{c}{(1)}&\multicolumn{1}{c}{(2)}&\multicolumn{1}{c}{(3)}&\multicolumn{1}{c}{(4)}\\
                 &\multicolumn{1}{c}{5y Disruption}&\multicolumn{1}{c}{5y Disruption}&\multicolumn{1}{c}{5y Disruption}&\multicolumn{1}{c}{5y Disruption}\\
\midrule
Female=1            &     0.00618\sym{***}&     0.00568\sym{***}&     0.00179\sym{***}&     0.00143\sym{**} \\
                    &      (8.35)         &      (8.31)         &      (4.49)         &      (3.28)         \\
\addlinespace
\textit{Surprise (References)}    &     -0.0127\sym{***}&                     &                     &                     \\
                    &    (-25.60)         &                     &                     &                     \\
\addlinespace
Female=1 $\times$ \textit{Surprise (References)}&    -0.00778\sym{***}&                     &                     &                     \\
                    &     (-8.41)         &                     &                     &                     \\
\end{tabular}
\end{adjustbox}
\end{table}

\begin{table}[H]\centering
\def\sym#1{\ifmmode^{#1}\else\(^{#1}\)\fi}
\caption{Linear regression model estimates for log of two-year journal impact factor on solo-authored papers in content and context surprise and prescience scores by gender,  controlling for number of awarded grants listed in a paper. We compute author-level clustered standard errors. Research field fixed effects and dummies for publishing year are omitted.}\label{tab2_cred}
\begin{adjustbox}{width=0.8\textwidth}
\begin{tabular}{l*{2}{D{.}{.}{-1}}}
\toprule
 &\multicolumn{1}{c}{(1)}&\multicolumn{1}{c}{(2)}\\
                 &\multicolumn{1}{c}{\textit{ln(2-year Journal Impact Factor)}}&\multicolumn{1}{c}{\textit{ln(2-year Journal Impact Factor)}}\\
\midrule
Female=1            &     -0.0665\sym{***}&     -0.0659\sym{***}\\
                    &     (-7.25)         &    (-10.20)         \\
\addlinespace
\textit{Surprise (References)}    &     0.00149         &                     \\
                    &      (0.25)         &                     \\
\addlinespace
Female=1 $\times$ \textit{Surprise (References)}&      0.0407\sym{***}&                     \\
                    &      (3.38)         &                     \\
\end{tabular}
\end{adjustbox}
\end{table}

\begin{table}[H]\centering
\def\sym#1{\ifmmode^{#1}\else\(^{#1}\)\fi}
\caption{Linear regression model estimates for two-step citation score on solo-authored papers in content and context surprise and prescience scores by gender, controlling for past citations of authors. We compute author-level clustered standard errors. Research field fixed effects and dummies for publishing year are omitted.}
\begin{adjustbox}{width=.8\textwidth}
\begin{tabular}{l*{4}{D{.}{.}{-1}}}
\toprule
                    &\multicolumn{1}{c}{(1)}&\multicolumn{1}{c}{(2)}&\multicolumn{1}{c}{(3)}&\multicolumn{1}{c}{(4)}\\
                    &\multicolumn{1}{c}{TwoStepsCreditScore}&\multicolumn{1}{c}{TwoStepsCreditScore}&\multicolumn{1}{c}{TwoStepsCreditScore}&\multicolumn{1}{c}{TwoStepsCreditScore}\\
\midrule
Female=1            &    -0.00200         &    -0.00110         &    -0.00200\sym{**} &    -0.00264\sym{***}\\
                    &     (-1.70)         &     (-1.00)         &     (-3.07)         &     (-3.43)         \\
\addlinespace
\textit{Surprise (References)}    &     -0.0124\sym{***}&                     &                     &                     \\
                    &    (-17.33)         &                     &                     &                     \\
\addlinespace
Female=1            &    -0.00200         &    -0.00110         &    -0.00200\sym{**} &    -0.00264\sym{***}\\
                    &     (-1.70)         &     (-1.00)         &     (-3.07)         &     (-3.43)         \\
\end{tabular}
\end{adjustbox}
\end{table}

\begin{table}[H]\centering
\def\sym#1{\ifmmode^{#1}\else\(^{#1}\)\fi}
\caption{Linear regression model estimates for five-year disruption score on solo-authored papers in content and context surprise and prescience scores by gender, controlling for past citations of authors. We compute author-level clustered standard errors. Research field fixed effects and dummies for publishing year are omitted.}
\begin{adjustbox}{width=.8\textwidth}
\begin{tabular}{l*{4}{D{.}{.}{-1}}}
\toprule
                    &\multicolumn{1}{c}{(1)}&\multicolumn{1}{c}{(2)}&\multicolumn{1}{c}{(3)}&\multicolumn{1}{c}{(4)}\\
                    &\multicolumn{1}{c}{5-year Disruption}&\multicolumn{1}{c}{5-year Disruption}&\multicolumn{1}{c}{5-year Disruption}&\multicolumn{1}{c}{5-year Disruption}\\
\midrule
Female=1            &     0.00619\sym{***}&     0.00570\sym{***}&     0.00182\sym{***}&     0.00144\sym{***}\\
                    &      (8.37)         &      (8.33)         &      (4.55)         &      (3.30)         \\
\addlinespace
\textit{Surprise (References)}    &     -0.0127\sym{***}&                     &                     &                     \\
                    &    (-25.48)         &                     &                     &                     \\
\addlinespace
Female=1 $\times$ \textit{Surprise (References)}&    -0.00779\sym{***}&                     &                     &                     \\
                    &     (-8.42)         &                     &                     &                     \\
\end{tabular}
\end{adjustbox}
\end{table}

\begin{table}[H]\centering
\def\sym#1{\ifmmode^{#1}\else\(^{#1}\)\fi}
\caption{Linear regression model estimates for two-year journal impact factor on solo-authored papers in content and context surprise scores by gender, controlling for past citations of authors. We compute author-level clustered standard errors. Research field fixed effects and dummies for publishing year are omitted.}
\begin{adjustbox}{width=.8\textwidth}
\begin{tabular}{l*{2}{D{.}{.}{-1}}}
\toprule
                    &\multicolumn{1}{c}{(1)}&\multicolumn{1}{c}{(2)}\\
                    &\multicolumn{1}{c}{\textit{ln(2-year Journal Impact Factor)}}&\multicolumn{1}{c}{\textit{ln(2-year Journal Impact Factor)}}\\
\midrule
Female=1            &     -0.0515\sym{***}&     -0.0521\sym{***}\\
                    &     (-5.63)         &     (-8.16)         \\
\addlinespace
\textit{Surprise (References)}    &      0.0105         &                     \\
                    &      (1.80)         &                     \\
\addlinespace
Female=1 $\times$ \textit{Surprise (References)}&      0.0364\sym{**} &                     \\
                    &      (3.03)         &                     \\
\end{tabular}
\end{adjustbox}
\end{table}

\section{Global effect of surprise and prescience on rewards by gender}

In this section, we report the predicted average rewards by genders when both surprise (prescience) in reference and concepts effects are considered within the same regression model on solo-authored papers:

\begin{equation}\label{global_eq}
\begin{split}
    \textit{Scholarly Reward}_i = \alpha + \beta Female_i + \gamma ReferenceSurprise_{i} + \delta Conceptsurprise_{i} + \\
    \tau(Female_i  \times ReferenceSurprise_{i})  +\eta(Female_i  \times Conceptsurprise_{i})  + \\ 
      \sum \delta_{j} controls_{i,j} + \epsilon_{i}
\end{split}
\end{equation}

\begin{figure}[H]
\includegraphics[width=1.0\linewidth]{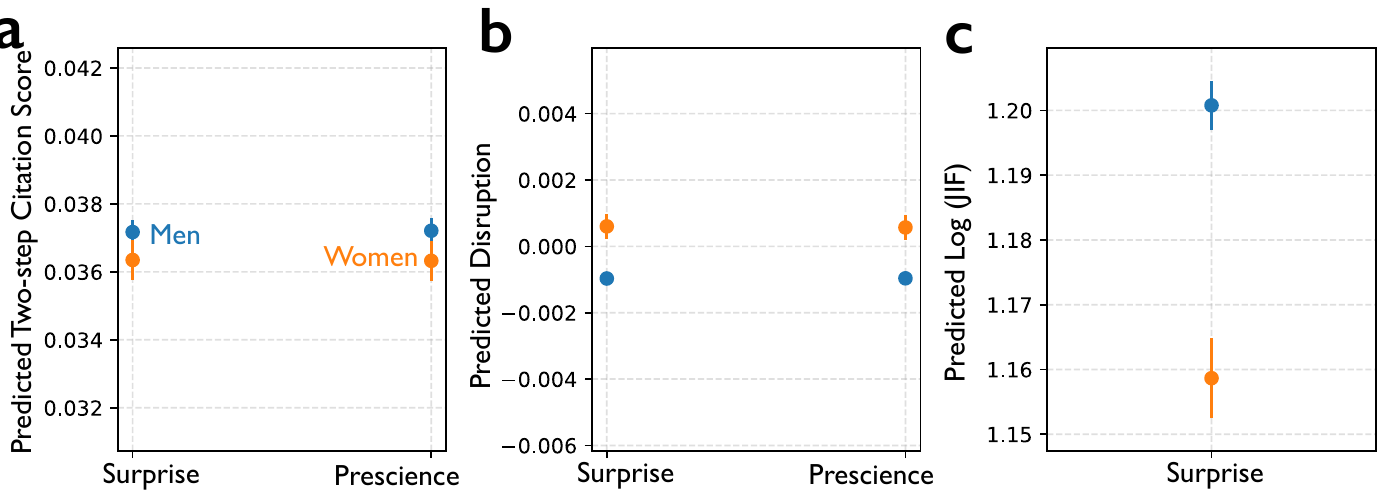}  
  \caption{Predicted average rewards in solo-authored papers by gender by author's gender in solo-authored papers, conditional at mean values of surprise or prescience. We control for career age, department size, year of publication, open access, and level-one fields.}
  \label{fig:fig_global1}
\end{figure}

\begin{figure}[H]
\includegraphics[width=1.0\linewidth]{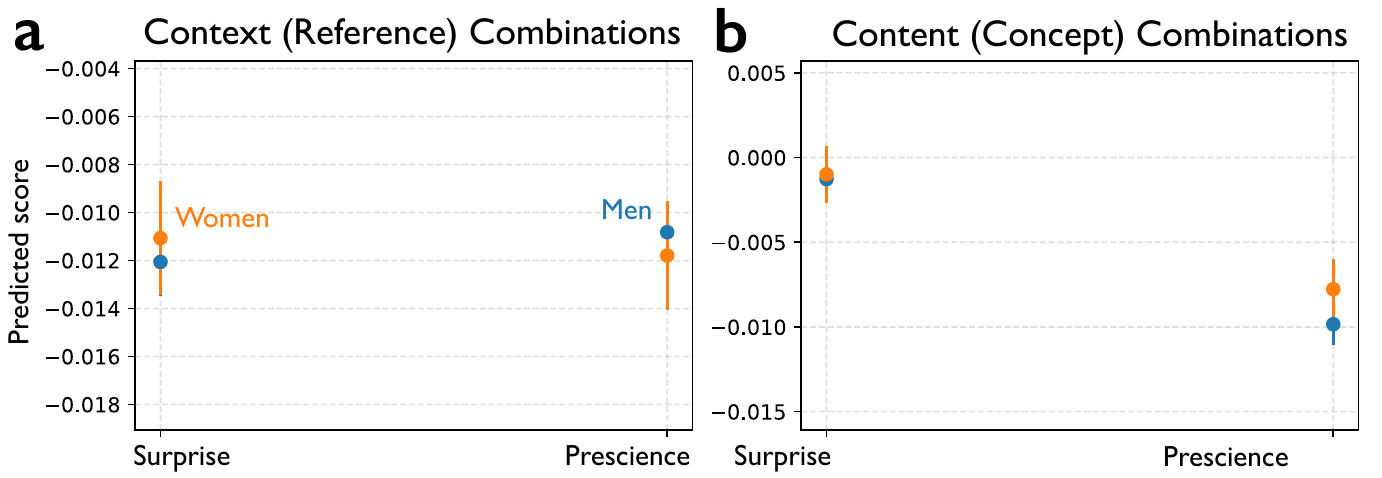}      
  \caption{Average marginal effect of surprise and prescience on two-step credit score by author's gender in solo-authored papers. We control for career age, department size, year of publication, open access, and level-one fields. }
  \label{fig:fig_global_ame_2steps}
\end{figure}

\begin{figure}[H]
\includegraphics[width=1.0\linewidth]{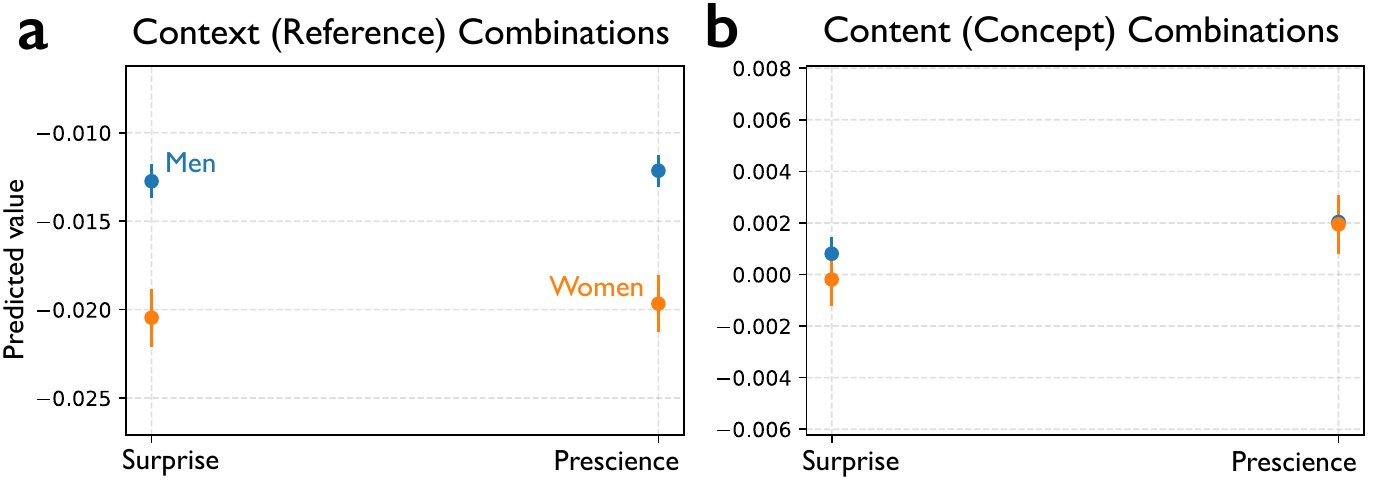}   
  \caption{Average marginal effect of surprise and prescience on disruption by author's gender in solo-authored papers. We control for career age, department size, year of publication, open access, and level-one fields. }
  \label{fig:fig_global_ame_cd}
\end{figure}

\begin{figure}[H]
\includegraphics[width=1.0\linewidth]{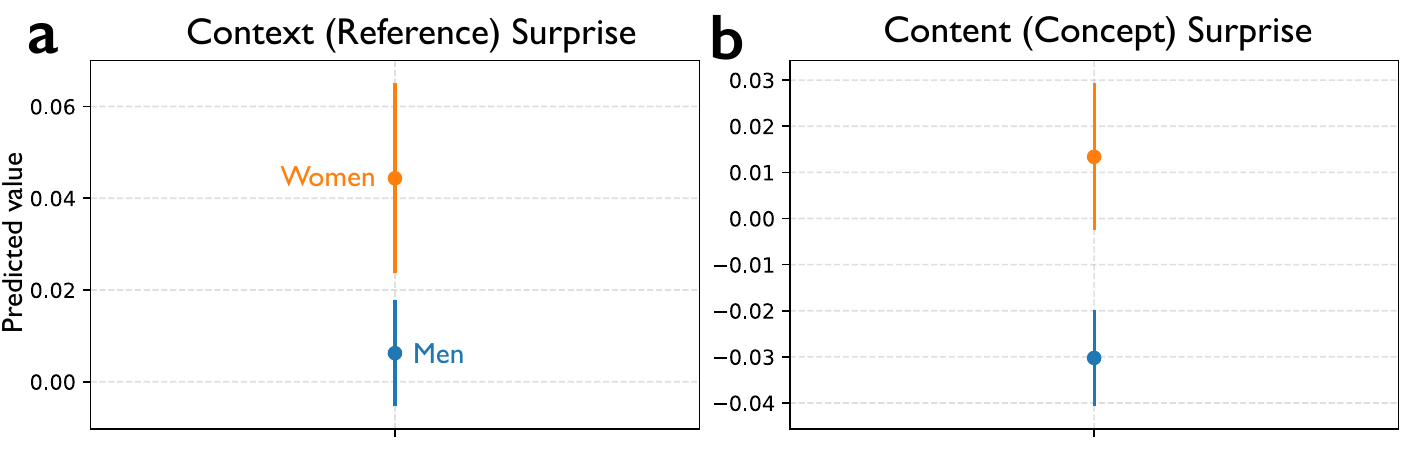}  
  \caption{Average marginal effect of surprise on the log of journal impact factor by author's gender in solo-authored papers. We control for career age, department size, year of publication, open access, and level-one fields. }
  \label{fig:fig_global_ame_jif}
\end{figure}

Standard errors are clustered at the author level. Figure \ref{fig:fig_global1} reports the predicted average rewards by gender estimated from model (\ref{global_eq}), and show that women's works tend to have significantly lower journal impact, but higher disruption. Figures \ref{fig:fig_global_ame_2steps}-\ref{fig:fig_global_ame_jif} show the average marginal effect of reference and concept surprise (prescience) on scholarly rewards by gender: marginal returns in disruption to increasing surprise (or prescience) of sourceare relatively higher for men's works, while women earn marginall more in journal impact. The results are in line with the baseline models discussed in the main text.

\section{Two-year forward citation count}\label{app3:cite2y}

We extend our analysis to forward citations counts of papers, to evaluate and study gender differences in how additional investments in surprise and prescience generate benefits in terms of \emph{direct} citations. We consider the two-year citations count -- i.e, the total number of citations to a focal paper after two years of publication. Below, we report model estimates on citations counts in solo and multi-authored papers (mixed and gender-homogeneous) relying on a negative binomial regression model for overly dispersed count data.

Among solo-authored papers (Table \ref{tab_logc_2}), greater prescience and surprise are associated with higher citation counts, with men’s work performing better on average. Women’s contributions exhibit significantly lower marginal returns to prescience (for both reference- and concept-based measures) and to reference surprise, while there is no significant gender difference in the marginal returns to concept surprise.

In Table \ref{tab_logc_6}, which reports results for multi-authored papers, a higher share of women is consistently associated with lower citation outcomes across all measures. However, teams with a higher women’s share earn relatively higher marginal returns to investing in more surprising concept combinations.

\begin{table}[H]\centering
\def\sym#1{\ifmmode^{#1}\else\(^{#1}\)\fi}
\caption{Negative binomial regression model estimates for two-year forward citation counts of solo-authored papers in content and context surprise and prescience scores by gender. We include  author-level clustered standard errors. Research field fixed effects and dummies for publishing year are omitted.}\label{tab_logc_2}
\begin{adjustbox}{width=\textwidth}
\begin{tabular}{lcccc}
\toprule
 &\multicolumn{1}{c}{(1)}&\multicolumn{1}{c}{(2)}&\multicolumn{1}{c}{(3)}&\multicolumn{1}{c}{(4)}\\
                 &\multicolumn{1}{c}{TwoYearsForwardCitations}&\multicolumn{1}{c}{TwoYearsForwardCitations}&\multicolumn{1}{c}{TwoYearsForwardCitations}&\multicolumn{1}{c}{TwoYearsForwardCitations}\\
\midrule

Female=1            &     -0.0152         &     -0.0368         &     -0.0512\sym{*}  &      0.0150         \\
                    &     (-0.49)         &     (-1.27)         &     (-2.20)         &      (0.61)         \\
\addlinespace
\textit{Surprise (References)}    &       1.319\sym{***}&                     &                     &                     \\
                    &     (69.76)         &                     &                     &                     \\
\addlinespace
Female=1 $\times$ \textit{Surprise (References)}&      -0.133\sym{**} &                     &                     &                     \\
                    &     (-3.23)         &                     &                     &                     \\
\addlinespace
DepSize&   0.0000431\sym{***}&   0.0000432\sym{***}&   0.0000436\sym{***}&   0.0000422\sym{***}\\
                    &     (18.03)         &     (17.39)         &     (17.09)         &     (17.15)         \\
\addlinespace
CareerAge          &     0.00836\sym{***}&     0.00825\sym{***}&     0.00686\sym{***}&     0.00702\sym{***}\\
                    &     (23.87)         &     (23.58)         &     (19.81)         &     (20.57)         \\
\addlinespace
\textit{OpenAccess}             &       0.298\sym{***}&       0.304\sym{***}&       0.319\sym{***}&       0.311\sym{***}\\
                    &     (29.19)         &     (29.54)         &     (30.24)         &     (29.65)         \\
\addlinespace
\textit{Prescience (References)}  &                     &       1.162\sym{***}&                     &                     \\
                    &                     &     (62.69)         &                     &                     \\
\addlinespace
Female=1 $\times$ \textit{Prescience (References)}&                     &     -0.0973\sym{*}  &                     &                     \\
                    &                     &     (-2.47)         &                     &                     \\
\addlinespace
\textit{Surprise (Concepts)}   &                     &                     &       0.335\sym{***}&                     \\
                    &                     &                     &     (19.36)         &                     \\
\addlinespace
Female=1 $\times$ \textit{Surprise (Concepts)}&                     &                     &     -0.0364         &                     \\
                    &                     &                     &     (-1.06)         &                     \\
\addlinespace
\textit{Prescience (Concepts)} &                     &                     &                     &       0.596\sym{***}\\
                    &                     &                     &                     &     (31.89)         \\
\addlinespace
Female=1 $\times$ \textit{Prescience (Concepts)}&                     &                     &                     &      -0.143\sym{***}\\
                    &                     &                     &                     &     (-3.78)         \\
\addlinespace
Constant            &       0.589\sym{***}&       0.684\sym{***}&       1.225\sym{***}&       1.051\sym{***}\\
                    &     (23.01)         &     (26.54)         &     (44.76)         &     (39.47)         \\
\midrule
Observations        &      226208         &      226208         &      226208         &      226208         \\
\midrule
lnalpha             &       0.271\sym{***}&       0.291\sym{***}&       0.362\sym{***}&       0.352\sym{***}\\
                    &     (41.96)         &     (45.40)         &     (60.91)         &     (59.43)         \\
\bottomrule
\multicolumn{5}{l}{\footnotesize \textit{t} statistics in parentheses}\\
\multicolumn{5}{l}{\footnotesize \sym{*} \(p<0.05\), \sym{**} \(p<0.01\), \sym{***} \(p<0.001\)}\\
\end{tabular}
\end{adjustbox}
\end{table}

\begin{table}[H]\centering
\def\sym#1{\ifmmode^{#1}\else\(^{#1}\)\fi}
\caption{Negative binomial regression model estimates for two-year forward citation counts in multi-authored papers in content and context surprise and prescience scores by gender. We include White standard errors. Research field fixed effects and dummies for publishing year are omitted.}\label{tab_logc_6}
\begin{adjustbox}{width=\textwidth}
\begin{tabular}{lcccc}
\toprule
 &\multicolumn{1}{c}{(1)}&\multicolumn{1}{c}{(2)}&\multicolumn{1}{c}{(3)}&\multicolumn{1}{c}{(4)}\\
                 &\multicolumn{1}{c}{TwoYearsForwardCitations}&\multicolumn{1}{c}{TwoYearsForwardCitations}&\multicolumn{1}{c}{TwoYearsForwardCitations}&\multicolumn{1}{c}{TwoYearsForwardCitations}\\
\midrule
\textit{Female Share}        &     -0.0864\sym{***}&      -0.107\sym{***}&      -0.160\sym{***}&      -0.132\sym{***}\\
                    &     (-5.92)         &     (-7.36)         &    (-16.99)         &    (-11.39)         \\
\addlinespace
Surprise (References)    &       0.887\sym{***}&                     &                     &                     \\
                    &    (136.19)         &                     &                     &                     \\
\addlinespace
\textit{Female Share} $\times$ Surprise (References)&     -0.0813\sym{***}&                     &                     &                     \\
                    &     (-3.31)         &                     &                     &                     \\
\addlinespace
DepartmentSize&   0.0000214\sym{***}&   0.0000214\sym{***}&   0.0000218\sym{***}&   0.0000206\sym{***}\\
                    &     (40.02)         &     (39.15)         &     (36.88)         &     (33.42)         \\
\addlinespace
CareerAge          &     0.00276\sym{***}&     0.00304\sym{***}&     0.00273\sym{***}&     0.00303\sym{***}\\
                    &     (21.28)         &     (23.05)         &     (19.44)         &     (20.95)         \\
\addlinespace
OpenAccess=1             &       0.313\sym{***}&       0.334\sym{***}&       0.343\sym{***}&       0.332\sym{***}\\
                    &    (118.48)         &    (123.63)         &    (128.24)         &    (123.65)         \\
\addlinespace
TeamSize         &      0.0583\sym{***}&      0.0585\sym{***}&      0.0555\sym{***}&      0.0543\sym{***}\\
                    &     (95.09)         &     (95.03)         &     (88.19)         &     (84.69)         \\
\addlinespace
TeamSize\_squared &  -0.0000118\sym{***}&  -0.0000118\sym{***}&  -0.0000119\sym{***}&  -0.0000113\sym{***}\\
                    &    (-29.04)         &    (-29.10)         &    (-27.89)         &    (-27.55)         \\
\addlinespace
num\_unknown         &     -0.0230\sym{***}&     -0.0230\sym{***}&     -0.0199\sym{***}&     -0.0204\sym{***}\\
                    &    (-26.46)         &    (-26.15)         &    (-21.63)         &    (-21.83)         \\
\addlinespace
Prescience (References)  &                     &       0.756\sym{***}&                     &                     \\
                    &                     &    (128.45)         &                     &                     \\
\addlinespace
\textit{Female Share} $\times$ Prescience (References)&                     &     -0.0418         &                     &                     \\
                    &                     &     (-1.66)         &                     &                     \\
\addlinespace
Surprise (concepts)   &                     &                     &       0.239\sym{***}&                     \\
                    &                     &                     &     (46.91)         &                     \\
\addlinespace
\textit{Female Share} $\times$ Surprise (concepts)&                     &                     &      0.0685\sym{**} &                     \\
                    &                     &                     &      (3.06)         &                     \\
\addlinespace
Prescience (concepts) &                     &                     &                     &       0.627\sym{***}\\
                    &                     &                     &                     &    (120.77)         \\
\addlinespace
\textit{Female Share} $\times$ Prescience (concepts)&                     &                     &                     &      0.0263         \\
                    &                     &                     &                     &      (1.53)         \\
\addlinespace
Constant            &       1.299\sym{***}&       1.385\sym{***}&       1.694\sym{***}&       1.441\sym{***}\\
                    &    (160.07)         &    (175.50)         &    (234.89)         &    (188.56)         \\
\midrule
Observations        &     4085543         &     4085543         &     4085543         &     4085543         \\
\midrule
lnalpha             &      0.0566\sym{***}&      0.0600\sym{***}&      0.0884\sym{***}&      0.0685\sym{***}\\
                   &     (23.31)         &     (24.37)         &     (35.89)         &     (26.32)         \\
\bottomrule
\multicolumn{5}{l}{\footnotesize \textit{t} statistics in parentheses}\\
\multicolumn{5}{l}{\footnotesize \sym{*} \(p<0.05\), \sym{**} \(p<0.01\), \sym{***} \(p<0.001\)}\\
\end{tabular}
\end{adjustbox}
\end{table}

\section{GAM model estimates in solo-authored papers}

\begin{figure}[H]
\includegraphics[width=1.0\linewidth]{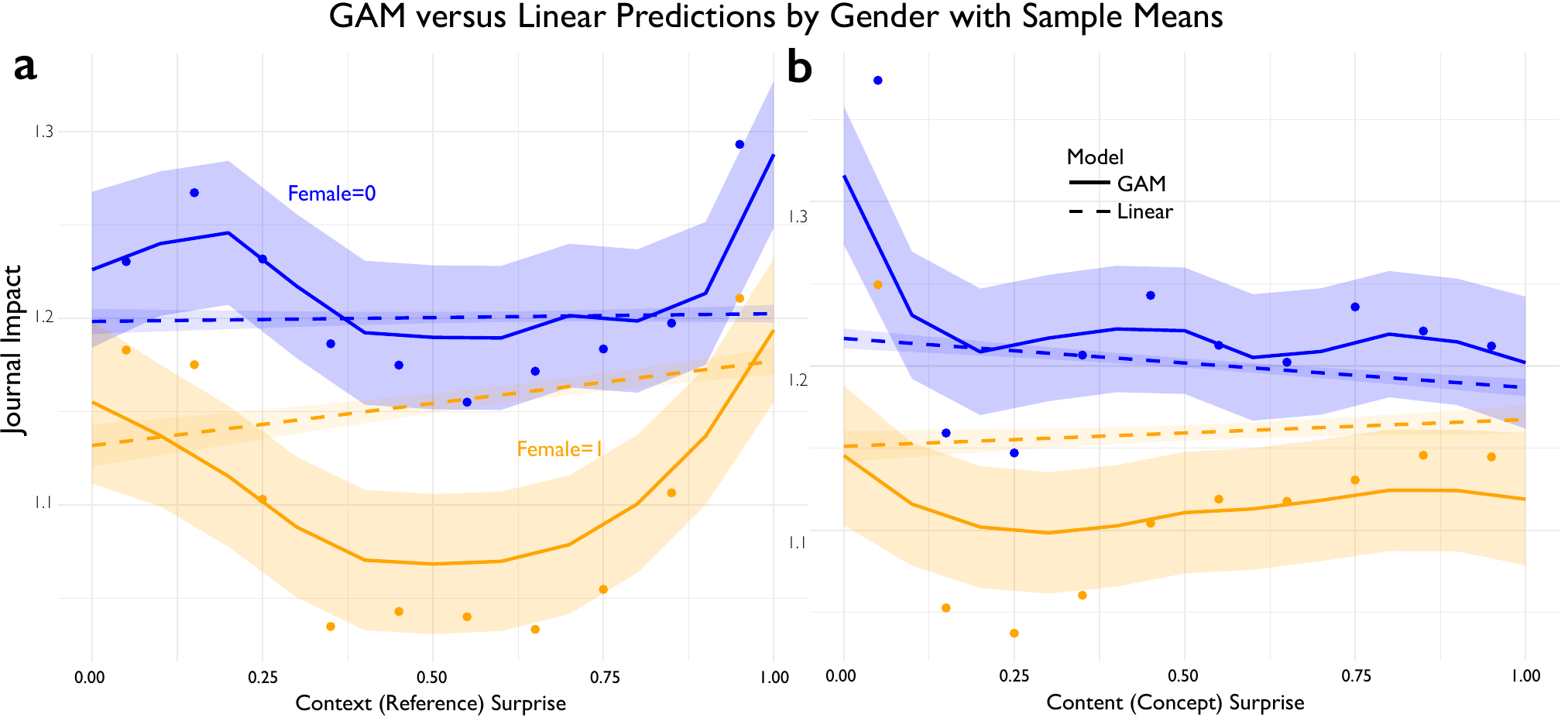}  
  \caption{Average fitted two-year journal impact by gender within 0.1 bins of surprise with GAM models. The models include a gender intercept shift and smooth functions of surprise interacted with gender, while conditioning linearly on all controls. Solid lines report the averaged GAM predictions with 95\% confidence intervals shown as shaded bands, dashed lines the corresponding linear regression margins, and points represent the observed sample means within 0.1 bins of the predictors.}
  \label{fig:fig_gam_jif}
\end{figure}

\begin{figure}[H]
\includegraphics[width=1.0\linewidth]{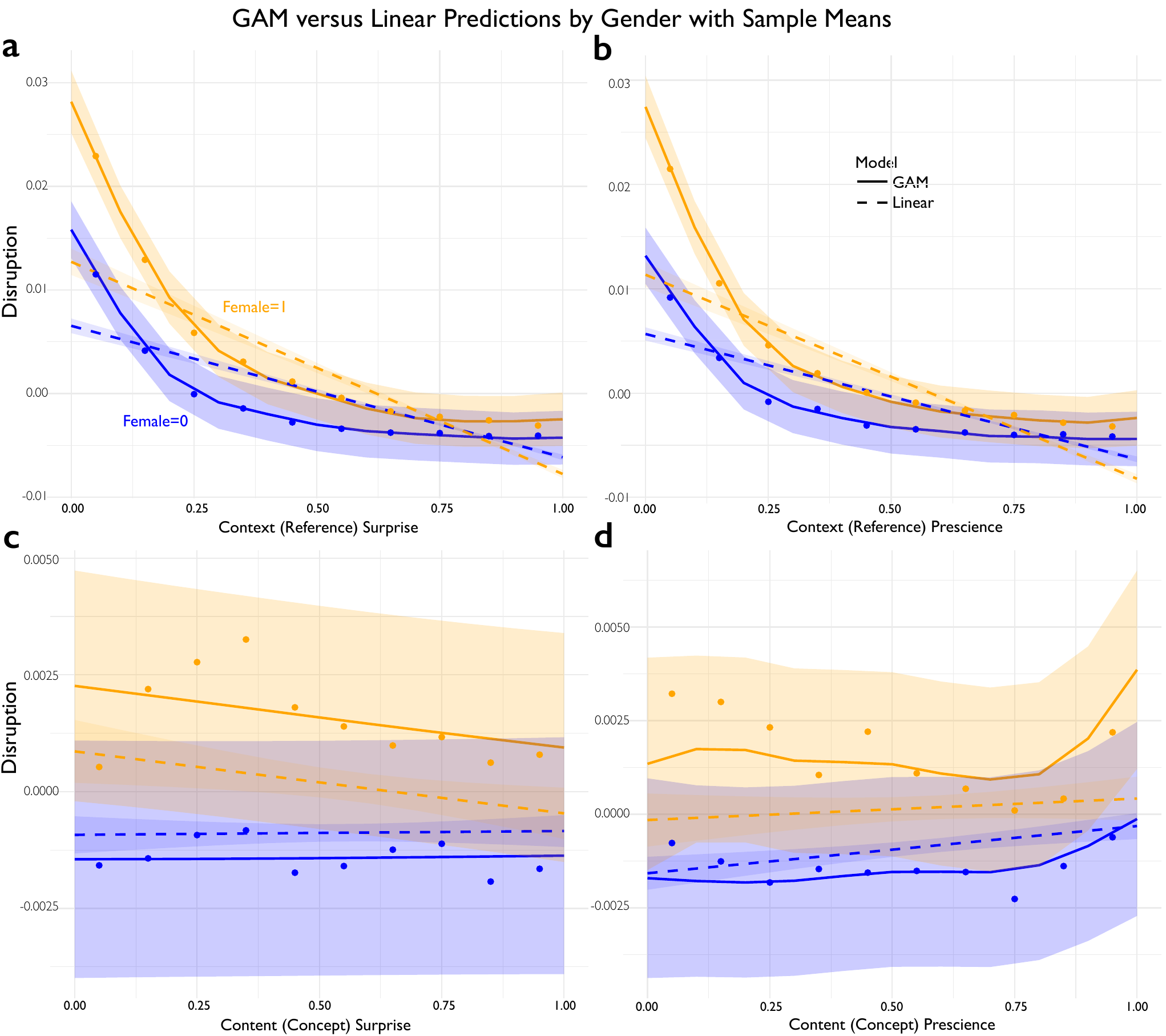}  
  \caption{Average fitted five-year disruption by gender within 0.1 bins of surprise (row 1) and prescience (row 2) with GAM models. The models include a gender intercept shift and smooth functions of surprise or prescience interacted with gender, while conditioning linearly on all controls. Solid lines report the averaged GAM predictions with 95\% confidence intervals shown as shaded bands, dashed lines the corresponding linear regression margins, and points represent the observed sample means within 0.1 bins of the predictors.}
  \label{fig:fig_gam_disr}
\end{figure}

\begin{figure}[H]
\includegraphics[width=1.0\linewidth]{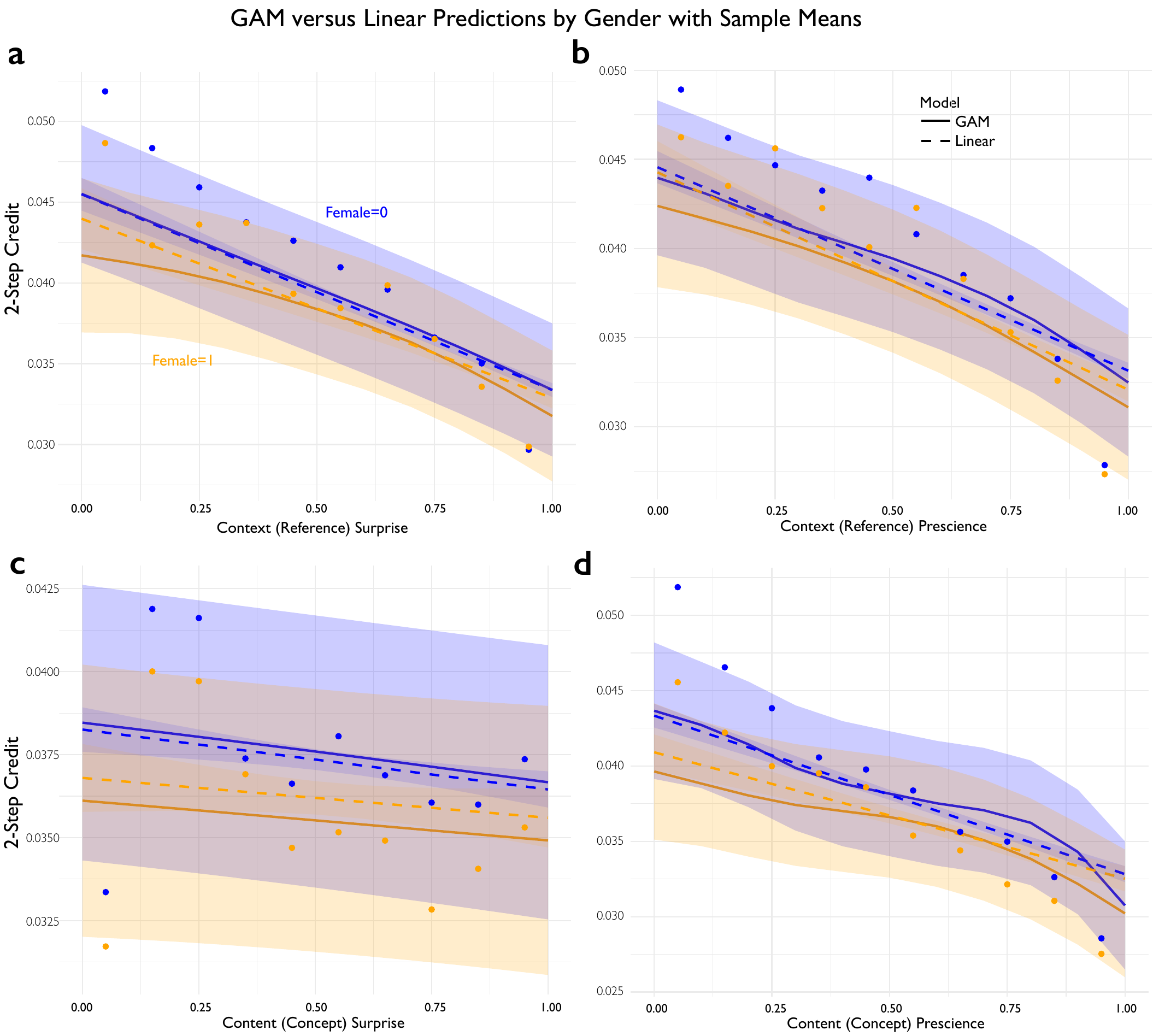} 
  \caption{Average fitted two-step credit score by gender within 0.1 bins of surprise (row 1) and prescience (row 2) with GAM models. The models include a gender intercept shift and smooth functions of surprise or prescience interacted with gender, while conditioning linearly on all controls. Solid lines report the averaged GAM predictions with 95\% confidence intervals shown as shaded bands, dashed lines the corresponding linear regression margins, and points represent the observed sample means within 0.1 bins of the predictors.}
  \label{fig:fig_gam_credit}
\end{figure}

We use a semi-parametric GAM model to examine non-linear associations between surprise or prescience and scholarly rewards by gender of solo-authored papers, while including linear controls and a gender-specific intercept shift. To ease computations in R, we choose to compute non-robust standard errors. Figures \ref{fig:fig_gam_jif}–\ref{fig:fig_gam_credit} show predicted rewards across 0.1 intervals of novelty. Journal impact exhibits a U-shaped pattern for reference surprise among women, while linear approximations capture mid-range novelty for men and extreme novelty for women (Figure \ref{fig:fig_gam_jif}). Reference surprise and prescience display slight non-linearities for disruption at low values, but linear predictions align with GAM confidence intervals across most of the observed range (Figure \ref{fig:fig_gam_disr}, row 1). Two-step credit shows similar alignment between linear and GAM predictions (Figure \ref{fig:fig_gam_credit}), as does disruption from concept-based novelty (Figure \ref{fig:fig_gam_disr}, row 2). 



\section{Non-linearity of gender in mixed multi-authored papers}\label{app:non_linear_ma}

We estimate linear regression models on multi-authored papers, including women’s share (\textit{FemaleShare}) and its square (\textit{FemaleShare}$^2$), and a dummy for gender-homogeneous teams (\textit{GenderHomogeneousTeams}, coded one for all-women, zero for all-men teams with high-confidence gender identification). To capture team-specific effects, we interact the dummy with female share (\textit{GenderHomogeneousTeam} $\times$ \textit{FemaleShare}). The dummy is largely collinear with the squared female share because mixed-team values are pushed toward the extremes.

Table \ref{tab1_ma_nv} shows that gender differences in surprise and prescience of source combinations (columns 1–2) primarily occur in gender-homogeneous teams (\textit{GenderHomogeneousTeam} $\times$ \textit{FemaleShare}). 
For concept-based innovations, the relationship with female share is non-linear. Concept surprise is concave, with diminishing marginal effects at higher female shares (FemaleShare$^2 < 0$), while prescience is convex: the negative effect at low female share diminishes as the share increases (FemaleShare$^2 > 0$).




Next, Table \ref{tab3_cred_nv} reports two-step credit scores. For reference combinations (columns 1-2), higher female share (\textit{FemaleShare}) is associated with increased downstream citations in mixed-gender teams, but lower values in gender-homogeneous teams (\textit{GenderHomogeneousTeam} $\times$ \textit{FemaleShare}). For concept combinations (columns 3-4), increasing female share corresponds to lower two-step credit, particularly in all-women teams. Non-linear terms suggest that the relationship between female share and novelty is not strictly linear and varies across the range of female representation. 


Turning to disruption (Table \ref{tab3_cd_nv}), higher women’s presence is associated with greater disruption for source-based novelty primarily in mixed-gender teams (\textit{FemaleShare}). For concept-based novelty, increased disruption is concentrated in all-women teams \textit{GenderHomogeneousTeam} $\times$ \textit{FemaleShare}, while the squared terms for female share are not significant, indicating no clear non-linear effects. These results suggest that the contribution of women to disruptive science is linear in team composition.

Finally, Table \ref{tab3_jif_nv} shows patterns in \emph{journal placement}. All-women teams exhibit a steeper increase in journal impact with rising concept surprise compared to all-men teams (\textit{GenderHomogeneousTeam} $\times$ \textit{FemaleShare} $\times$ \textit{Surprise}), consistent with the solo-authored results in the main text. Non-linear effects of female share are generally modest, suggesting that the three-way interactions with gender-homogeneous teams and surprise primarily drive the observed journal placement patterns.

\begin{table}[H]\centering
\def\sym#1{\ifmmode^{#1}\else\(^{#1}\)\fi}
\caption{Linear regression model estimates for gender differences on content and context surprise and prescience scores, on multi-authored papers. We include White standard errors. Research field fixed effects and dummies for publishing year are omitted.} \label{tab1_ma_nv}
\begin{adjustbox}{width=\textwidth}
\begin{tabular}{lcccc}
\toprule
                    &\multicolumn{1}{c}{(1)}&\multicolumn{1}{c}{(2)}&\multicolumn{1}{c}{(3)}&\multicolumn{1}{c}{(4)}\\
                    &\multicolumn{1}{c}{Surprise (References)}&\multicolumn{1}{c}{Prescience (References)}&\multicolumn{1}{c}{Surprise (concepts)}&\multicolumn{1}{c}{Prescience (concepts)}\\
\midrule
GenderHomogeneousTeam=1&    -0.00794\sym{***}&    -0.00122         &    0.000134         &     -0.0140\sym{***}\\
                    &     (-6.40)         &     (-0.90)         &      (0.09)         &     (-9.48)         \\
\addlinespace
\textit{Female Share}        &     0.00335         &   -0.000637         &    0.000752         &     -0.0678\sym{***}\\
                    &      (0.62)         &     (-0.11)         &      (0.12)         &    (-10.99)         \\
\addlinespace
GenderHomogeneousTeam=1 $\times$ \textit{Female Share}&     0.00696\sym{***}&     0.00362\sym{**} &     0.00791\sym{***}&      0.0105\sym{***}\\
                    &      (5.48)         &      (2.66)         &      (5.41)         &      (7.64)         \\
\addlinespace
\textit{Female Share}\_square &     0.00994         &      0.0120         &     -0.0144\sym{*}  &      0.0498\sym{***}\\
                    &      (1.73)         &      (1.93)         &     (-2.17)         &      (7.65)         \\
\addlinespace
DepSize&  0.00000154\sym{***}&  0.00000117\sym{***}& 0.000000395\sym{***}&  0.00000196\sym{***}\\
                    &     (52.75)         &     (37.00)         &     (11.94)         &     (61.91)         \\
\addlinespace
CareerAge          &   -0.000158\sym{***}&   -0.000550\sym{***}&   -0.000153\sym{***}&   -0.000469\sym{***}\\
                    &    (-14.05)         &    (-46.14)         &    (-12.39)         &    (-40.06)         \\
\addlinespace
num\_authors         &     0.00112\sym{***}&    0.000925\sym{***}&     0.00101\sym{***}&     0.00262\sym{***}\\
                    &     (17.93)         &     (14.40)         &     (15.00)         &     (26.33)         \\
\addlinespace
num\_authors\_squared &-0.000000431\sym{***}&-0.000000368\sym{***}&-0.000000160\sym{***}&-0.000000419\sym{***}\\
                    &    (-32.49)         &    (-28.24)         &    (-15.80)         &    (-27.95)         \\
\addlinespace
num\_unknown         &  -0.0000126         & -0.00000114         &   -0.000581\sym{***}&    -0.00151\sym{***}\\
                    &     (-0.17)         &     (-0.02)         &     (-7.81)         &    (-13.62)         \\
\addlinespace
Constant            &       0.550\sym{***}&       0.541\sym{***}&       0.523\sym{***}&       0.592\sym{***}\\
                    &    (350.75)         &    (321.46)         &    (293.83)         &    (324.31)         \\
\midrule
Observations        &     4085543         &     4085543         &     4085543         &     4085543         \\
\bottomrule
\multicolumn{5}{l}{\footnotesize \textit{t} statistics in parentheses}\\
\multicolumn{5}{l}{\footnotesize \sym{*} \(p<0.05\), \sym{**} \(p<0.01\), \sym{***} \(p<0.001\)}\\
\end{tabular}
\end{adjustbox}
\end{table}


 \begin{table}[H]\centering
\def\sym#1{\ifmmode^{#1}\else\(^{#1}\)\fi}
\caption{Linear regression model estimates for 2-steps-out credit score for surprise and prescience scores in multi-authored papers. We include White standard errors. Research field fixed effects and dummies for publishing year are omitted.}\label{tab3_cred_nv}
\begin{adjustbox}{width=\textwidth}
\begin{tabular}{lcccc}
\toprule
 &\multicolumn{1}{c}{(1)}&\multicolumn{1}{c}{(2)}&\multicolumn{1}{c}{(3)}&\multicolumn{1}{c}{(4)}\\
                    &\multicolumn{1}{c}{TwoStepsCreditScore}&\multicolumn{1}{c}{TwoStepsCreditScore}&\multicolumn{1}{c}{TwoStepsCreditScore}&\multicolumn{1}{c}{TwoStepsCreditScore}\\
\midrule

GenderHomogeneousTeam=1&     0.00566\sym{***}&     0.00627\sym{***}&    -0.00118\sym{*}  &   -0.000844         \\
                    &      (7.99)         &     (11.64)         &     (-2.40)         &     (-1.41)         \\
\addlinespace
\textit{Female Share}        &     0.00888\sym{**} &     0.00973\sym{***}&     -0.0104\sym{***}&     -0.0132\sym{***}\\
                    &      (2.66)         &      (3.82)         &     (-4.56)         &     (-4.69)         \\
\addlinespace
GenderHomogeneousTeam=1 $\times$ \textit{Female Share}&    -0.00425\sym{***}&    -0.00517\sym{***}&    -0.00307\sym{***}&    -0.00508\sym{***}\\
                    &     (-4.78)         &     (-7.60)         &     (-5.26)         &     (-7.10)         \\
\addlinespace
\textit{Surprise (References)}     &    -0.00571\sym{***}&                     &                     &                     \\
                    &     (-6.33)         &                     &                     &                     \\
\addlinespace
GenderHomogeneousTeam=1 $\times$ \textit{Surprise (References)} &    -0.00936\sym{***}&                     &                     &                     \\
                    &    (-10.24)         &                     &                     &                     \\
\addlinespace
\textit{Female Share} $\times$ \textit{Surprise (References)} &     -0.0241\sym{***}&                     &                     &                     \\
                    &     (-5.57)         &                     &                     &                     \\
\addlinespace
GenderHomogeneousTeam=1 $\times$ \textit{Female Share} $\times$ \textit{Surprise (References)} &     0.00223         &                     &                     &                     \\
                    &      (1.95)         &                     &                     &                     \\
\addlinespace
\textit{Female Share}\_square &    -0.00515         &    -0.00503         &      0.0132\sym{***}&      0.0175\sym{***}\\
                    &     (-1.39)         &     (-1.77)         &      (5.21)         &      (5.63)         \\
\addlinespace
\textit{Female Share}\_square $\times$ \textit{Surprise (References)} &      0.0225\sym{***}&                     &                     &                     \\
                    &      (4.71)         &                     &                     &                     \\
\addlinespace
\textit{Prescience (References)}   &                     &    -0.00174\sym{*}  &                     &                     \\
                    &                     &     (-2.37)         &                     &                     \\
\addlinespace
GenderHomogeneousTeam=1 $\times$ \textit{Prescience (References)} &                     &     -0.0114\sym{***}&                     &                     \\
                    &                     &    (-15.28)         &                     &                     \\
\addlinespace
\textit{Female Share} $\times$ \textit{Prescience (References)} &                     &     -0.0286\sym{***}&                     &                     \\
                    &                     &     (-8.06)         &                     &                     \\
\addlinespace
GenderHomogeneousTeam=1 $\times$ \textit{Female Share} $\times$ \textit{Prescience (References)} &                     &     0.00395\sym{***}&                     &                     \\
                    &                     &      (4.21)         &                     &                     \\
\addlinespace
\textit{Female Share}\_square $\times$ \textit{Prescience (References)} &                     &      0.0251\sym{***}&                     &                     \\
                    &                     &      (6.37)         &                     &                     \\
\addlinespace
\textit{Surprise (concepts)}    &                     &                     &    -0.00358\sym{***}&                     \\
                    &                     &                     &     (-5.21)         &                     \\
\addlinespace
GenderHomogeneousTeam=1 $\times$ \textit{Surprise (concepts)} &                     &                     &    0.000828         &                     \\
                    &                     &                     &      (1.19)         &                     \\
\addlinespace
\textit{Female Share} $\times$ \textit{Surprise (concepts)} &                     &                     &     0.00469         &                     \\
                    &                     &                     &      (1.42)         &                     \\
\addlinespace
GenderHomogeneousTeam=1 $\times$ \textit{Female Share} $\times$ \textit{Surprise (concepts)} &                     &                     &    0.000277         &                     \\
                    &                     &                     &      (0.32)         &                     \\
\addlinespace
\textit{Female Share}\_square $\times$ \textit{Surprise (concepts)} &                     &                     &    -0.00508         &                     \\
                    &                     &                     &     (-1.39)         &                     \\
\addlinespace
\textit{Prescience (concepts)}  &                     &                     &                     &     -0.0125\sym{***}\\
                    &                     &                     &                     &    (-16.24)         \\
\addlinespace
GenderHomogeneousTeam=1 $\times$ \textit{Prescience (concepts)} &                     &                     &                     &  -0.0000244         \\
                    &                     &                     &                     &     (-0.03)         \\
\addlinespace
\textit{Female Share} $\times$ \textit{Prescience (concepts)} &                     &                     &                     &     0.00752\sym{*}  \\
                    &                     &                     &                     &      (2.03)         \\
\addlinespace
GenderHomogeneousTeam=1 $\times$ \textit{Female Share} $\times$ \textit{Prescience (concepts)} &                     &                     &                     &     0.00359\sym{***}\\
                    &                     &                     &                     &      (3.69)         \\
\addlinespace
\textit{Female Share}\_square $\times$ \textit{Prescience (concepts)} &                     &                     &                     &     -0.0107\sym{**} \\
                    &                     &                     &                     &     (-2.58)         \\
\addlinespace
DepSize&-0.000000419\sym{***}&-0.000000424\sym{***}&-0.000000424\sym{***}&-0.000000405\sym{***}\\
                    &    (-53.51)         &    (-54.13)         &    (-54.19)         &    (-52.05)         \\
\addlinespace
CareerAge          &  -0.0000140\sym{***}&  -0.0000176\sym{***}&  -0.0000116\sym{***}&  -0.0000172\sym{***}\\
                    &     (-6.61)         &     (-8.29)         &     (-5.47)         &     (-8.13)         \\
\addlinespace
OpenAccess=1             &    -0.00242\sym{***}&    -0.00279\sym{***}&    -0.00287\sym{***}&    -0.00269\sym{***}\\
                    &    (-48.14)         &    (-55.63)         &    (-56.78)         &    (-53.65)         \\
\addlinespace
num\_authors         &   -0.000266\sym{***}&   -0.000261\sym{***}&   -0.000249\sym{***}&   -0.000228\sym{***}\\
                    &    (-17.10)         &    (-16.95)         &    (-16.81)         &    (-16.15)         \\
\addlinespace
num\_authors\_squared &    1.77e-08\sym{***}&    1.86e-08\sym{***}&    1.94e-08\sym{***}&    1.63e-08\sym{***}\\
                    &     (11.81)         &     (12.35)         &     (12.92)         &     (11.17)         \\
\addlinespace
num\_unknown         &    0.000216\sym{***}&    0.000208\sym{***}&    0.000194\sym{***}&    0.000182\sym{***}\\
                    &     (12.92)         &     (12.60)         &     (12.20)         &     (11.99)         \\
\addlinespace
Constant            &      0.0305\sym{***}&      0.0286\sym{***}&      0.0300\sym{***}&      0.0356\sym{***}\\
                    &     (42.87)         &     (52.27)         &     (59.57)         &     (58.96)         \\
\midrule
Observations        &     2451029         &     2451029         &     2451029         &     2451029         \\
\multicolumn{5}{l}{\footnotesize \textit{t} statistics in parentheses}\\
\multicolumn{5}{l}{\footnotesize \sym{*} \(p<0.05\), \sym{**} \(p<0.01\), \sym{***} \(p<0.001\)}\\
\end{tabular}
\end{adjustbox}
\end{table}

\begin{table}[H]\centering
\def\sym#1{\ifmmode^{#1}\else\(^{#1}\)\fi}
\caption{Linear regression model estimates for disruption score of publication on surprise and prescience scores by gender in multi-authored papers, with White standard errors. Research field fixed effects and dummies for publishing year are omitted.}\label{tab3_cd_nv}
\begin{adjustbox}{width=\textwidth}
\begin{tabular}{l*{4}{D{.}{.}{-1}}}
\toprule
                    &\multicolumn{1}{c}{(1)}&\multicolumn{1}{c}{(2)}&\multicolumn{1}{c}{(3)}&\multicolumn{1}{c}{(4)}\\
                &\multicolumn{1}{c}{5-year Disruption}&\multicolumn{1}{c}{5-year Disruption}&\multicolumn{1}{c}{5-year Disruption}&\multicolumn{1}{c}{5-year Disruption}\\
\midrule
GenderHomogeneousTeam=1&     0.00350\sym{***}&     0.00223\sym{***}&  0.00000510         &   0.0000722         \\
                    &      (6.03)         &      (5.38)         &      (0.02)         &      (0.23)         \\
\addlinespace
\textit{Female Share}        &     0.00732\sym{**} &     0.00423\sym{*}  &     0.00123         &     0.00136         \\
                    &      (2.78)         &      (2.23)         &      (1.09)         &      (0.99)         \\
\addlinespace
GenderHomogeneousTeam=1 $\times$ \textit{Female Share}&  -0.0000199         &   0.0000890         &    0.000721\sym{**} &    0.000769\sym{*}  \\
                    &     (-0.03)         &      (0.19)         &      (2.76)         &      (2.51)         \\
\addlinespace
\textit{Surprise (References)}     &    0.000807         &                     &                     &                     \\
                    &      (1.08)         &                     &                     &                     \\
\addlinespace
GenderHomogeneousTeam=1 $\times$ \textit{Surprise (References)} &    -0.00426\sym{***}&                     &                     &                     \\
                    &     (-5.62)         &                     &                     &                     \\
\addlinespace
\textit{Female Share} $\times$ \textit{Surprise (References)} &    -0.00654         &                     &                     &                     \\
                    &     (-1.90)         &                     &                     &                     \\
\addlinespace
GenderHomogeneousTeam=1 $\times$ \textit{Female Share} $\times$ \textit{Surprise (References)} &    0.000270         &                     &                     &                     \\
                    &      (0.32)         &                     &                     &                     \\
\addlinespace
\textit{Female Share}\_square &    -0.00514         &    -0.00271         &    -0.00149         &    -0.00149         \\
                    &     (-1.83)         &     (-1.32)         &     (-1.26)         &     (-1.04)         \\
\addlinespace
\textit{Female Share}\_square $\times$ \textit{Surprise (References)} &     0.00383         &                     &                     &                     \\
                    &      (1.04)         &                     &                     &                     \\
\addlinespace
\textit{Prescience (References)}   &                     &   -0.000432         &                     &                     \\
                    &                     &     (-0.77)         &                     &                     \\
\addlinespace
GenderHomogeneousTeam=1 $\times$ \textit{Prescience (References)} &                     &    -0.00251\sym{***}&                     &                     \\
                    &                     &     (-4.41)         &                     &                     \\
\addlinespace
\textit{Female Share} $\times$ \textit{Prescience (References)} &                     &    -0.00179         &                     &                     \\
                    &                     &     (-0.69)         &                     &                     \\
\addlinespace
GenderHomogeneousTeam=1 $\times$ \textit{Female Share} $\times$ \textit{Prescience (References)} &                     &    0.000112         &                     &                     \\
                    &                     &      (0.17)         &                     &                     \\
\addlinespace
\textit{Female Share}\_square $\times$ \textit{Prescience (References)} &                     &   -0.000144         &                     &                     \\
                    &                     &     (-0.05)         &                     &                     \\
\addlinespace
\textit{Surprise (concepts)}    &                     &                     &    -0.00135\sym{***}&                     \\
                    &                     &                     &     (-3.61)         &                     \\
\addlinespace
GenderHomogeneousTeam=1 $\times$ \textit{Surprise (concepts)} &                     &                     &     0.00141\sym{***}&                     \\
                    &                     &                     &      (3.73)         &                     \\
\addlinespace
\textit{Female Share} $\times$ \textit{Surprise (concepts)} &                     &                     &     0.00369\sym{*}  &                     \\
                    &                     &                     &      (2.19)         &                     \\
\addlinespace
GenderHomogeneousTeam=1 $\times$ \textit{Female Share} $\times$ \textit{Surprise (concepts)} &                     &                     &    -0.00115\sym{**} &                     \\
                    &                     &                     &     (-2.91)         &                     \\
\addlinespace
\textit{Female Share}\_square $\times$ \textit{Surprise (concepts)} &                     &                     &    -0.00255         &                     \\
                    &                     &                     &     (-1.43)         &                     \\
\addlinespace
\textit{Prescience (concepts)}  &                     &                     &                     &    0.000368         \\
                    &                     &                     &                     &      (0.85)         \\
\addlinespace
GenderHomogeneousTeam=1 $\times$ \textit{Prescience (concepts)} &                     &                     &                     &     0.00123\sym{**} \\
                    &                     &                     &                     &      (2.87)         \\
\addlinespace
\textit{Female Share} $\times$ \textit{Prescience (concepts)} &                     &                     &                     &     0.00333         \\
                    &                     &                     &                     &      (1.75)         \\
\addlinespace
GenderHomogeneousTeam=1 $\times$ \textit{Female Share} $\times$ \textit{Prescience (concepts)} &                     &                     &                     &    -0.00118\sym{**} \\
                    &                     &                     &                     &     (-2.68)         \\
\addlinespace
\textit{Female Share}\_square $\times$ \textit{Prescience (concepts)} &                     &                     &                     &    -0.00245         \\
                    &                     &                     &                     &     (-1.23)         \\
\addlinespace
DepSize&   -1.93e-08\sym{***}&   -2.00e-08\sym{***}&   -2.23e-08\sym{***}&   -2.45e-08\sym{***}\\
                    &     (-6.90)         &     (-7.16)         &     (-7.91)         &     (-8.75)         \\
\addlinespace
CareerAge          & -0.00000617\sym{***}& -0.00000714\sym{***}& -0.00000565\sym{***}& -0.00000494\sym{***}\\
                    &     (-5.46)         &     (-6.31)         &     (-5.00)         &     (-4.36)         \\
\addlinespace
OpenAccess=1             &   -0.000562\sym{***}&   -0.000611\sym{***}&   -0.000653\sym{***}&   -0.000682\sym{***}\\
                    &    (-19.23)         &    (-20.80)         &    (-21.96)         &    (-23.09)         \\
\addlinespace
num\_authors         &   -0.000109\sym{***}&   -0.000107\sym{***}&   -0.000105\sym{***}&   -0.000108\sym{***}\\
                    &    (-10.04)         &     (-9.94)         &     (-9.69)         &     (-9.87)         \\
\addlinespace
num\_authors\_squared &    9.76e-09\sym{***}&    9.98e-09\sym{***}&    1.06e-08\sym{***}&    1.11e-08\sym{***}\\
                    &      (7.33)         &      (7.50)         &      (7.81)         &      (8.13)         \\
\addlinespace
num\_unknown         &   0.0000863\sym{***}&   0.0000842\sym{***}&   0.0000806\sym{***}&   0.0000818\sym{***}\\
                    &      (8.53)         &      (8.34)         &      (7.99)         &      (8.06)         \\
\addlinespace
impact\_factor\_2y    &   -0.000269\sym{***}&   -0.000273\sym{***}&   -0.000276\sym{***}&   -0.000279\sym{***}\\
                    &    (-15.20)         &    (-15.43)         &    (-15.12)         &    (-15.10)         \\
\addlinespace
2y\_cites            &  -0.0000107         &  -0.0000106         &  -0.0000114         &  -0.0000118         \\
                    &     (-1.50)         &     (-1.49)         &     (-1.55)         &     (-1.58)         \\
\addlinespace
Constant            &    -0.00794\sym{***}&    -0.00699\sym{***}&    -0.00639\sym{***}&    -0.00732\sym{***}\\
                    &    (-13.55)         &    (-16.24)         &    (-22.27)         &    (-21.34)         \\
\midrule
Observations        &     3864455         &     3864455         &     3864455         &     3864455         \\
\bottomrule
\multicolumn{5}{l}{\footnotesize \textit{t} statistics in parentheses}\\
\multicolumn{5}{l}{\footnotesize \sym{*} \(p<0.05\), \sym{**} \(p<0.01\), \sym{***} \(p<0.001\)}\\
\end{tabular}
\end{adjustbox}
\end{table}

\begin{table}[H]\centering
\def\sym#1{\ifmmode^{#1}\else\(^{#1}\)\fi}
\caption{Linear regression model estimates for log of journal impact factor on surprise on multi-authored papers, with White standard errors. Research field fixed effects and dummies for publishing year are omitted.}\label{tab3_jif_nv}
\begin{adjustbox}{width=\textwidth}
\begin{tabular}{lcccc}
\toprule
                    &\multicolumn{1}{c}{(1)}&\multicolumn{1}{c}{(2)}&\multicolumn{1}{c}{(3)}&\multicolumn{1}{c}{(4)}\\
                    &\multicolumn{1}{c}{\textit{ln(2-year Journal Impact Factor)}}&\multicolumn{1}{c}{\textit{ln(2-year Journal Impact Factor)}}&\multicolumn{1}{c}{\textit{ln(5y Impact Factor)}}&\multicolumn{1}{c}{\textit{ln(5y Impact Factor)}}\\
\midrule

GenderHomogeneousTeam=1&     -0.0182\sym{*}  &     -0.0152\sym{*}  &     -0.0171\sym{*}  &     -0.0160\sym{*}  \\
                    &     (-2.27)         &     (-2.17)         &     (-2.13)         &     (-2.27)         \\
\addlinespace
\textit{Female Share}        &     0.00549         &    0.000454         &     0.00680         &    -0.00108         \\
                    &      (0.17)         &      (0.02)         &      (0.20)         &     (-0.04)         \\
\addlinespace
GenderHomogeneousTeam=1 $\times$ \textit{Female Share}&      0.0447\sym{***}&      0.0801\sym{***}&      0.0441\sym{***}&      0.0733\sym{***}\\
                    &      (6.39)         &     (14.56)         &      (6.32)         &     (13.32)         \\
\addlinespace
\textit{Surprise (References)}     &       0.281\sym{***}&                     &       0.303\sym{***}&                     \\
                    &     (29.58)         &                     &     (31.71)         &                     \\
\addlinespace
GenderHomogeneousTeam=1 $\times$ \textit{Surprise (References)} &     -0.0363\sym{***}&                     &     -0.0413\sym{***}&                     \\
                    &     (-3.78)         &                     &     (-4.28)         &                     \\
\addlinespace
\textit{Female Share} $\times$ \textit{Surprise (References)} &      -0.142\sym{**} &                     &      -0.161\sym{***}&                     \\
                    &     (-3.28)         &                     &     (-3.72)         &                     \\
\addlinespace
GenderHomogeneousTeam=1 $\times$ \textit{Female Share} $\times$ \textit{Surprise (References)} &      0.0782\sym{***}&                     &      0.0682\sym{***}&                     \\
                    &      (7.92)         &                     &      (6.92)         &                     \\
\addlinespace
\textit{Female Share}\_square $\times$ \textit{Surprise (References)} &      0.0347         &                     &      0.0694         &                     \\
                    &      (0.76)         &                     &      (1.53)         &                     \\
\addlinespace
DepSize&   0.0000118\sym{***}&   0.0000120\sym{***}&   0.0000120\sym{***}&   0.0000123\sym{***}\\
                    &    (167.38)         &    (170.28)         &    (171.76)         &    (174.79)         \\
\addlinespace
CareerAge          &     0.00102\sym{***}&    0.000962\sym{***}&     0.00106\sym{***}&     0.00100\sym{***}\\
                    &     (44.82)         &     (41.94)         &     (46.36)         &     (43.23)         \\
\addlinespace
OpenAccess=1             &       0.184\sym{***}&       0.193\sym{***}&       0.177\sym{***}&       0.187\sym{***}\\
                    &    (309.22)         &    (323.28)         &    (296.51)         &    (311.53)         \\
\addlinespace
num\_authors         &      0.0163\sym{***}&      0.0164\sym{***}&      0.0167\sym{***}&      0.0167\sym{***}\\
                    &     (28.68)         &     (28.79)         &     (28.65)         &     (28.77)         \\
\addlinespace
num\_authors\_squared & -0.00000158\sym{***}& -0.00000166\sym{***}& -0.00000152\sym{***}& -0.00000161\sym{***}\\
                    &    (-31.08)         &    (-31.58)         &    (-30.92)         &    (-31.48)         \\
\addlinespace
num\_unknown         &     -0.0122\sym{***}&     -0.0120\sym{***}&     -0.0127\sym{***}&     -0.0126\sym{***}\\
                    &    (-20.06)         &    (-19.77)         &    (-20.55)         &    (-20.24)         \\
\addlinespace
\textit{Surprise (concepts)}    &                     &      0.0943\sym{***}&                     &      0.0973\sym{***}\\
                    &                     &     (11.14)         &                     &     (11.45)         \\
\addlinespace
GenderHomogeneousTeam=1 $\times$ \textit{Surprise (concepts)} &                     &     -0.0486\sym{***}&                     &     -0.0509\sym{***}\\
                    &                     &     (-5.68)         &                     &     (-5.92)         \\
\addlinespace
\textit{Female Share} $\times$ \textit{Surprise (concepts)} &                     &      -0.150\sym{***}&                     &      -0.167\sym{***}\\
                    &                     &     (-3.92)         &                     &     (-4.34)         \\
\addlinespace
GenderHomogeneousTeam=1 $\times$ \textit{Female Share} $\times$ \textit{Surprise (concepts)} &                     &      0.0299\sym{***}&                     &      0.0293\sym{***}\\
                    &                     &      (3.46)         &                     &      (3.39)         \\
\addlinespace
\textit{Female Share}\_square $\times$ \textit{Surprise (concepts)} &                     &       0.125\sym{**} &                     &       0.139\sym{***}\\
                    &                     &      (3.11)         &                     &      (3.45)         \\
\addlinespace
Constant            &       0.983\sym{***}&       1.090\sym{***}&       1.018\sym{***}&       1.135\sym{***}\\
                    &    (112.74)         &    (139.34)         &    (115.59)         &    (143.27)         \\
\midrule
Observations        &     3861357         &     3861357         &     3864495         &     3864495         \\
\bottomrule
\multicolumn{5}{l}{\footnotesize \textit{t} statistics in parentheses}\\
\multicolumn{5}{l}{\footnotesize \sym{*} \(p<0.05\), \sym{**} \(p<0.01\), \sym{***} \(p<0.001\)}\\
\end{tabular}
\end{adjustbox}
\end{table}



\end{document}